\documentclass[english,journal=jctcce,manuscript=article,articletitle=true,layout=twocolumn]{achemso}

\usepackage[english]{babel}
\usepackage[T1]{fontenc}

\usepackage{amsmath}
\usepackage{amssymb}
\usepackage{graphicx}
\usepackage{float}
\usepackage[colorlinks=true, allcolors=blue]{hyperref}
\usepackage{cleveref}
\usepackage[version=4]{mhchem}
\usepackage{xcolor}
\usepackage{rotating}

\newcommand*\Erkale{\textsc{Erkale}}
\newcommand*\HelFEM{\textsc{HelFEM}}

\title{Insight on Gaussian Basis Set Truncation Errors in Weak to Intermediate Magnetic Fields with an Approximate Hamiltonian}

\author{Hugo {\AA}str{\"o}m}
\affiliation{University of Helsinki, Department of Chemistry, 
Faculty of Science, P.O. Box
  55 (A.I. Virtanens plats 1), FI-00014 University of Helsinki, Finland}
\author{Susi Lehtola}
\email{susi.lehtola@alumni.helsinki.fi}
\affiliation{University of Helsinki, Department of Chemistry, 
Faculty of Science, P.O. Box
  55 (A.I. Virtanens plats 1), FI-00014 University of Helsinki, Finland}

\begin{document}


\begin{abstract}
  Strong magnetic fields such as those found on white dwarfs have
  significant effects on the electronic structure of atoms and
  molecules.  However, the vast majority of molecular studies in the
  literature in such fields are carried out with Gaussian basis sets
  designed for zero field, leading to large basis set truncation
  errors [Lehtola et al, Mol. Phys. 2020, 118, e1597989].  In this
  work, we aim to identify the failures of the Gaussian basis sets in
  atomic calculations to guide the design of new basis sets for strong
  magnetic fields.  We achieve this by performing fully numerical
  electronic structure calculations at the complete basis set (CBS)
  limit for the ground state and low lying excited states of the atoms
  $1 \leq Z\leq18$ in weak to intermediate magnetic fields. We also
  carry out finite-field calculations for a variety of Gaussian basis
  sets, introducing a real-orbital approximation for the magnetic-field
  Hamiltonian. Our primary focus is on the aug-cc-pVTZ basis set,
  which has been used in many works in the literature. A study of the
  differences in total energies of the fully numerical CBS limit
  calculations and the approximate Gaussian basis calculations is
  carried out to provide insight into basis set truncation
  errors. Examining a variety of states over the range of magnetic
  field strengths from $B=0$ to $B=0.6 B_0$, we observe significant
  differences for the aug-cc-pVTZ basis set, while much smaller errors
  are afforded by the benchmark-quality AHGBSP3-9 basis set [Lehtola,
    J. Chem. Phys. 2020, 152, 134108]. This suggests that there is
  considerable room to improve Gaussian basis sets for calculations at
  finite magnetic fields.
\end{abstract}

\newcommand*\citeref[1]{ref.~\citenum{#1}}
\newcommand*\citerefs[1]{refs.~\citenum{#1}}

\section{Introduction}
\label{sec:introduction}

The behavior of atoms and molecules in strong magnetic fields is of
interest in astrochemistry and astrophysics, as magnetic neutron stars
and white dwarfs exhibit magnetic fields with strengths of the order of 1
$B_0\approx2.35\times10^{5}$ T. Such fields are well-known to cause significant
changes in the electronic structure.\cite{Angel1977_AJ_1,
  Schmidt1995_AJ_274, Jordan1998__, Wickramasinghe2000_PASP_873,
  Liebert2003_AJ_2521} Moreover, as these field strengths are several orders
of magnitude larger than what can be achieved in experiments on Earth,
a computational approach is required to study their effects.

Many methods for performing quantum chemical electronic structure
calculations at finite magnetic fields have been developed in recent
years.\cite{Tellgren2008_JCP_154114, Tellgren2012_PCCP_9492,
  Furness2015_JCTC_4169, Reynolds2015_PCCP_14280, Irons2017_JCTC_3636,
  Reimann2017_JCTC_4089, Hampe2017_JCP_154105, Reynolds2018_JCP_14106,
  Sen2018_JCP_184112, Sun2018_JCTC_348, Holzer2019_JCP_214112,
  Sun2019_JCTC_3162, Hampe2019_JCTC_4036, WilliamsYoung2020_WCMS_1436,
  Holzer2021_FC_746162, Pausch2022_JCTC_3747, Pausch2022_JPCL_4335,
  Monzel2022_JCP_54106, Blaschke2022_JCP_44115, Speake2022_JCTC_7412,
  Culpitt2023_JCP_114115, Holzer2023_JCTC_3131} These studies have
revealed the spectrum of white dwarfs, \cite{Greenstein1984_AJ_47,
  Greenstein1985_AJ_25, Hollands2023_MNRAS_3560} as well as a vast
richness of new chemistry, such as the paramagnetic bonding mechanism
of \ce{H2} and \ce{He2} \cite{Lange2012_S_327} and the lattice
structure of \ce{He} atoms in strong magnetic fields
\cite{Tellgren2012_PCCP_9492}.

A central aspect of all electronic structure calculations is the
choice of the one-electron basis set. In the weak field region,
isotropic Gaussian-type orbitals (GTOs) are a good choice due to their
long history in quantum chemistry: a richness of GTO basis sets has
been developed for a variety of purposes.\cite{Davidson1986_CR_681,
  Jensen2013_WIRCMS_273, Hill2013_IJQC_21} GTOs enjoy an overwhelming
popularity in the literature also in finite-field calculations at
various levels of theory, including Hartree--Fock (HF)
\cite{Jones1999_PRA_2875, Tellgren2008_JCP_154114,
  Tellgren2012_PCCP_9492, Sen2018_JCP_184112, Sun2018_JCTC_348},
density-functional theory (DFT) \cite{Furness2015_JCTC_4169,
  Reimann2017_JCTC_4089}, coupled cluster (CC)
theory,\cite{Hampe2017_JCP_154105, Hampe2019_JCTC_4036,
  Hampe2020_PCCP_23522, Stopkowicz2015_JCP_74110} as well as
configuration interaction (CI) theory.\cite{Detmer1997_PRA_1825,
  Detmer1998_PRA_1767, Schmelcher1999_PRA_3424,
  Becken1999_JPBAMOP_1557, Becken2000_JPBAMOP_545,
  Becken2001_PRA_53412, AlHujaj2004_PRA_33411, AlHujaj2004_PRA_23411,
  Lange2012_S_327}

However, the choice of the one-electron basis set requires special
attention when studying atoms and molecules in magnetic fields: as was
already mentioned above, the field affects the electronic structure.
One of these effects is that when the magnetic field is turned on, the
atomic orbitals that are spherical at zero field become
cylindrical. Isotropic GTOs are therefore not well-suited for
describing the electronic structure in strong magnetic fields, as we
have demonstrated by the existence of large basis set truncation
errors in diatomic molecules.\cite{Lehtola2020_MP_1597989}

An alternative is to use anisotropic GTOs;\cite{Aldrich1979_PSS_343,
  Schmelcher1988_PRA_672} however, they introduce new types of
challenges. As the magnetic field interaction confines movement in the
direction orthogonal to the field (see \cref{sec:theory}), the
anisotropic GTO basis set splits the exponents in the directions
parallel and orthogonal to the field, which complicates the
optimization of the exponents.\cite{Kubo2007_JPCA_5572,
  Zhu2014_PRA_22504, Zhu2017_JCP_244108} Moreover, because the basis
set is formed by the product of these two sets of exponents, the
number of basis functions explodes if the basis is required to be
accurate for a range of magnetic field strengths. The use of
anisotropic GTOs also requires dedicated
approaches,\cite{Aldrich1979_PSS_343, Schmelcher1988_PRA_672} and such
basis functions are supported in few programs.

The wide support for isotropic GTOs in quantum chemical packages
motivate their continued use at intermediate field strengths: as long
as the field is not too strong, the cylindrical distortion to the
atomic orbitals at the finite field can be recovered by including
additional polarization functions, analogously to the manner in which
the linear combination of atomic orbitals (LCAO) works at zero fields
to model polyatomic systems in which the atomic symmetry is similarly
lifted.

This begs the question: can standard isotropic GTO basis sets be
modified to better suit calculations in external magnetic fields? The
first step towards the answer is to identify the failures in standard
Gaussian basis sets, which is the focus of this work.  We will
introduce an approximation for performing finite-field calculations
with Gaussian basis sets that is compatible with established
Gaussian-basis methodology, that is, the use of real-orbital orbitals
and orbital coefficients. Employing this approximation and a
computational approach the senior author has recently
developed,\cite{Lehtola2019_IJQC_25945, Lehtola2020_MP_1597989} we
will study shortcomings of existing Gaussian basis sets by examining
differences in total energies observed for the various low-lying
electronic states of atoms.

We will also carry out fully numerical
calculations\cite{Lehtola2019_IJQC_25968} for atoms in finite fields
with the Hartree--Fock (HF) method with complex atomic orbitals,
enabling us to determine the complete basis set (CBS) limit for a
number of single-determinant HF states over a range of magnetic
fields. Similar calculations have been previously reported on the
series of neutral atoms from H to Ne and their singly positive ions in
a larger range of magnetic fields;\cite{Ivanov1998_PRA_3793,
  Ivanov1999_PRA_3558, Ivanov2001_JPBAMOP_2031, Ivanov2001_EPJD_279,
  Ivanov2000_PRA_22505, Ivanov2001_AQC_361} in this work, we will
examine the whole series of atoms from H to Ar and focus on weak
to intermediate fields.

Equipped with the CBS limit data for the complex wave functions for a
number of configurations, yielding numerically exact HF energies, we
will determine the field-dependent differences in total energies in a
wide variety of isotropic GTO basis sets for these configurations,
enabling us to assess both the accuracy of the GTO basis and the
employed real-orbital approximation. Especially, we rely on the large
benchmark-quality GTO basis sets of \citeref{Lehtola2020_JCP_134108}
for insight onto the limitations of isotropic GTO basis sets. These
large basis sets enable us to identify states that are poorly
described by commonly-used basis sets optimized for field-free
calculations, and thereby allow the identification of the kinds of
exponents that should be included in future isotropic basis sets for
finite-field calculations.

The outline of this work is as follows. We outline the theory behind
the methods used in this study---the Hamiltonian in a finite magnetic
field, and the numerical methods employed in this work---in
\cref{sec:theory}; especially, the real-orbital approximation is
discussed in \cref{sec:real-approx}. We discuss the computational
details in \cref{sec:computational-details}, followed by the results
of the calculations in \cref{sec:results}. We end the article with a
summary and discussion in \cref{sec:summary}. Atomic units are used
throughout the work: the magnitude of the magnetic field is given in
units of $B_0$ and energy in units of $E_h$.

\section{Theory}
\label{sec:theory}

We have previously discussed electronic structure calculations in the
presence of an external magnetic field in
\citeref{Lehtola2020_MP_1597989}. Following the same outline, we
employ a Hamiltonian of the form
\begin{equation}\label{eq:hamilt}
H=H_0+\frac{1}{2}BL_z+BS_z+\frac{1}{8}B^2(x^2+y^2),
\end{equation}
where the terms linear in $B$ are the orbital and spin Zeeman terms,
respectively, which are responsible for the paramagnetic response that
can either increase or decrease the energy of the system relative to
zero magnetic field. The quadratic term in \cref{eq:hamilt} leads to
diamagnetic response that always increases the energy of the system
relative to the zero field case. It also acts as a confinining
potential for the orbitals in the $(x,y)$ plane, which leads to the
orbitals ballooning in the direction of the magnetic field, which is
chosen to coincide with the $z$ axis in \cref{eq:hamilt}.

The fully numerical calculations are pursued in atomic orbital basis
sets of the form
\begin{equation}
  \label{eq:fem-ao}
  \psi_{nlm}(\mathbf{r})=r^{-1}B_{n}(r)Y_{l}^{m}(\theta,\varphi),
\end{equation}
where $B_\mathrm{n}(r)$ are piecewise polynomial basis functions of
the finite element method (FEM), and
$Y_\mathrm{l}^\mathrm{m}(\theta,\varphi)$ are complex spherical
harmonics; we refer to the earlier literature on discussion on the FEM
approach.\cite{Lehtola2019_IJQC_25945, Lehtola2019_IJQC_25968,
  Lehtola2020_MP_1597989} The evaluation of the magnetic field terms
in the Hamiltonian of \cref{eq:hamilt} with respect to basis functions
of the type of \cref{eq:fem-ao} has been described in
\citeref{Lehtola2020_MP_1597989}.

The GTOs calculations, on the other hand, are pursued with basis
functions of the type
\begin{equation}
    \label{eq:gto-ao}
\psi_{nlm}(\mathbf{r})= \left[r^l e^{-\alpha_{nl} {\bf r}^2}\right]
Y_{lm}(\theta, \varphi),
\end{equation}
where $Y_{lm}$ are spherical harmonics of the real form. In accordance
with standard practices of finite field calculations, we therefore use
uncontracted GTO basis sets to allow better flexibility to the wave
function to adapt to the finite magnetic field.

We note that the use of London atomic orbitals
(LAOs),\cite{London1937_JPlR_397, Pople1962_JCP_53} also known as
gauge-including atomic orbitals (GIAOs),\cite{Ditchfield1974_MP_789}
is important in the general case at finite magnetic field. In the
LAO/GIAO approach, one includes a magnetic gauge factor in the
definition of the atomic-orbital basis functions
\begin{equation}
    \label{eq:giao}
\psi_{nlm}(\mathbf{r})= \exp [-i{\bf B} \times {\bf R} \cdot {\bf r} /(2c) ] \psi^0_{nlm}(\mathbf{r}),
\end{equation}
where $\psi^0_{nlm}$ is the zero-field basis function centered at
${\bf R}$, ${\bf B}$ is the magnetic field, and $c$ is the speed of
light. However, the gauge factor in \cref{eq:giao} yields unity in the
case of linear molecules in a parallel field---such as the case of the
diatomic molecules previously studied in
\citeref{Lehtola2020_MP_1597989}---as well as in the present case of
atoms where the basis functions are located at the origin. The
calculations of this work and \citeref{Lehtola2020_MP_1597989}
are therefore of LAO/GIAO quality.

\subsection{Real-Orbital Approximation}
\label{sec:real-approx}

Even the calculations in the GTO basis sets are carried out with
real-orbital basis functions, while the magnetic field interaction
matrix elements are defined in terms of complex GTOs, for the purposes
of this study we chose to disregard this difference and carry out
approximate calculations, instead.

The real-orbital approximation employed in this work consists of
reusing the magnetic field interaction matrix elements derived for
complex-valued spherical harmonic $Y_{l}^{m}$ basis functions in the
basis of the real-orbital $Y_{lm}$ basis functions that are actually
used in the calculations.

While this approximation may seem coarse, it avoids the need to deal
with complex basis functions or complex expansion coefficients
altogether, allowing the reuse of field-free machinery. The
approximation is also exact in a number of cases. The energy is exact
for all states with only $\sigma$ orbitals, since $Y_{l 0} =
Y_{l}^{0}$. The energy is also exact for states occupying
$\pi,\delta,\varphi,\dots$ orbitals, if both the $|m|$ and $-|m|$
magnetic subchannels are equally occupied, because the resulting
density is cylindrically symmetric. An analogous approximation was
used in \citeref{Lehtola2019_IJQC_25944} to implement linear molecule
symmetry in \Erkale{}, as it is exact for the $\Sigma$ states that
were considered in that work.

We will find below that this real-orbital approximation does afford an
excellent level of accuracy for many states, and that it captures the
most important effects of the magnetic field on the basis functions.

\section{Computational Details}
\label{sec:computational-details}

The fully numerical calculations are performed with the \HelFEM{}
program.\cite{Lehtola2019_IJQC_25945, Lehtola2019_IJQC_25944,
  Lehtola2020_MP_1597989, Lehtola2018__} We employed five radial
elements with shape functions determined by 15-node Legendre
interpolating polynomials (LIPs) defined by Gauss--Lobatto quadrature
nodes and a practical infinity $r_\infty = 40 a_0$, which was found to
afford the CBS limit for the studied systems. The sole exception was
the field-free ($B=0$) calculations, for which the high-lying excited
states are extremely diffuse, and the calculations used seven radial
elements and $r_\infty=100a_0$, instead.

All calculations in this work are performed at the unrestricted HF
(UHF) level of theory, where all spatial and spin restrictions on the
atomic orbitals are let go. The resulting atomic UHF configurations
can be identified by their symmetry: the configuration is fully
specified by the number of alpha and beta electrons for each value of
$m$, which is the quantum number resulting from Noether's theorem that
describes the orbital's symmetry around the magnetic field axis, as
the corresponding angle does not appear in the
Hamiltonian.\cite{Lehtola2019_IJQC_25968} The value $m=0$ corresponds
to $\sigma$ orbitals, $m=\pm1$ to $\pi$ orbitals, $m=\pm 2$ to
$\delta$ orbitals, $m=\pm3$ to $\phi$ orbitals, etc.

To keep the notation more compact, we will denote the configurations
with the following notation
\begin{equation}
\prod_{m\in\{\sigma,\pi,\delta,\phi\}} m_{+/-}^{n_\alpha,n_\beta}, \label{eq:statelabel}
\end{equation}
where $+/-$ indicates the sign of the $m$ and $n_\alpha$ and $n_\beta$
are the number of $\alpha$ and $\beta$ electrons occupying orbitals
with this value of $m$. As an example, the zero-field UHF ground state
of \ce{F} with orbital occupations
$1\sigma^{2}2\sigma^{2}3\sigma^{1}1\pi_+^{2}1\pi_-^{2}$ would be
written out in our compact notation as
$\sigma^{3,2}\pi_+^{1,1}\pi_-^{1,1}$. (Note, however, that the
$\alpha$ and $\beta$ spatial orbitals may differ in the UHF
calculations!)

The work began by identifying the atomic configurations of interest in
the range of studied fields, $B \in [0,0.60B_0]$.  An automated Python
program was employed to find the lowest 3 states at each of the field
strengths $B = [0,0.10B_0,\dots,0.60B_0]$. The logic employed was to
generate trial configurations from the current-lowest UHF
configuration by moving one electron at a time, allowing changes in
$m$ as well as flips of the electron's spin.

Because at this stage the states' total energies are not of
importance---only the relative ranking of the configurations
is---while the number of candidate configurations is large, this part
of the study employed a smaller numerical basis set than the CBS limit
calculations described above. Approximate total energies were computed
for each group of candidate configurations by beginning from the
smallest possible value of $l$ able to describe all the configurations
in the group, $l_\mathrm{min}=\max_{i \sigma}|m_{i\sigma}|$, where
$m_{i \sigma}$ is the $m$ quantum number of occupied orbital $i$ with
spin $\sigma$. The set of three lowest configurations was then
converged by increasing the truncation with the $l$ quantum number by
2, that is, until the same set of three lowest-lying configurations is
obtained with $l_\text{max}=l_\text{max}^0$ and
$l_\text{max}=l_\text{max}^0+2$. (If the lowest-lying configuration
changes at this stage, the procedure automatically restarts from the
generation of a new set of candidate configurations.)

In the second stage, calculations were performed at the CBS limit for
all of the low-lying configurations determined in the initial
step. The procedure to converge the numerical basis was the same as in
the first part, but now the procedure is carried out for each state
separately until its total energy converges to the to the threshold of
$10^{-6} E_h$; the most difficult states required $l_\text{max}=21$ to
reach this criterion for the CBS limit. Because HF energies are
well-known to exhibit exponential convergence in the basis set, we are
confident that our energies are accurate to $\mu E_h$ precision; this
exponential convergence is also demonstrated graphically in the
Supporting Information (SI).

We used a superposition of atomic potentials (SAP)
\cite{Lehtola2019_JCTC_1593, Lehtola2020_JCP_144105} as the initial
guess in the FEM calculations. The only exception is the
$\sigma^{2,0}$ state of He, for which the SAP guess was found to lead
to saddle point convergence at stronger fields, $B \gtrsim
0.3B_0$, and for which the core guess was employed instead.

The Gaussian-basis calculations were performed with the \Erkale{}
program \cite{Lehtola2012_JCC_1572}. Calculations were carried out
with the double-$\zeta$ (D) to quintuple-$\zeta$ (5)
correlation-consistent cc-pVXZ and aug-cc-pVXZ basis sets
\cite{Dunning1989_JCP_1007, Kendall1992_JCP_6796, Woon1993_JCP_1358,
  Peterson1994_JCP_7410}, which have been commonly used in the
literature at finite magnetic fields.

By the request of a reviewer, we also performed calculations with the
def2-TZVP\cite{Weigend2005_PCCP_305} and the
6-311++G(3df,3pd)\cite{Krishnan1980_JCP_650} basis set for
completeness, because these basis sets have also been used in some
studies of the literature, even though the latter is an obsolete basis
set which should not be used.\cite{Grev1989_JCP_7305,
  Moran2006_JACS_9342} The results for these basis sets are not
discussed in the main text, but they are available in the Supporting
Information.

Importantly, both \HelFEM{} and \Erkale{} are free and open-source
software,\cite{Lehtola2022_WIRCMS_1610} and are publicly available on
GitHub.  As was already mentioned above in \cref{sec:theory}, all GTO
basis sets are employed in fully uncontracted form.

As was already mentioned above in \cref{sec:introduction}, an
approximate Gaussian basis set limit was determined with the benchmark
quality hydrogenic Gaussian basis sets (HGBS) of
\citeref{Lehtola2020_JCP_134108}. The HGBS basis sets are modular and
determined with one-electron calculations, only: the basis set for
angular momentum $l$ of the element $Z$ is determined with
calculations on the one-electron ions $Z^{(Z-1)+}$, $\dots$, \ce{He+},
and H.

Requiring that the energy of the lowest-energy orbital of each angular
momentum is converged to the relative accuracy $10^{-n}$ with respect
the number of even-tempered exponents on the shell yields the HGBS-$n$
basis set.\cite{Lehtola2020_JCP_134108} Augmented versions of the
HGBS-$n$ basis sets---the AHGBS-$n$ basis sets---are obtained by
extending the consideration to also the fictitious single-electron ion
with nuclear charge $Z=1/2$.\cite{Lehtola2020_JCP_134108} Finally,
polarized versions of the basis sets are obtained by adding higher
angular momentum shells, the basis with $m$ added polarization shells
being denoted as (A)HGBSP$m$-$n$ basis
sets.\cite{Lehtola2020_JCP_134108}

In this work, we will examine the (A)HGBSP$m$-$n$ basis sets with $m
\in\{1,2,3\}$ and $n\in\{5,7,9\}$.  Comparing data at fixed $m$ and
growing $n$ demonstrates convergence with respect to the radial
expansion, whereas comparison of data for fixed $n$ and growing $m$
demonstrates converegence with respect to the angular momentum
expansion. The utility of the HGBS basis sets is exactly their modular
nature: the basis sets can be straightforwardly determined for an
arbitrary element, an arbitrary precision, and an arbitrary number of
polarization shells.

We found many of the Gaussian-basis calculations to be sensitive to
saddle point convergence, even with the explicit handling of the
orbital symmetry with respect to the magnetic field axis. This issue
was diagnosed from discontinuities in plots of the total energy as a
function of the strength of the magnetic field.  We were able to
circumvent most of this issue by calculating each state of each atom
on a hysteresis curve. Starting from the converged field-free
calculation, we ran calculations in increasing field strength by
reading in the orbitals from the previous calculation as the initial
guess. This gave us one set of solutions. In the next step, we
repeated the calculations in the opposite direction: starting from the
strongest field, we ran a new set of calculations in decreasing field
strength by reading in the orbitals from the previous field strength
as the initial guess.  The reported energies at each field strength
were then obtained by choosing the minimal energy of these two
calculations.

Finally, given the CBS limit energy from the \HelFEM{} calculation and
the GTO energy from \Erkale{} for a given state and field strength
$B$, $E^\text{CBS}(B)$ and $E^\mathrm{GTO}(B)$, respectively, we
calculate differences in energy as
\begin{equation}
\Delta E^\text{GTO} (B)=E^\mathrm{GTO}(B)-E^\text{CBS}(B). \label{eq:bste}
\end{equation}
In the case where the approximation discussed in
\cref{sec:real-approx} is exact, the difference measured by
\cref{eq:bste} is a metric of basis set truncation error (BSTE) and
$\Delta E^\text{GTO}$ is positive, $\Delta E^\text{GTO} > 0$.

Note, however, that a positive $\Delta E^\text{GTO}$ may also be
obtained when the approximation is not exact. Indeed, we observe that
this difference can take either sign for the configurations examined
in this work. Such differences are usually observed already at zero
field, indicating that the differences between the FEM and GTO
calculations arise already from the differences in handling the
orbital symmetries in these two programs, as was discussed above in
\cref{sec:real-approx}.

To simplify the analysis, we will furthermore average the energy
difference over the magnetic field:
\begin{equation}
\Delta E^\text{GTO} =\frac 1 N \sum_{i=1}^N |\Delta E^\text{GTO} (B_i)|, \label{eq:mbste}
\end{equation}
where the average is performed with respect to the $N=7$ considered
values for the magnetic field strength $B_i \in\{0,0.10B_0,\dots,0.60
B_0\}$.  A comparison of the mean absolute energy differences (MAEDs)
defined by \cref{eq:mbste} allows a straightforward identification of
states that are ill-described by the studied GTO basis.

\section{Results}
\label{sec:results}

The results for all basis sets are available in the Supporting
Information (SI). Due to the large amount of data, we will limit the
discussion to results obtained with the aug-cc-pVTZ basis set, which
generally provides a good balance between cost and accuracy at zero
field and is therefore considered an attractive choice in most GTO
calculations. This basis set was also used in our previous work on
diatomic molecules in \citeref{Lehtola2020_MP_1597989}, and has been
employed in studies by other authors as well in the
literature.\cite{Lange2012_S_327, Tellgren2012_PCCP_9492,
  Stopkowicz2015_JCP_74110, Reimann2017_JCTC_4089, Reimann2019_MP_97}

The AHGBSP3-9 basis set\cite{Lehtola2020_JCP_134108} is the largest
GTO basis considered in this work, and we use it to represent a
feasible limit for GTO basis sets in the discussion.  The comparison
of the aug-cc-pVTZ and AHGBSP3-9 results then affords insights onto
the limitations of GTO basis sets in finite field calculations when
GIAOs/LAOs are employed. As we believe AHGBSP3-9 to be close to the
CBS limit for GTOs, an optimized GTO basis set for finite fields
should be able to get close to the AHGBSP3-9 values with considerably
fewer basis functions, showing a marked improvement on the aug-cc-pVTZ
values which are limited by the basis set designed for field-free
calculations.


To aid the discussion on the real-orbital approximation of
\cref{sec:real-approx}, we will show differences in total energies in
{\color{blue} blue}, if the energy difference between the FEM and
AHGBSP3-9 calculations are positive at all studied magnetic field
strenghts, and in {\color{red} red} if the difference is negative for
at least one field strength; we note that the latter usually happens
already at zero field.

\paragraph{H  \label{sec:H}}

\begin{figure*}
\begin{center}
\includegraphics[width=0.45\linewidth]{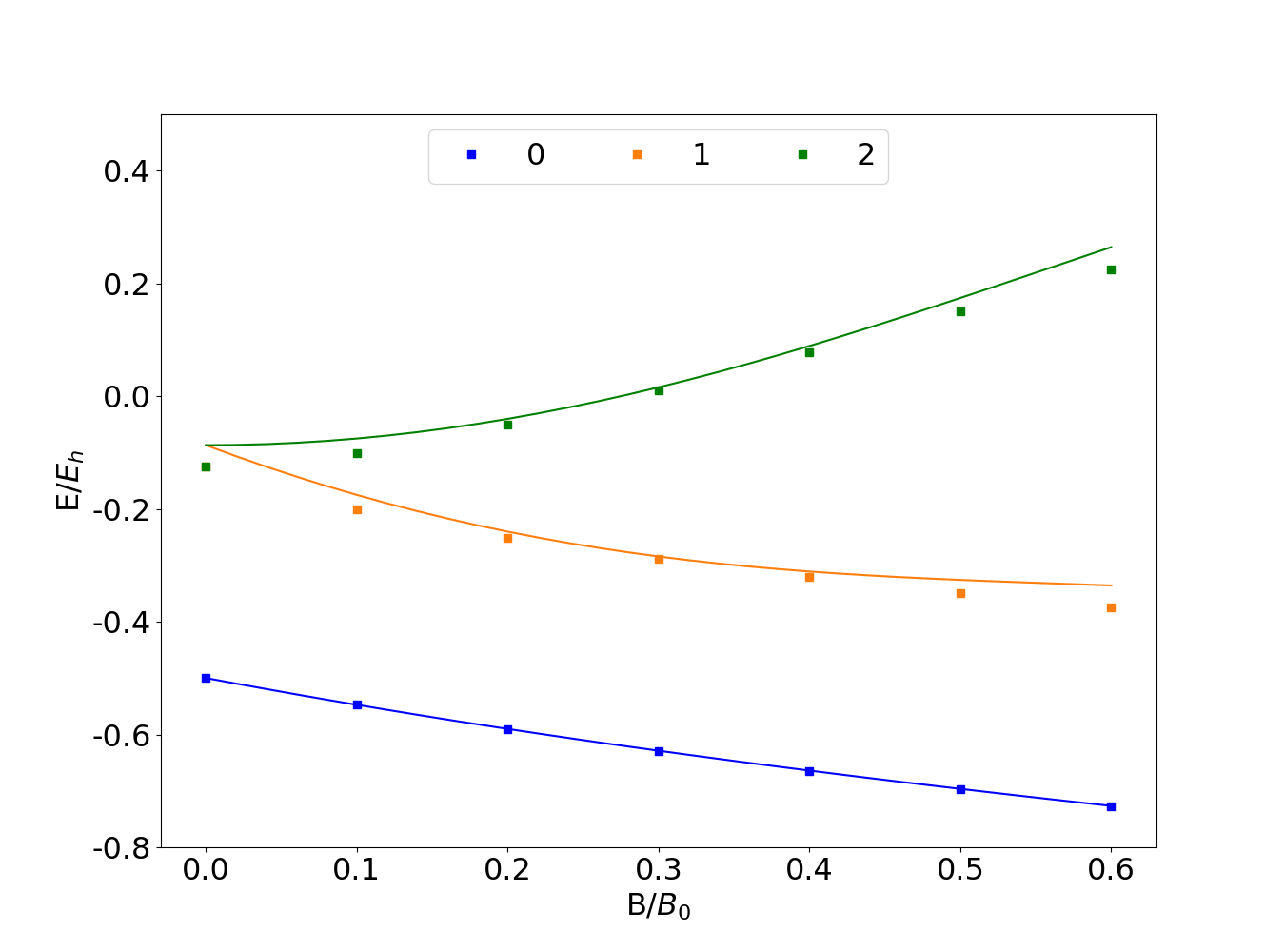}
\includegraphics[width=0.45\linewidth]{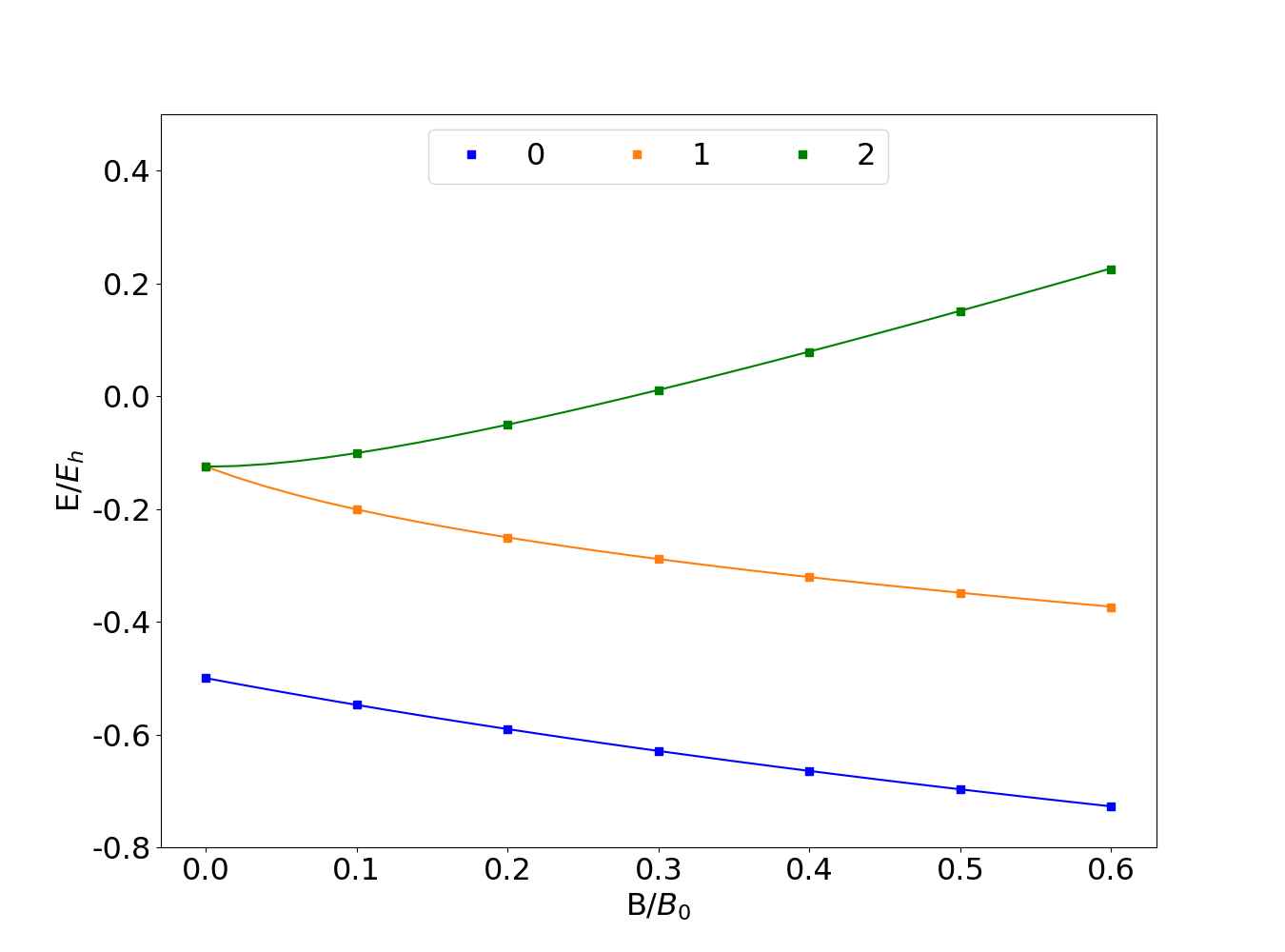}
\caption{Total energy of the H atom as a function of the magnetic field strength $B$ in the aug-cc-pVTZ (left) and AHGBSP3-9 (right) basis sets.}
\label{fig:H}
\end{center}
\end{figure*}

\begin{table}
\centering
\small
\begin{tabular}{llcc}
\hline
 & state & aug-cc-pVTZ & AHGBSP3-9\\ 
\hline \hline
0 & $\sigma^{1,0}$ & \color{blue}{ $ 0.447 $ } & \color{blue}{ $ 0.007 $ }\\ 
1 & $\pi_-^{1,0}$ & \color{blue}{ $ 21.719 $ } & \color{blue}{ $ 0.430 $ }\\ 
2 & $\pi_+^{1,0}$ & \color{blue}{ $ 21.719 $ } & \color{blue}{ $ 0.430 $ }\\ 
\hline
\end{tabular}
\caption{MAEDs between GTO and FEM energies in m$E_h$ for H in the fully uncontracted aug-cc-pVTZ and AHGBSP3-9 basis sets.}
\label{tab:H-mean-differ}
\end{table}

The energies of the low-lying states of the H atom are shown as a
function of the field strength in \cref{fig:H}. The mean differences
between the FEM and GTO energies are shown in
\cref{tab:H-mean-differ}.  The $1\sigma$ state of H, which is also the
ground state throughout the range of field strengths considered in
this work, is qualitatively well described by both aug-cc-pVTZ and
AHGBSP3-9, the latter affording much lower MAEDs.

The BSTEs for the $\Pi$ states of H are significant in the aug-cc-pVTZ
basis set. We also see from \cref{fig:H} that the BSTE for the $\Pi$
states in the aug-cc-pVTZ basis have a minimum around $B = 0.3B_0$,
which likely arises from fortuitious error cancellation. The energy
differences for the $\Pi$ states are negligible in the AHGBSP3-9 basis
set.

We also note that even the aug-cc-pV5Z basis set exhibits significant
differences for the $\Pi$ states, which are visually discernible in
the plots included in the SI, while the AHGBSP3-9 data appear spot-on.
Thes results suggest that the description of the $\Pi$ states for H
could be significantly improved for finite field calculations in
standard basis sets by adding more $p$ and higher functions to improve
the description of the $\pi$ orbital .

\paragraph{He \label{sec:He}}

\begin{figure*}
\begin{center}
\includegraphics[width=0.45\linewidth]{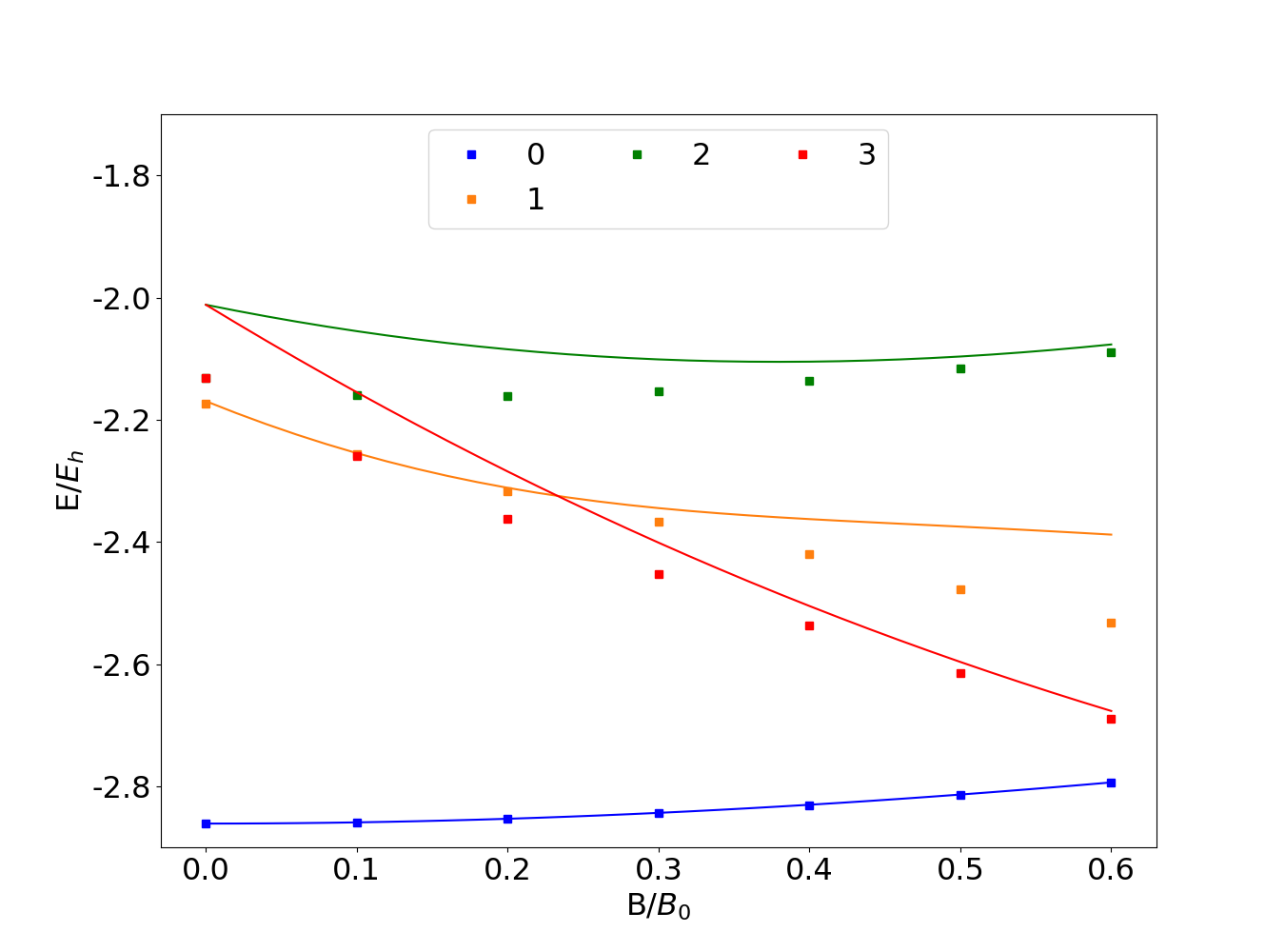}
\includegraphics[width=0.45\linewidth]{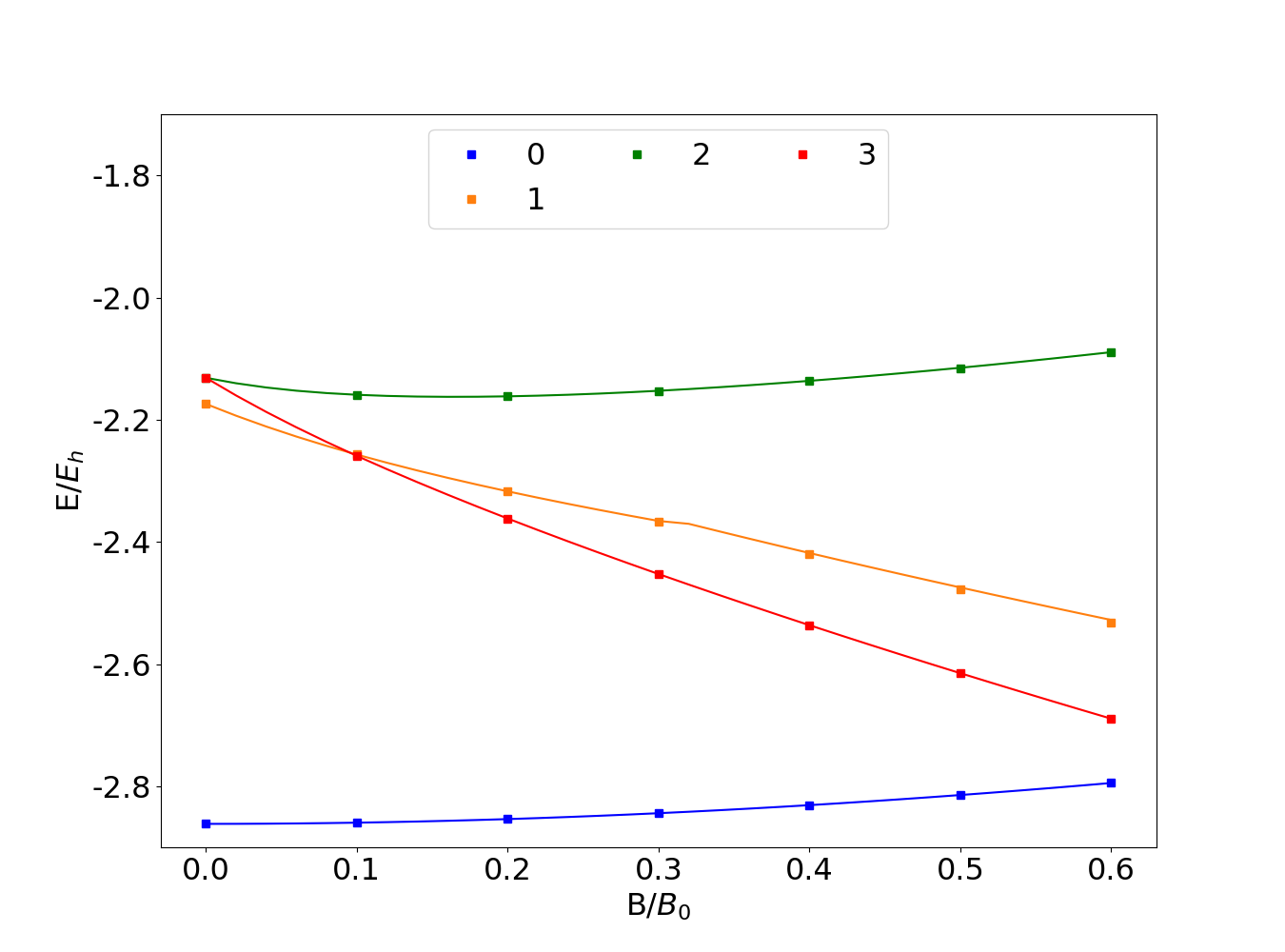}
\caption{Total energy of the He atom as a function of the magnetic field strength $B$ in the aug-cc-pVTZ (left) and AHGBSP3-9 (right) basis sets.}
\label{fig:He}
\end{center}
\end{figure*}

\begin{table}
\centering
\small
\begin{tabular}{llcc}
\hline
 & state & aug-cc-pVTZ & AHGBSP3-9\\ 
\hline \hline
0 & $\sigma^{1,1}$ & \color{blue}{ $ 0.635 $ } & \color{blue}{ $ 0.000 $ }\\ 
1 & $\sigma^{2,0}$ & \color{blue}{ $ 48.331 $ } & \color{blue}{ $ 1.401 $ }\\ 
2 & $\sigma^{1,0}\pi_+^{1,0}$ & \color{blue}{ $ 59.467 $ } & \color{blue}{ $ 0.266 $ }\\ 
3 & $\sigma^{1,0}\pi_-^{1,0}$ & \color{blue}{ $ 59.467 $ } & \color{blue}{ $ 0.266 $ }\\ 
\hline
\end{tabular}
\caption{MAEDs between GTO and FEM energies in m$E_h$ for He in the fully uncontracted aug-cc-pVTZ and AHGBSP3-9 basis sets.}
\label{tab:He-mean-differ}
\end{table}

The energies of the low-lying states of the He atom are shown as a
function of the field strength in \cref{fig:He}. The mean differences
between the FEM and GTO energies are shown in
\cref{tab:He-mean-differ}. Similarly to H, He does not exhibit ground
state crossings at the observed range of field strengths. The
$1\sigma^2$ ground state configuration is again qualitatively well
described by both GTO basis sets, with AHGBSP3-9 affording much
smaller errors.

The $\Pi$ states exhibit large energy differences in the aug-cc-pVTZ
basis set in the weak field regime, but the differences decrease at
stronger fields, indicating that the spatial shape of the orbitals can
be better described in the aug-cc-pVTZ basis at the relatively
stronger fields.

The AHGBSP3-9 basis set, in contrast, again affords much lower MAEDs
for the $\Pi$ states, as well. This again indicates room to improve on
the standard basis sets for finite-field calculations.

The lowering of the $\sigma^{2,0}$ triplet state in increasing field
strength is described extremely poorly by the aug-cc-pVTZ basis set,
but better recovered by AHGBSP3-9. Also the aug-cc-pV5Z basis set
gives poor results for the $\sigma^{2,0}$ state, as shown by the data
in the SI.

\paragraph{Li \label{sec:Li}}

\begin{figure*}
\begin{center}
\includegraphics[width=0.45\linewidth]{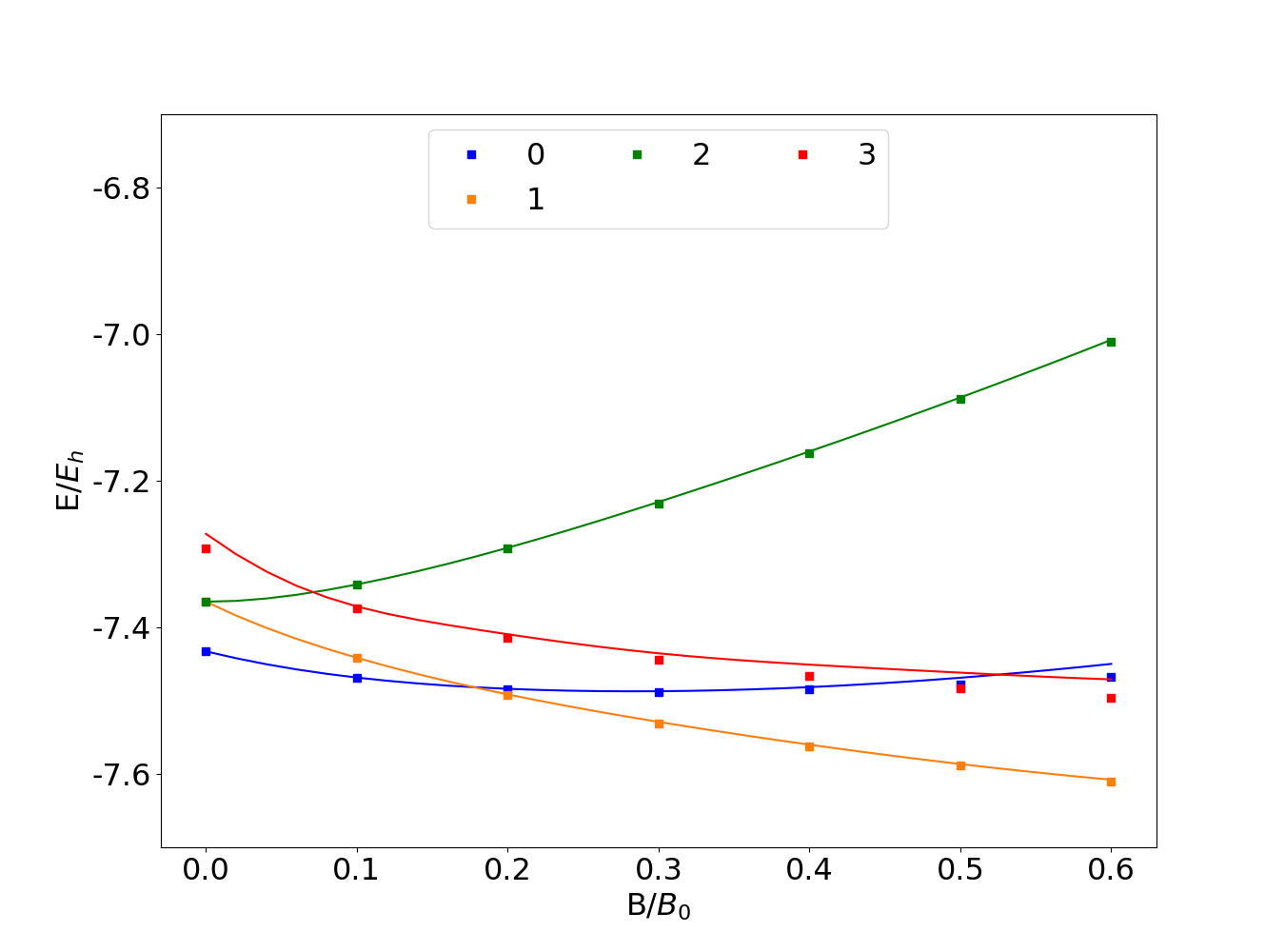}
\includegraphics[width=0.45\linewidth]{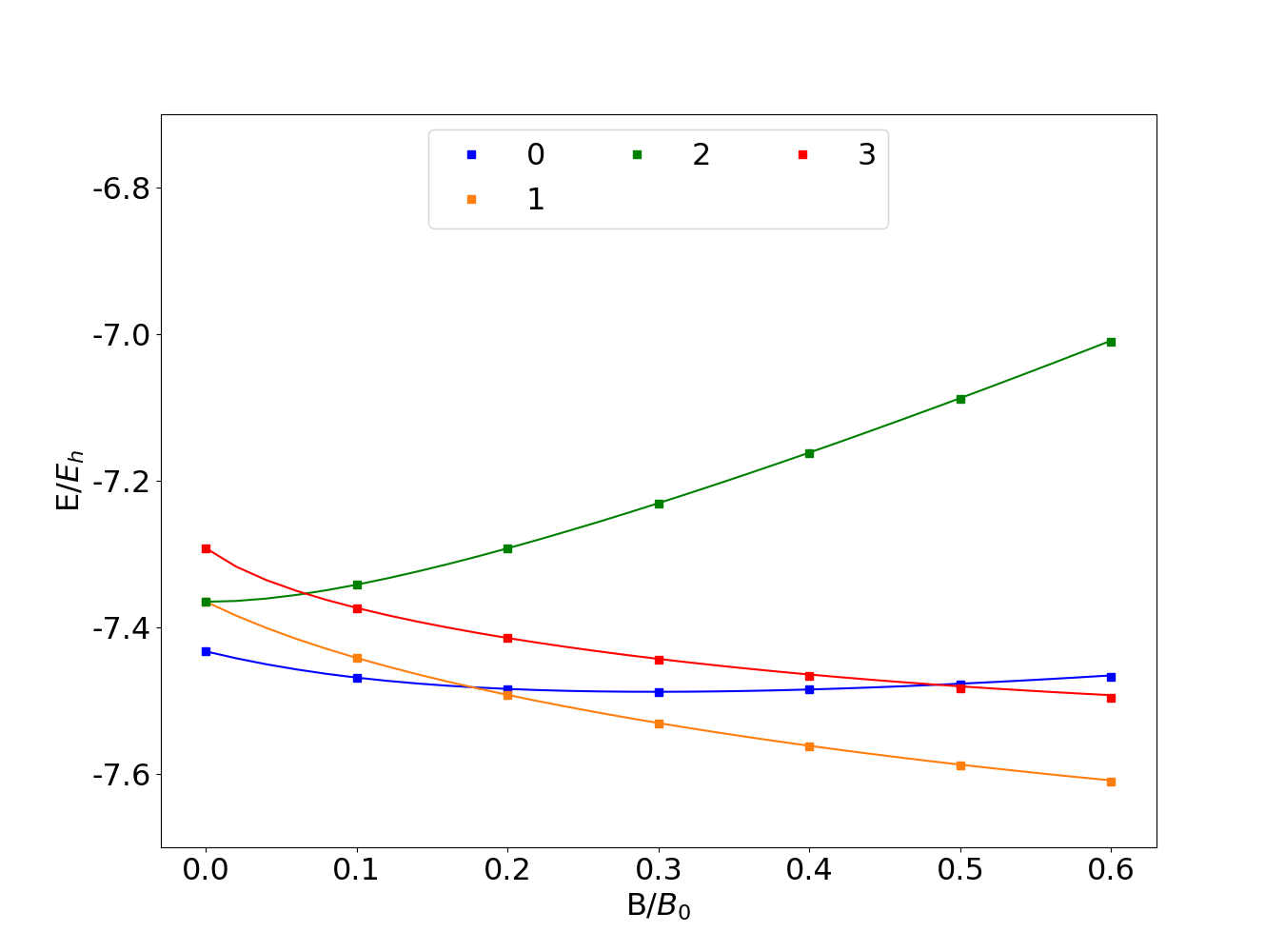}
\caption{Total energy of the Li atom as a function of the magnetic field strength $B$ in the aug-cc-pVTZ (left) and AHGBSP3-9 (right) basis sets.}
\label{fig:Li}
\end{center}
\end{figure*}

\begin{table}
\centering
\small
\begin{tabular}{llcc}
\hline
 & state & aug-cc-pVTZ & AHGBSP3-9\\ 
\hline \hline
0 & $\sigma^{2,1}$ & \color{blue}{ $ 4.331 $ } & \color{blue}{ $ 0.313 $ }\\ 
1 & $\sigma^{1,1}\pi_-^{1,0}$ & \color{blue}{ $ 1.160 $ } & \color{blue}{ $ 0.312 $ }\\ 
2 & $\sigma^{1,1}\pi_+^{1,0}$ & \color{blue}{ $ 1.160 $ } & \color{blue}{ $ 0.312 $ }\\ 
3 & $\sigma^{1,1}\delta_-^{1,0}$ & \color{blue}{ $ 14.008 $ } & \color{blue}{ $ 1.427 $ }\\ 
\hline
\end{tabular}
\caption{MAEDs between GTO and FEM energies in m$E_h$ for Li in the fully uncontracted aug-cc-pVTZ and AHGBSP3-9 basis sets.}
\label{tab:Li-mean-differ}
\end{table}

The energies of the low-lying states of the Li atom are shown as a
function of the field strength in \cref{fig:Li}. The mean differences
between the FEM and GTO energies are shown in
\cref{tab:Li-mean-differ}. We observe that Li is the first element to
exhibit a ground state crossing at the studied field strengths: the
UHF ground state configuration changes from $\sigma^{2,1}$ to
$\sigma^{1,1}\pi_-^{1,0}$ around $B \approx 0.2B_0$.

Interestingly, a large BSTE is observed for the $\sigma^{2,1}$ state
at larger fields for the aug-cc-pVTZ basis set, while the state is
again much better described by AHGBSP3-9. However, the remaining mean
energy difference for AHGBSP3-9 is still surprisingly large, even
though the FEM and AHGBSP3-9 energies are in perfect agreement at zero
field, as can be seen from the data in the SI. The growing difference
in the total energy as a function of the magnetic field from 0 $\mu
E_h$ to 1.38 m$E_h$ at $B=0.6B_0$ for the $\sigma^{2,1}$ state is
explained by the weak binding of the outermost electron, which is
thereby strongly affected by the magnetic field, and undergoes a large
deformation.

We also see that the $\Delta$ state with an occupied $\delta$ orbital
is poorly described by the aug-cc-pVTZ basis set, while it is well
recovered by the AHGBSP3-9 basis set.

\paragraph{Be \label{sec:Be}}

\begin{figure*}
\begin{center}
\includegraphics[width=0.45\linewidth]{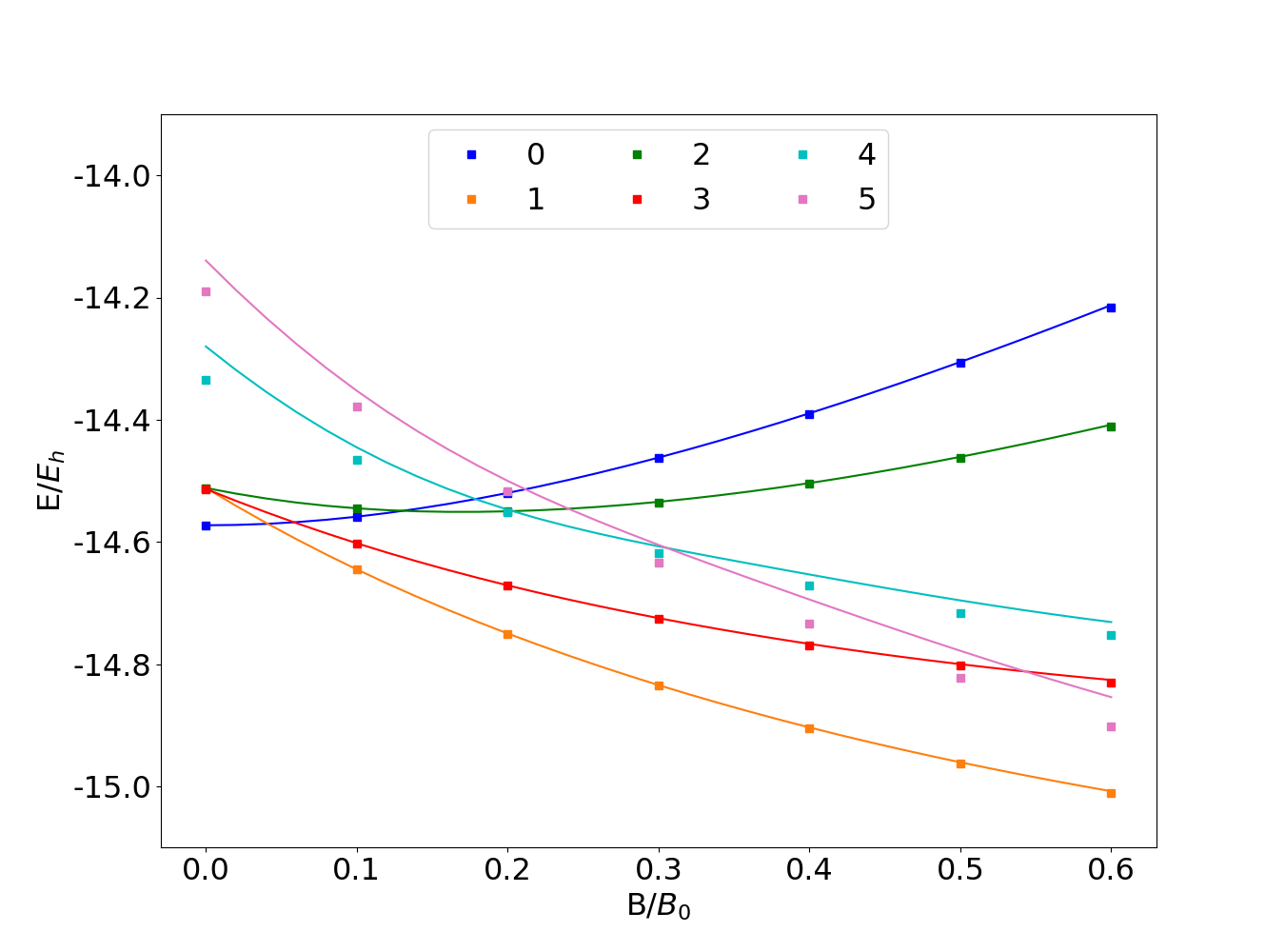}
\includegraphics[width=0.45\linewidth]{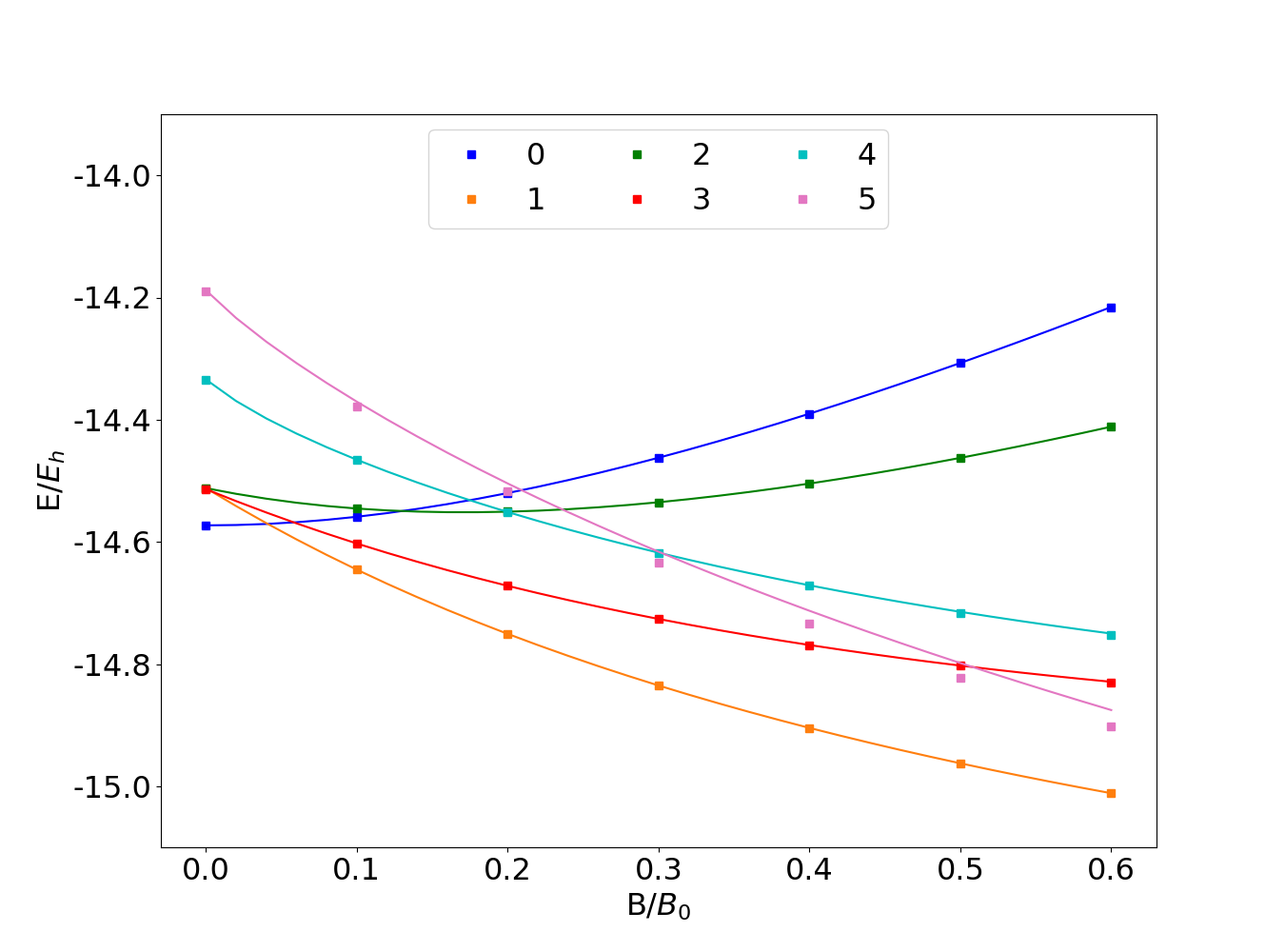}
\caption{Total energy of the Be atom as a function of the magnetic field strength $B$ in the aug-cc-pVTZ (left) and AHGBSP3-9 (right) basis sets.}
\label{fig:Be}
\end{center}
\end{figure*}

\begin{table}
\centering
\small
\begin{tabular}{llcc}
\hline
 & state & aug-cc-pVTZ & AHGBSP3-9\\ 
\hline \hline
0 & $\sigma^{2,2}$ & \color{blue}{ $ 0.902 $ } & \color{blue}{ $ 0.034 $ }\\ 
1 & $\sigma^{2,1}\pi_-^{1,0}$ & \color{blue}{ $ 1.150 $ } & \color{blue}{ $ 0.057 $ }\\ 
2 & $\sigma^{2,1}\pi_+^{1,0}$ & \color{blue}{ $ 1.150 $ } & \color{blue}{ $ 0.057 $ }\\ 
3 & $\sigma^{3,1}$ & \color{blue}{ $ 1.651 $ } & \color{blue}{ $ 0.221 $ }\\ 
4 & $\sigma^{2,1}\delta_-^{1,0}$ & \color{blue}{ $ 21.312 $ } & \color{blue}{ $ 0.782 $ }\\ 
5 & $\sigma^{1,1}\pi_-^{1,0}\delta_-^{1,0}$ & \color{blue}{ $ 36.338 $ } & \color{blue}{ $ 16.129 $ }\\ 
\hline
\end{tabular}
\caption{MAEDs between GTO and FEM energies in m$E_h$ for Be in the fully uncontracted aug-cc-pVTZ and AHGBSP3-9 basis sets.}
\label{tab:Be-mean-differ}
\end{table}

The energies of the low-lying states of the Be atom are shown as a
function of the field strength in \cref{fig:Be}. The mean differences
between the FEM and GTO energies are shown in
\cref{tab:Be-mean-differ}.

We see a ground state crossing between $\sigma^{2,2}$ and
$\sigma^{2,1}\pi_-^{1,0}$ around $B \approx 0.05B_0$. All the $\Sigma$
and $\Pi$ states are well described in both GTO basis sets. The
$\sigma^{2,1}\delta_-^{1,0}$ state is ill-described in aug-cc-pVTZ,
but is well recovered by AHGBSP3-9, indicating that higher
polarization functions can recover the state well.

However, the $\sigma^{1,1}\pi_-^{1,0}\delta_-^{1,0}$ state is
ill-described even by AHGBSP3-9 with a MAED of over 10 m$E_h$. This
large difference likely arises mostly from the difference of the
real-orbital approximation (\cref{sec:real-approx}) used in the GTO
calculations and the complex-orbital FEM calculations, instead of
incompleteness of the GTO basis set.

\paragraph{B \label{sec:B}}

\begin{figure*}
\begin{center}
\includegraphics[width=0.45\linewidth]{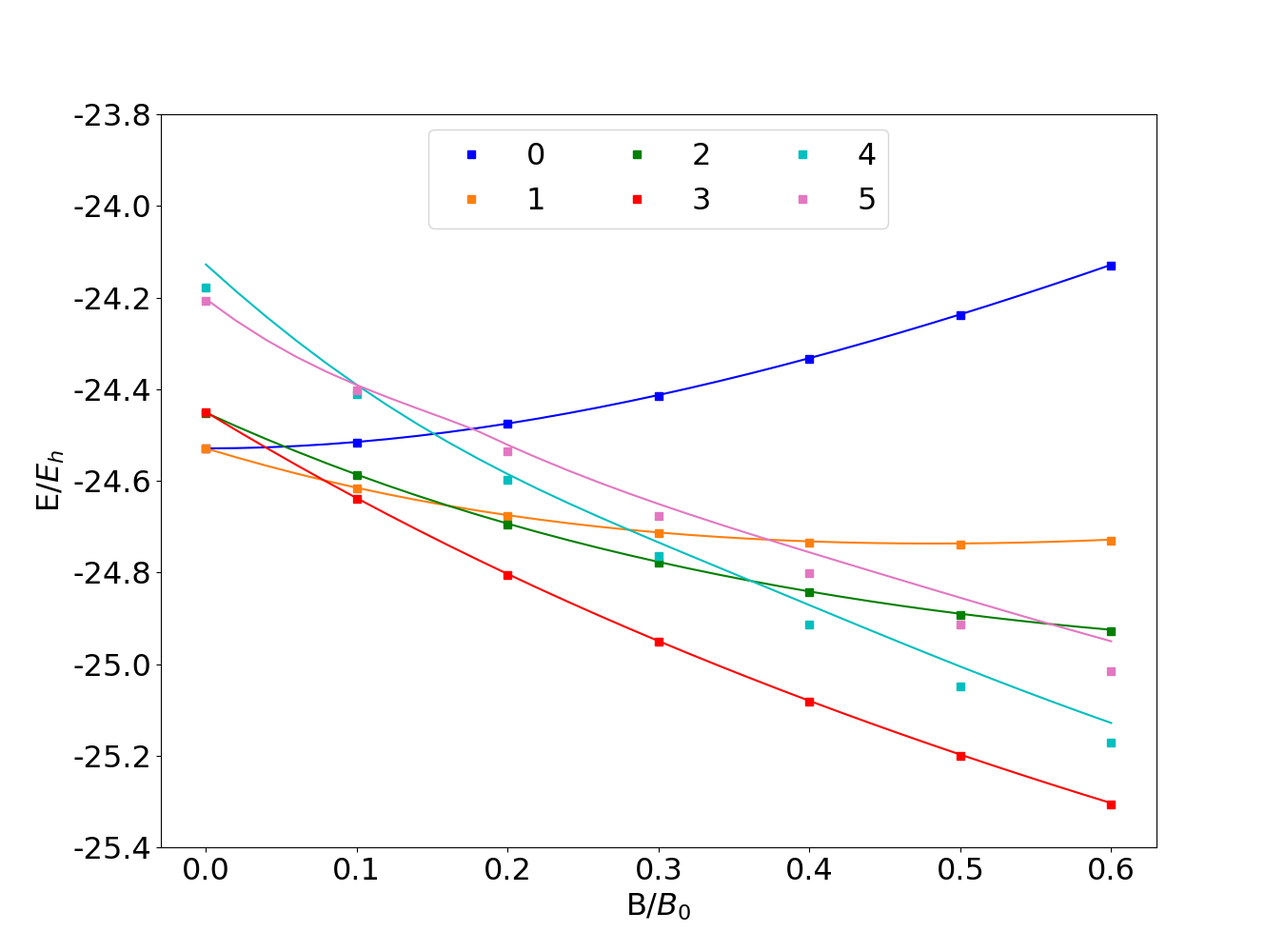}
\includegraphics[width=0.45\linewidth]{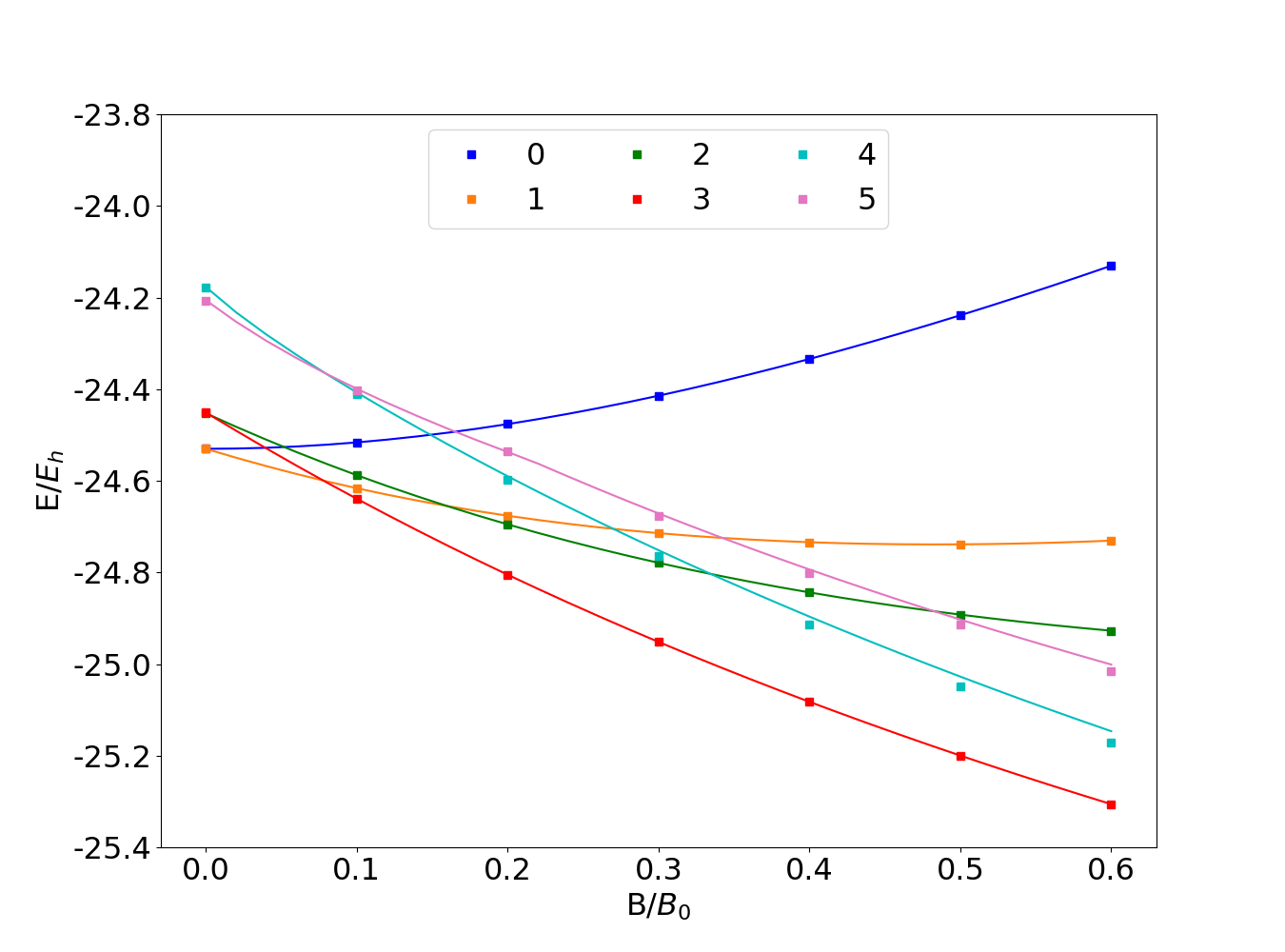}
\caption{Total energy of the B atom as a function of the magnetic field strength $B$ in the aug-cc-pVTZ (left) and AHGBSP3-9 (right) basis sets.}
\label{fig:B}
\end{center}
\end{figure*}

\begin{table}
\centering
\small
\begin{tabular}{llcc}
\hline
 & state & aug-cc-pVTZ & AHGBSP3-9\\ 
\hline \hline
0 & $\sigma^{2,2}\pi_+^{1,0}$ & \color{blue}{ $ 1.523 $ } & \color{blue}{ $ 0.015 $ }\\ 
1 & $\sigma^{2,2}\pi_-^{1,0}$ & \color{blue}{ $ 1.523 $ } & \color{blue}{ $ 0.015 $ }\\ 
2 & $\sigma^{2,1}\pi_+^{1,0}\pi_-^{1,0}$ & \color{blue}{ $ 1.780 $ } & \color{blue}{ $ 0.022 $ }\\ 
3 & $\sigma^{3,1}\pi_-^{1,0}$ & \color{blue}{ $ 1.855 $ } & \color{blue}{ $ 0.055 $ }\\ 
4 & $\sigma^{2,1}\pi_-^{1,0}\delta_-^{1,0}$ & \color{blue}{ $ 34.409 $ } & \color{blue}{ $ 12.586 $ }\\ 
5 & $\sigma^{2,1}\pi_-^{2,0}$ & \color{red}{ $ 32.096 $ } & \color{red}{ $ 6.319 $ }\\ 
\hline
\end{tabular}
\caption{MAEDs between GTO and FEM energies in m$E_h$ for B in the fully uncontracted aug-cc-pVTZ and AHGBSP3-9 basis sets.}
\label{tab:B-mean-differ}
\end{table}

The energies of the low-lying states of the B atom are shown as a
function of the field strength in \cref{fig:B}. The mean differences
between the FEM and GTO energies are shown in
\cref{tab:B-mean-differ}.

There is a ground state crossing between $\sigma^{2,2}\pi_-^{1,0}$ and
$\sigma^{3,1}\pi_-^{1,0}$ around $B\approx 0.075B_0$. All the $\Pi$
states are well described in both GTO basis sets, except the
$\sigma^{2,1}\pi_-^{2,0}$ state that has a large error in aug-cc-pVTZ
of over 30 m$E_h$, which is reduced considerably to 6 m$E_h$ in the
AHGBSP3-9 basis set. We note that this state was one of the states
prone to saddle point convergence (see
\cref{sec:computational-details}).

The state with an occupied $\delta$ orbital also exhibits large MAEDs,
which we again tentatively attribute to the use of the real-orbital
approximation (\cref{sec:real-approx}).

\paragraph{C \label{sec:C}}

\begin{figure*}
\begin{center}
\includegraphics[width=0.45\linewidth]{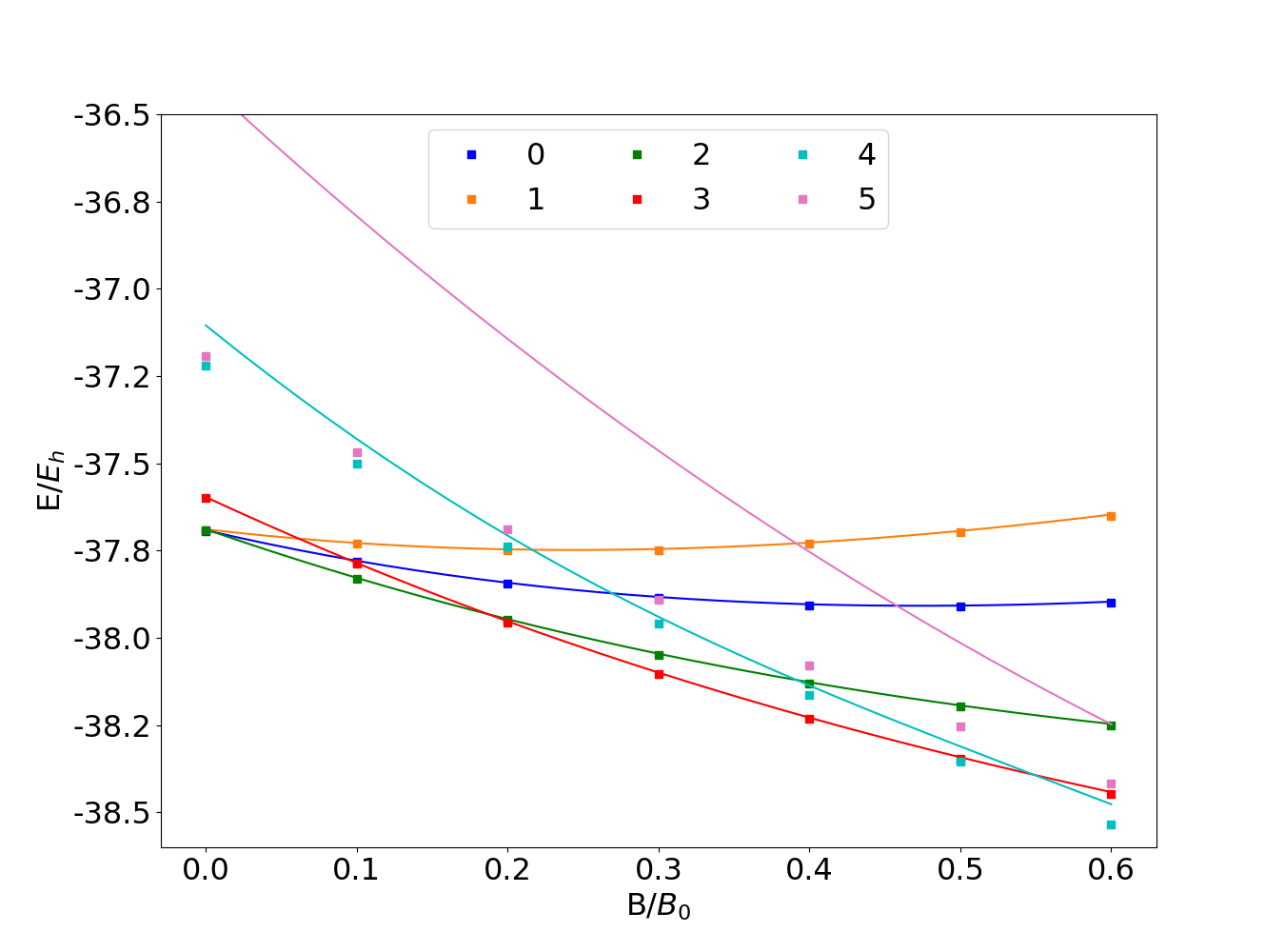}
\includegraphics[width=0.45\linewidth]{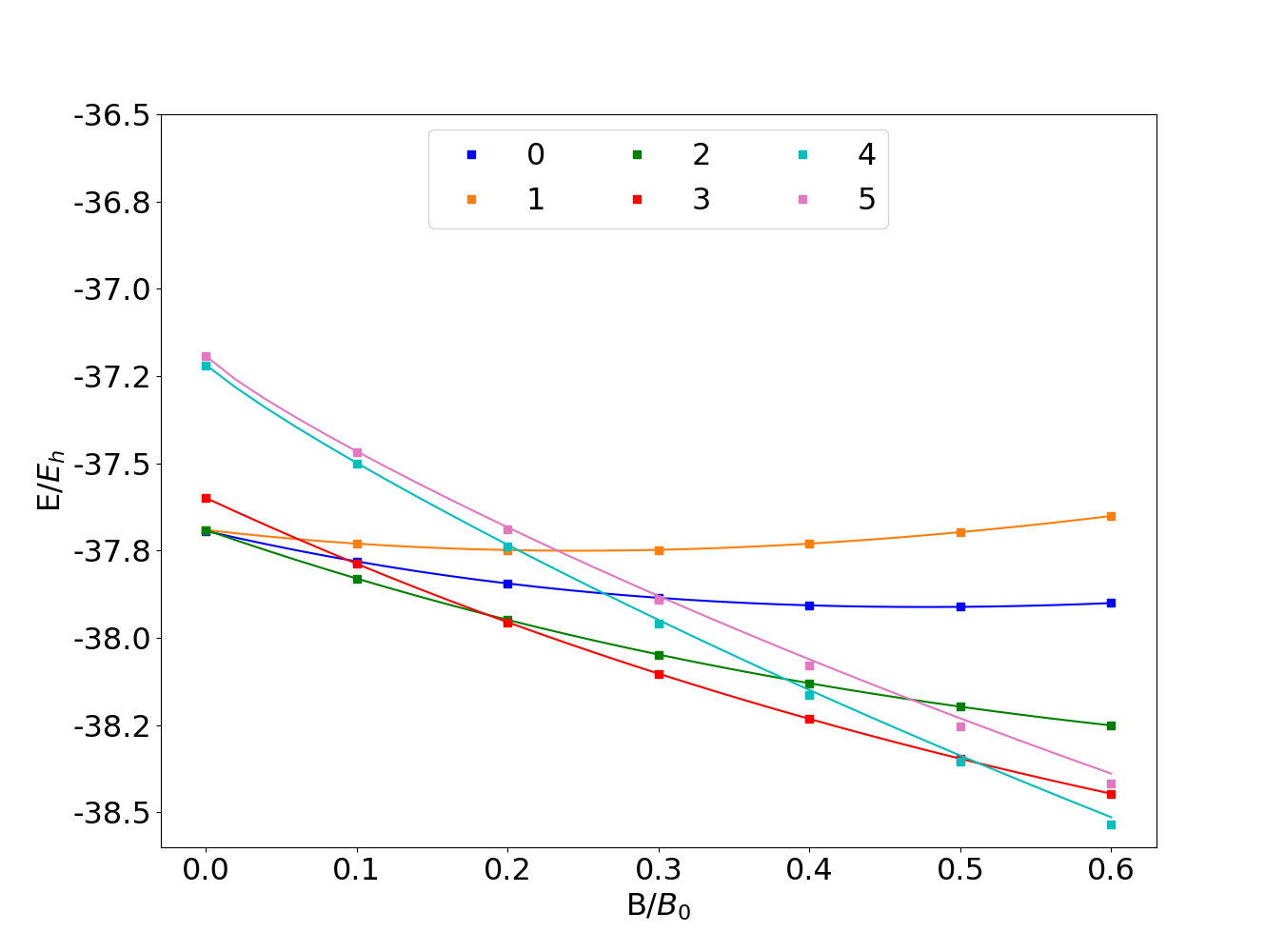}
\caption{Total energy of the C atom as a function of the magnetic field strength $B$ in the aug-cc-pVTZ (left) and AHGBSP3-9 (right) basis sets.}
\label{fig:C}
\end{center}
\end{figure*}

\begin{table}
\centering
\small
\begin{tabular}{llcc}
\hline
 & state & aug-cc-pVTZ & AHGBSP3-9\\ 
\hline \hline
0 & $\sigma^{2,2}\pi_+^{1,0}\pi_-^{1,0}$ & \color{blue}{ $ 2.761 $ } & \color{blue}{ $ 0.006 $ }\\ 
1 & $\sigma^{3,2}\pi_+^{1,0}$ & \color{blue}{ $ 2.714 $ } & \color{blue}{ $ 0.016 $ }\\ 
2 & $\sigma^{3,2}\pi_-^{1,0}$ & \color{blue}{ $ 2.714 $ } & \color{blue}{ $ 0.016 $ }\\ 
3 & $\sigma^{3,1}\pi_+^{1,0}\pi_-^{1,0}$ & \color{blue}{ $ 3.288 $ } & \color{blue}{ $ 0.015 $ }\\ 
4 & $\sigma^{3,1}\pi_-^{1,0}\delta_-^{1,0}$ & \color{blue}{ $ 51.333 $ } & \color{blue}{ $ 9.391 $ }\\ 
5 & $\sigma^{3,1}\pi_-^{1,0}\phi_-^{1,0}$ & \color{blue}{ $ 452.152 $ } & \color{blue}{ $ 12.204 $ }\\ 
\hline
\end{tabular}
\caption{MAEDs between GTO and FEM energies in m$E_h$ for C in the fully uncontracted aug-cc-pVTZ and AHGBSP3-9 basis sets.}
\label{tab:C-mean-differ}
\end{table}

The energies of the low-lying states of the C atom are shown as a
function of the field strength in \cref{fig:C}. The mean differences
between the FEM and GTO energies are shown in
\cref{tab:C-mean-differ}.

C is the first element with more than one observed ground state
crossing: we see a change from $\sigma^{3,2}\pi_-^{1,0}$ to
$\sigma^{3,1}\pi_+^{1,0}\pi_-^{1,0}$ around $B\approx 0.2B_0$, and
further to $\sigma^{3,1}\pi_-^{1,0}\delta_-^{1,0}$ around $B \approx
0.5B_0$.

All the low lying $\Pi$ states are reasonably well described by the
aug-cc-pVTZ basis. However, the energy differences are two orders of
magnitude smaller in the AHGBSP3-9 basis set.

We see that the state with an occupied $\delta$ orbital is
ill-described by the aug-cc-pVTZ basis set, that it is better
described by the AHGBSP3-9 basis set, and that the remaining MAED
would likely be much smaller without the use of the real-orbital
approximation in the GTO calculations.

The state with the occupied $\varphi$ orbital exhibits very large
energy differences in aug-cc-pVTZ. The state is drastically better
described in the AHGBSP3-9 basis, reducing the mean difference by
hundreds of millihartrees from the aug-cc-pVTZ value. However, a
negative energy difference is observed for this state at zero field
with the AHGBSP3-9 basis set, which can only arise from the use of the
real-orbital approximation of \cref{sec:real-approx}.

\paragraph{N \label{sec:N}}

\begin{figure*}
\begin{center}
\includegraphics[width=0.45\linewidth]{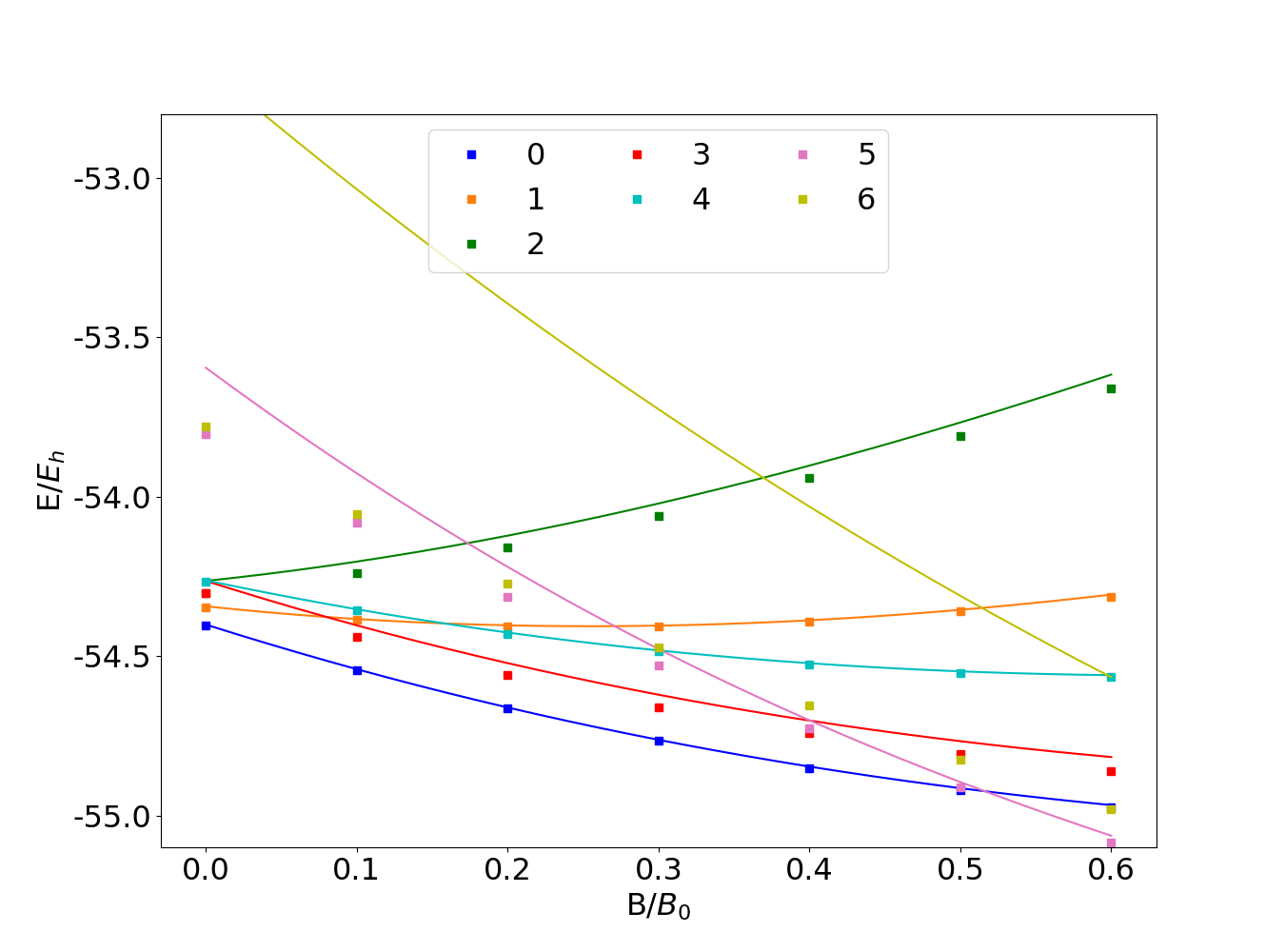}
\includegraphics[width=0.45\linewidth]{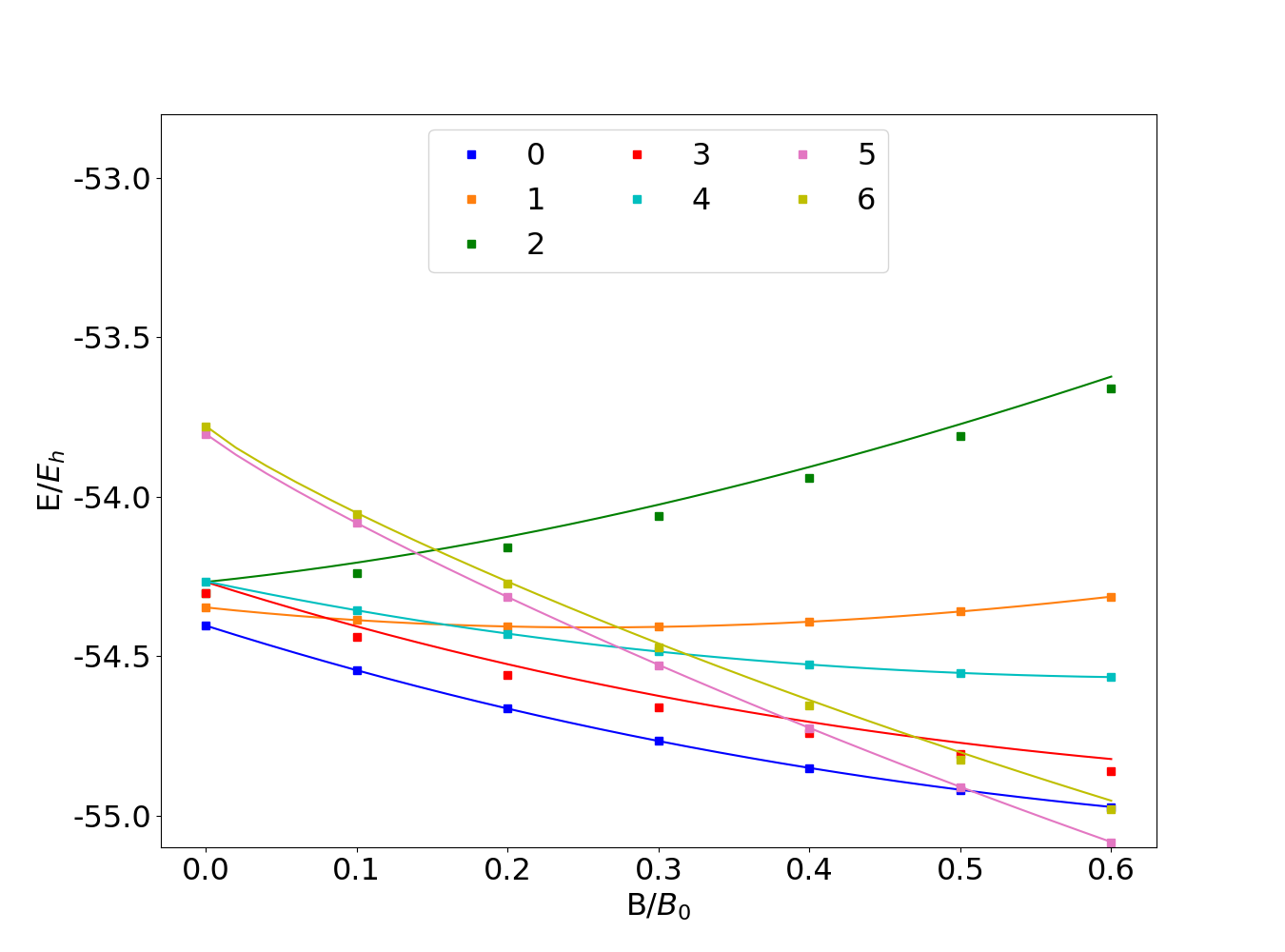}
\caption{Total energy of the N atom as a function of the magnetic field strength $B$ in the aug-cc-pVTZ (left) and AHGBSP3-9 (right) basis sets.}
\label{fig:N}
\end{center}
\end{figure*}

\begin{table}
\centering
\small
\begin{tabular}{llcc}
\hline
 & state & aug-cc-pVTZ & AHGBSP3-9\\ 
\hline \hline
0 & $\sigma^{3,2}\pi_+^{1,0}\pi_-^{1,0}$ & \color{blue}{ $ 4.147 $ } & \color{blue}{ $ 0.005 $ }\\ 
1 & $\sigma^{2,3}\pi_+^{1,0}\pi_-^{1,0}$ & \color{blue}{ $ 4.438 $ } & \color{blue}{ $ 0.008 $ }\\ 
2 & $\sigma^{3,2}\pi_+^{1,1}$ & \color{blue}{ $ 39.321 $ } & \color{blue}{ $ 35.137 $ }\\ 
3 & $\sigma^{3,2}\pi_-^{1,1}$ & \color{blue}{ $ 39.321 $ } & \color{blue}{ $ 35.137 $ }\\ 
4 & $\sigma^{3,3}\pi_-^{1,0}$ & \color{blue}{ $ 4.129 $ } & \color{blue}{ $ 0.011 $ }\\ 
5 & $\sigma^{3,1}\pi_+^{1,0}\pi_-^{1,0}\delta_-^{1,0}$ & \color{blue}{ $ 81.788 $ } & \color{blue}{ $ 0.801 $ }\\ 
6 & $\sigma^{3,1}\pi_+^{1,0}\pi_-^{1,0}\phi_-^{1,0}$ & \color{blue}{ $ 760.601 $ } & \color{blue}{ $ 12.395 $ }\\ 
\hline
\end{tabular}
\caption{MAEDs between GTO and FEM energies in m$E_h$ for N in the fully uncontracted aug-cc-pVTZ and AHGBSP3-9 basis sets.}
\label{tab:N-mean-differ}
\end{table}

The energies of the low-lying states of the N atom are shown as a
function of the field strength in \cref{fig:N}. The mean differences
between the FEM and GTO energies are shown in
\cref{tab:N-mean-differ}.

N only has one observed ground state crossing: around $B\approx0.5B_0$
the ground state changes from $\sigma^{3,2}\pi_+^{1,0}\pi_-^{1,0}$ to
$\sigma^{3,1}\pi_+^{1,0}\pi_-^{1,0}\delta_-^{1,0}$.

The $\Pi$ states $\sigma^{3,2}\pi_\pm^{1,1}$ are poorly described by
both GTO basis sets, with mean differences of 39.32 m$E_h$ and 35.14
m$E_h$ respectively; this is likely again an artefact of the
real-orbital approximation used for the GTO calculations.
The other $\Pi$ states are well described by both GTO basis sets.

The large MAED for the state with the occupied $\delta$ orbital arises
mainly at the weak field regime in the aug-cc-pVTZ basis set; the
state is much better recovered at stronger fields. The small MAED of
AHGBSP3-9 indicates that this state can be recovered by a GTO
expansion.

Similarly to the case of carbon discussed above, also here the
description of low-lying state with an occupied $\varphi$ orbital is
drastically improved by the AHGBSP3-9 basis, even though a negative
energy difference arising from the real-orbital approximation is again
observed.

\paragraph{O \label{sec:O}}

\begin{figure*}
\begin{center}
\includegraphics[width=0.45\linewidth]{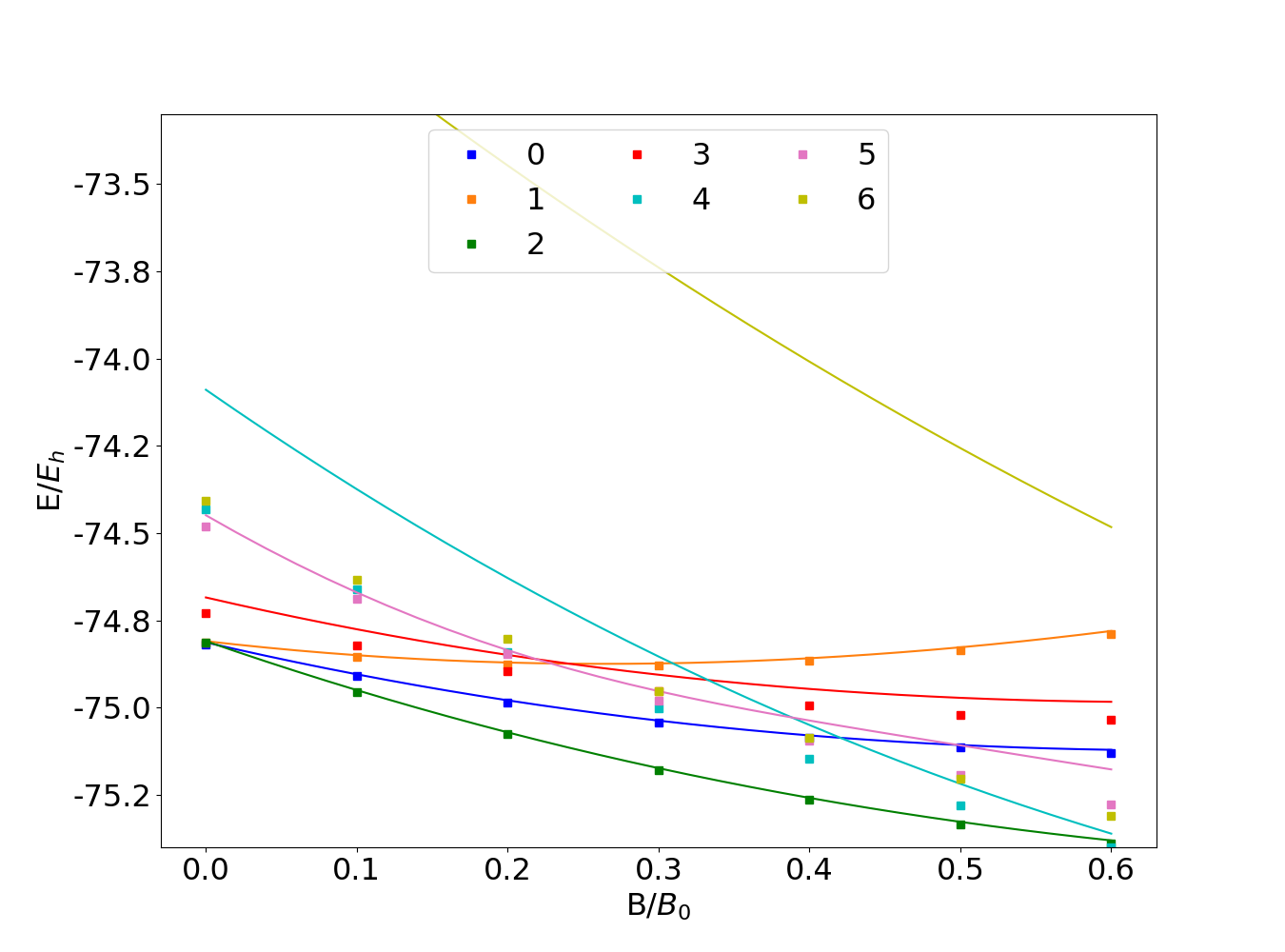}
\includegraphics[width=0.45\linewidth]{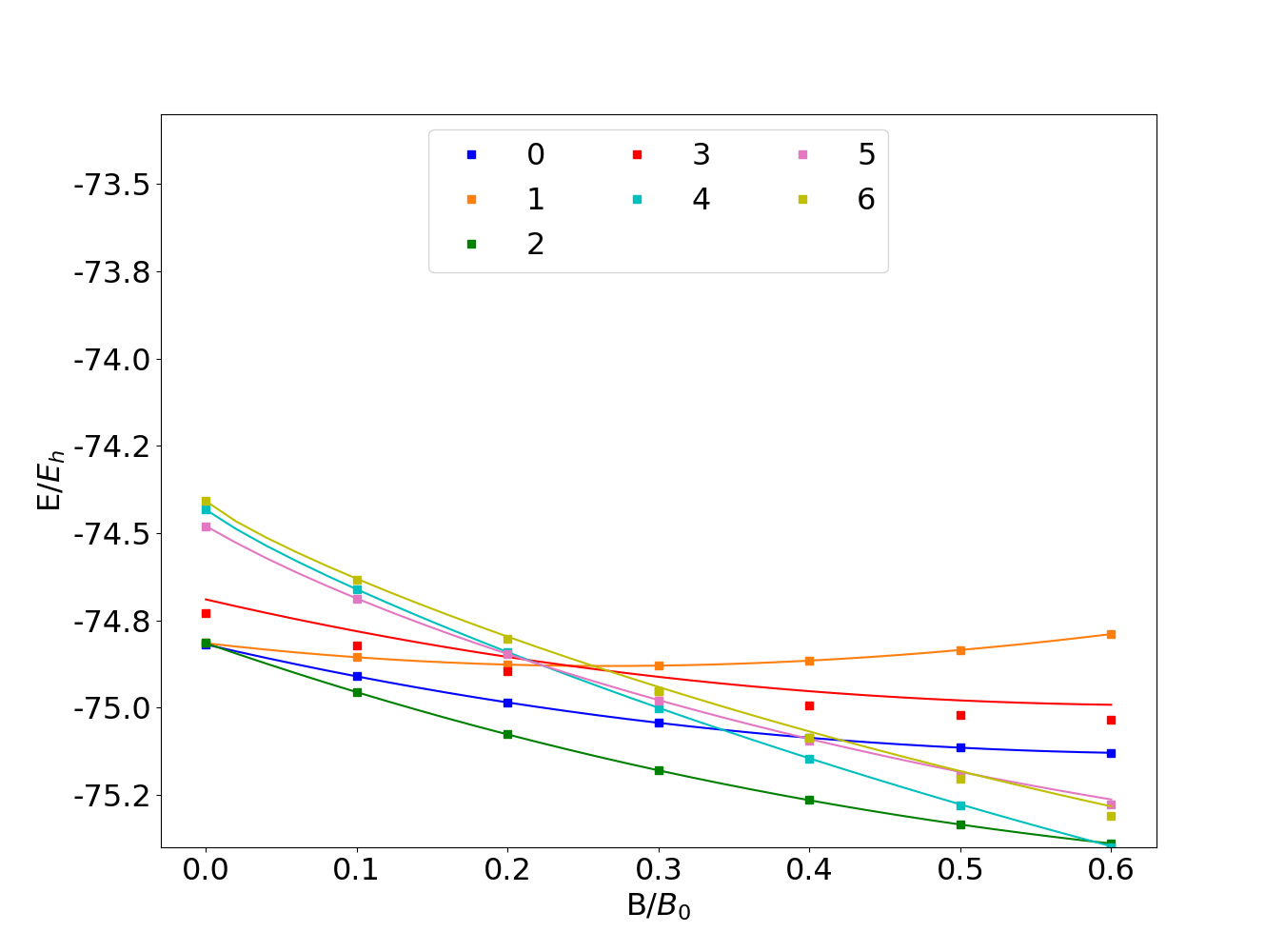}
\caption{Total energy of the O atom as a function of the magnetic field strength $B$ in the aug-cc-pVTZ (left) and AHGBSP3-9 (right) basis sets.}
\label{fig:O}
\end{center}
\end{figure*}

\begin{table}
\centering
\small
\begin{tabular}{llcc}
\hline
 & state & aug-cc-pVTZ & AHGBSP3-9\\ 
\hline \hline
0 & $\sigma^{3,3}\pi_+^{1,0}\pi_-^{1,0}$ & \color{blue}{ $ 6.769 $ } & \color{blue}{ $ 0.004 $ }\\ 
1 & $\sigma^{3,2}\pi_+^{1,1}\pi_-^{1,0}$ & \color{red}{ $ 6.426 $ } & \color{red}{ $ 0.340 $ }\\ 
2 & $\sigma^{3,2}\pi_+^{1,0}\pi_-^{1,1}$ & \color{red}{ $ 6.426 $ } & \color{red}{ $ 0.340 $ }\\ 
3 & $\sigma^{3,3}\pi_-^{1,1}$ & \color{blue}{ $ 47.676 $ } & \color{blue}{ $ 41.149 $ }\\ 
4 & $\sigma^{3,2}\pi_+^{1,0}\pi_-^{1,0}\delta_-^{1,0}$ & \color{blue}{ $ 169.531 $ } & \color{blue}{ $ 0.901 $ }\\ 
5 & $\sigma^{3,2}\pi_+^{1,0}\pi_-^{2,0}$ & \color{red}{ $ 46.803 $ } & \color{red}{ $ 4.010 $ }\\ 
6 & $\sigma^{3,2}\pi_+^{1,0}\pi_-^{1,0}\phi_-^{1,0}$ & \color{blue}{ $ 1219.923 $ } & \color{blue}{ $ 12.693 $ }\\ 
\hline
\end{tabular}
\caption{MAEDs between GTO and FEM energies in m$E_h$ for O in the fully uncontracted aug-cc-pVTZ and AHGBSP3-9 basis sets.}
\label{tab:O-mean-differ}
\end{table}

The energies of the low-lying states of the O atom are shown as a
function of the field strength in \cref{fig:O}. The mean differences
between the FEM and GTO energies are shown in
\cref{tab:O-mean-differ}.

We observe a ground state crossing around $B\approx0.6B_0$ from
$\sigma^{3,2}\pi_+^{1,0}\pi_-^{1,1}$ to
$\sigma^{3,2}\pi_+^{1,0}\pi_-^{1,0}\delta_-^{1,0}$.  The $\Pi$ states
$\sigma^{3,3}\pi_+^{1,0}\pi_-^{1,0}$,
$\sigma^{3,2}\pi_+^{1,1}\pi_-^{1,0}$ and
$\sigma^{3,2}\pi_+^{1,0}\pi_-^{1,1}$ are reasonably well described in
aug-cc-pVTZ. However, we still see a significant improvement going to
the AHGBSP3-9 basis, while aug-cc-pV5Z exhibits similarly large errors
to aug-cc-pVTZ.

The $\sigma^{3,3}\pi_-^{1,1}$ state exhibits large errors of similar
magnitude in both GTO basis sets, again suggesting that this state is
not captured by the real-orbital approximation of
\cref{sec:real-approx}.

The state with the occupied $\delta$ orbital has a very large MAED in
aug-cc-pVTZ, and even though the difference becomes smaller in
increasing field strength, it remains significant at $B=0.6B_0$. The
description of the state is drastically better in AHGBSP3-9.

The $\sigma^{3,2}\pi_+^{1,0}\pi_-^{2,0}$ state and the state with an
occupied $\varphi$ orbital again show drastic improvement going from
aug-cc-pVTZ to the AHGBSP3-9 basis set, indicating room to improve
upon standard GTO basis sets at finite magnetic fields.

\paragraph{F \label{sec:F}}

\begin{figure*}
\begin{center}
\includegraphics[width=0.45\linewidth]{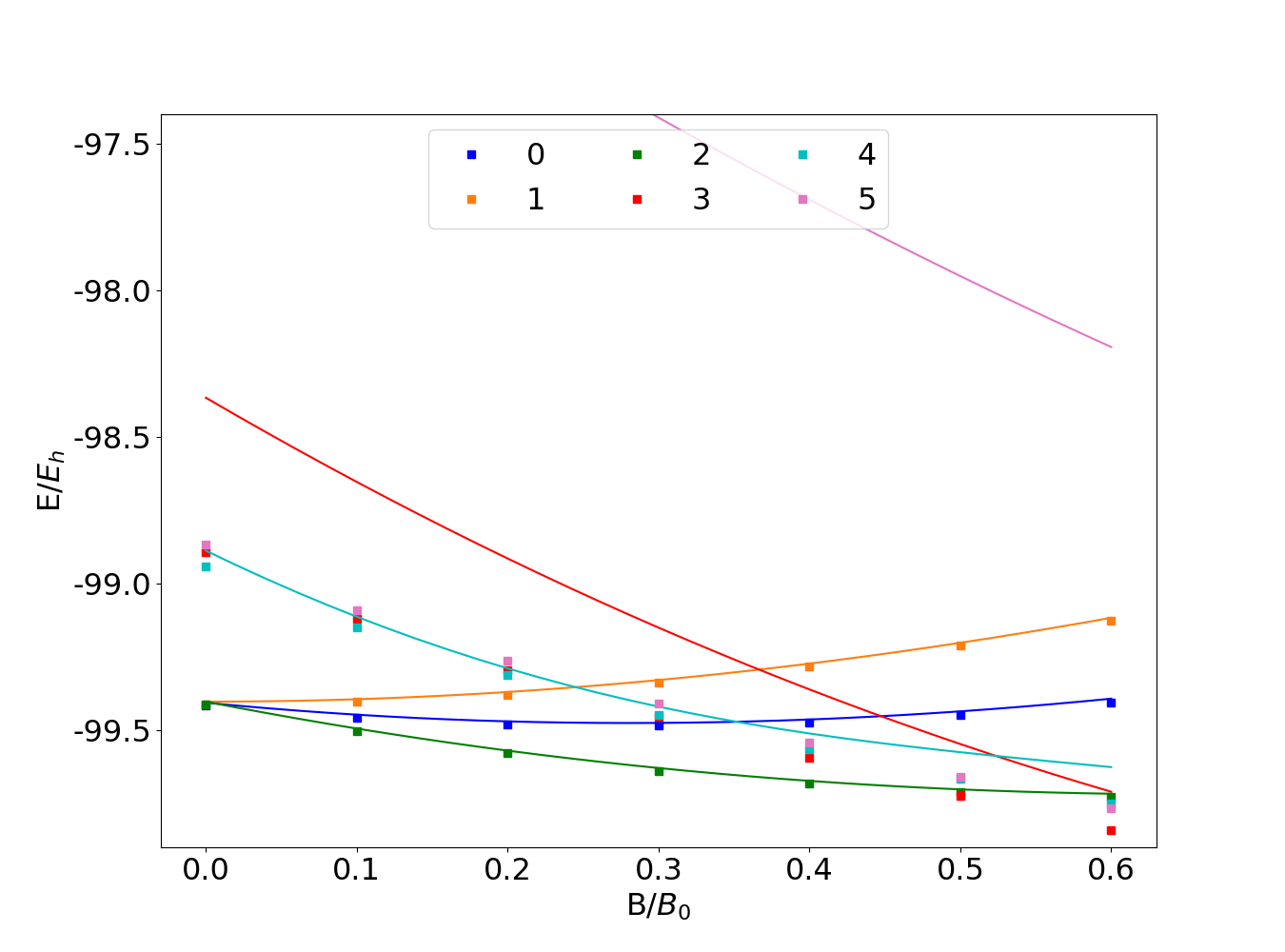}
\includegraphics[width=0.45\linewidth]{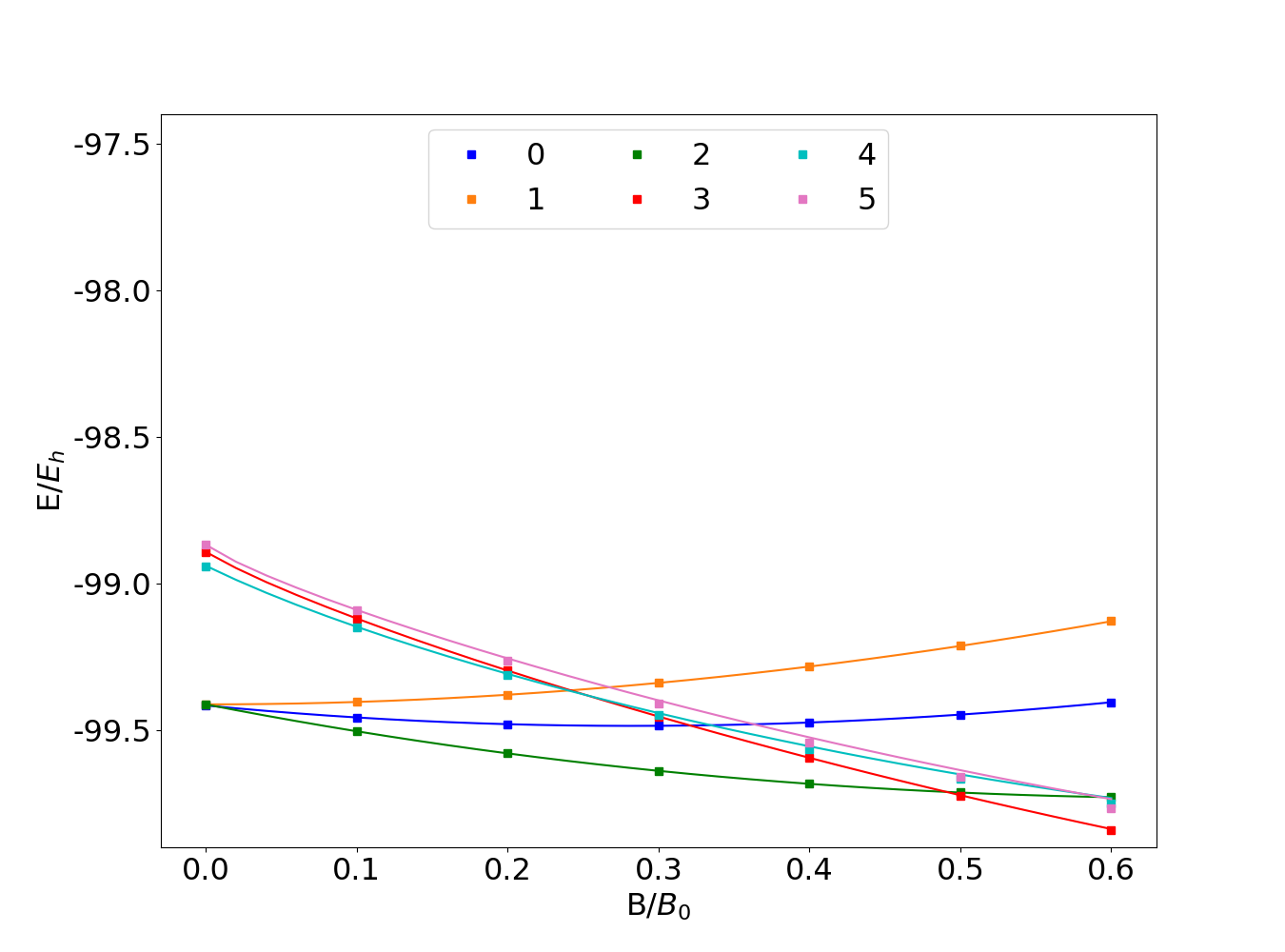}
\caption{Total energy of the F atom as a function of the magnetic field strength $B$ in the aug-cc-pVTZ (left) and AHGBSP3-9 (right) basis sets.}
\label{fig:F}
\end{center}
\end{figure*}

\begin{table}
\centering
\small
\begin{tabular}{llcc}
\hline
 & state & aug-cc-pVTZ & AHGBSP3-9\\ 
\hline \hline
0 & $\sigma^{3,2}\pi_+^{1,1}\pi_-^{1,1}$ & \color{blue}{ $ 10.293 $ } & \color{blue}{ $ 0.002 $ }\\ 
1 & $\sigma^{3,3}\pi_+^{1,1}\pi_-^{1,0}$ & \color{red}{ $ 9.658 $ } & \color{red}{ $ 0.398 $ }\\ 
2 & $\sigma^{3,3}\pi_+^{1,0}\pi_-^{1,1}$ & \color{red}{ $ 9.658 $ } & \color{red}{ $ 0.398 $ }\\ 
3 & $\sigma^{3,2}\pi_+^{1,0}\pi_-^{1,1}\delta_-^{1,0}$ & \color{red}{ $ 316.702 $ } & \color{red}{ $ 0.900 $ }\\ 
4 & $\sigma^{3,2}\pi_+^{1,0}\pi_-^{2,1}$ & \color{blue}{ $ 58.373 $ } & \color{blue}{ $ 8.087 $ }\\ 
5 & $\sigma^{3,2}\pi_+^{1,0}\pi_-^{1,1}\phi_-^{1,0}$ & \color{red}{ $ 1999.482 $ } & \color{red}{ $ 12.994 $ }\\ 
\hline
\end{tabular}
\caption{MAEDs between GTO and FEM energies in m$E_h$ for F in the fully uncontracted aug-cc-pVTZ and AHGBSP3-9 basis sets.}
\label{tab:F-mean-differ}
\end{table}

The energies of the low-lying states of the F atom are shown as a
function of the field strength in \cref{fig:F}. The mean differences
between the FEM and GTO energies are shown in
\cref{tab:F-mean-differ}.

We observe a ground state change from
$\sigma^{3,3}\pi_+^{1,0}\pi_-^{1,1}$ to
$\sigma^{3,2}\pi_+^{1,0}\pi_-^{1,1}\delta_-^{1,0}$ around
$B\approx0.5B_0$.  Similarly to the case of oxygen discussed above,
the states $\sigma^{3,2}\pi_+^{1,1}\pi_-^{1,1}$,
$\sigma^{3,3}\pi_+^{1,1}\pi_-^{1,0}$ and
$\sigma^{3,3}\pi_+^{1,0}\pi_-^{1,1}$ are relatively well described by
aug-cc-pVTZ, but their errors are orders of magnitude smaller in the
AHGBSP3-9 basis set.

The MAED of the state with the occupied $\delta$ orbital decreases in
increasing field strength in aug-cc-pVTZ, but the difference remains
significant at $B=0.6B_0$; AHGBSP3-9 affords a MAED for this state
that is over two orders of magnitude smaller.

The $\sigma^{3,2}\pi_+^{1,0}\pi_-^{2,1}$ state is likewise
ill-described in the aug-cc-pVTZ basis, but better described by
AHGBSP3-9.

We again notice that AHGBSP3-9 offers a drastic improvement in
accuracy for the state with an occupied $\varphi$ orbital that is very
poorly described even in the aug-cc-pV5Z basis.

\paragraph{Ne \label{sec:Ne}}

\begin{figure*}
\begin{center}
\includegraphics[width=0.45\linewidth]{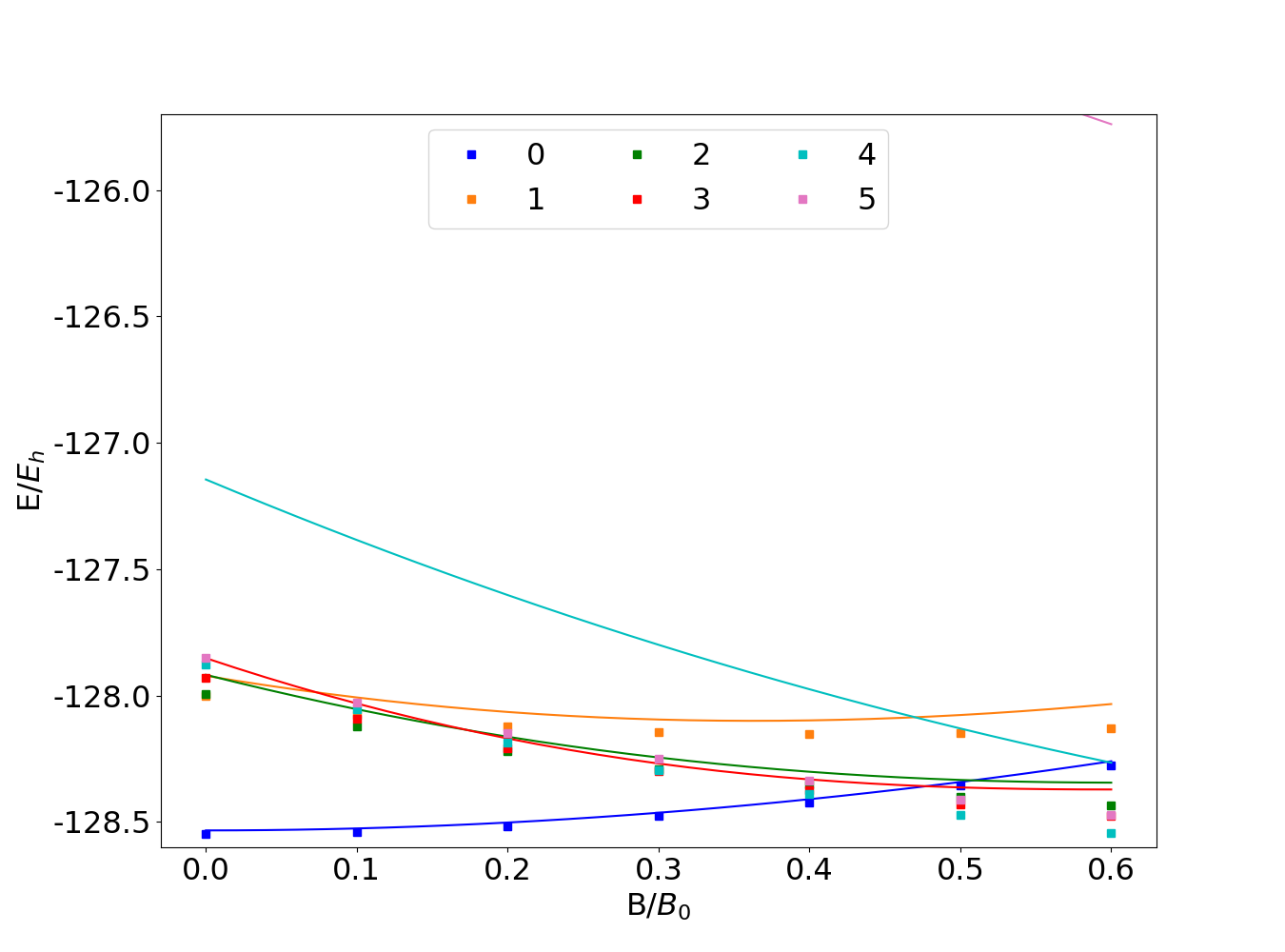}
\includegraphics[width=0.45\linewidth]{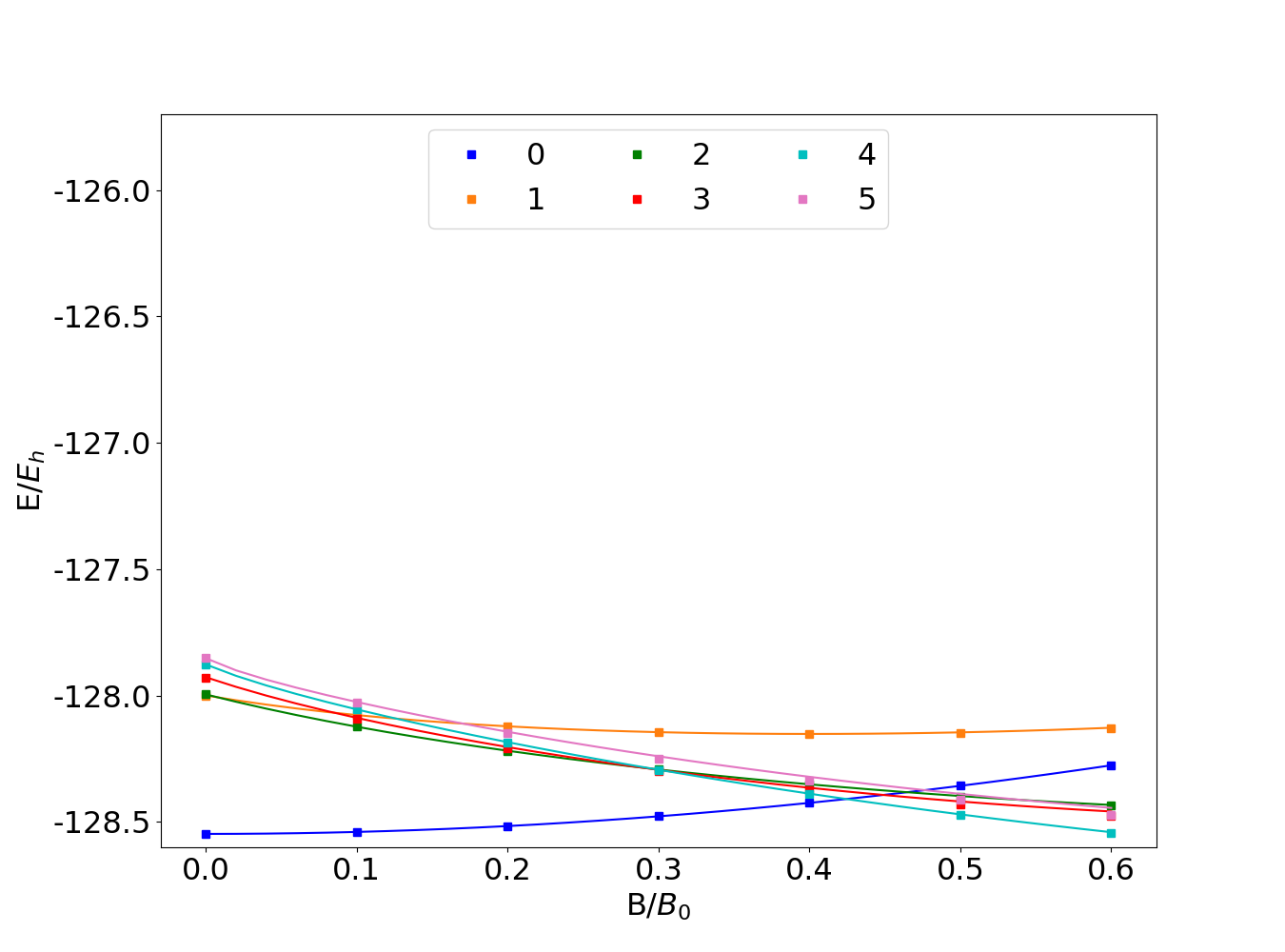}
\caption{Total energy of the Ne atom as a function of the magnetic field strength $B$ in the aug-cc-pVTZ (left) and AHGBSP3-9 (right) basis sets.}
\label{fig:Ne}
\end{center}
\end{figure*}

\begin{table}
\centering
\small
\begin{tabular}{llcc}
\hline
 & state & aug-cc-pVTZ & AHGBSP3-9\\ 
\hline \hline
0 & $\sigma^{3,3}\pi_+^{1,1}\pi_-^{1,1}$ & \color{blue}{ $ 14.502 $ } & \color{blue}{ $ 0.001 $ }\\ 
1 & $\sigma^{4,2}\pi_+^{1,1}\pi_-^{1,1}$ & \color{blue}{ $ 68.049 $ } & \color{blue}{ $ 0.879 $ }\\ 
2 & $\sigma^{4,3}\pi_+^{1,0}\pi_-^{1,1}$ & \color{red}{ $ 65.035 $ } & \color{red}{ $ 0.943 $ }\\ 
3 & $\sigma^{3,3}\pi_+^{1,0}\pi_-^{2,1}$ & \color{blue}{ $ 59.289 $ } & \color{blue}{ $ 6.809 $ }\\ 
4 & $\sigma^{3,3}\pi_+^{1,0}\pi_-^{1,1}\delta_-^{1,0}$ & \color{red}{ $ 501.584 $ } & \color{red}{ $ 0.861 $ }\\ 
5 & $\sigma^{3,3}\pi_+^{1,0}\pi_-^{1,1}\phi_-^{1,0}$ & \color{red}{ $ 3186.481 $ } & \color{red}{ $ 12.446 $ }\\ 
\hline
\end{tabular}
\caption{MAEDs between GTO and FEM energies in m$E_h$ for Ne in the fully uncontracted aug-cc-pVTZ and AHGBSP3-9 basis sets.}
\label{tab:Ne-mean-differ}
\end{table}

The energies of the low-lying states of the Ne atom are shown as a
function of the field strength in \cref{fig:Ne}. The mean differences
between the FEM and GTO energies are shown in \cref{tab:Ne-mean-differ}.

Ne has one ground state crossing: around $B\approx0.42B_0$ we observe
a change from $\sigma^{3,3}\pi_+^{1,1}\pi_-^{1,1}$ to
$\sigma^{3,3}\pi_+^{1,0}\pi_-^{1,1}\delta_-^{1,0}$. We observe that
only $\sigma^{3,3}\pi_+^{1,1}\pi_-^{1,1}$ is reasonably well described
by the aug-cc-pVTZ basis, and that the MAED is five orders of
magnitude smaller in the AHGBSP3-9 basis set.

The $\Pi$ states $\sigma^{4,2}\pi_+^{1,1}\pi_-^{1,1}$ and
$\sigma^{4,3}\pi_+^{1,0}\pi_-^{1,1}$ are poorly described by
aug-cc-pVTZ, but they are well recovered by AHGBSP3-9.  The
$\sigma^{3,3}\pi_+^{1,0}\pi_-^{2,1}$ state is poorly described by
aug-cc-pVTZ and well recovered by AHGBSP3-9 as well.

Similarly to several cases discussed above, the MAED for the state
with an occupied $\delta$ orbital is large in the aug-cc-pVTZ basis,
and AHGBSP3-9 again offers a drastic improvement. Likewise, the
description of the state with an occupied $\varphi$ orbital is
drastically improved by AHGBSP3-9, suggesting room to improve standard
basis sets.

\paragraph{Na \label{sec:Na}}

\begin{figure*}
\begin{center}
\includegraphics[width=0.45\linewidth]{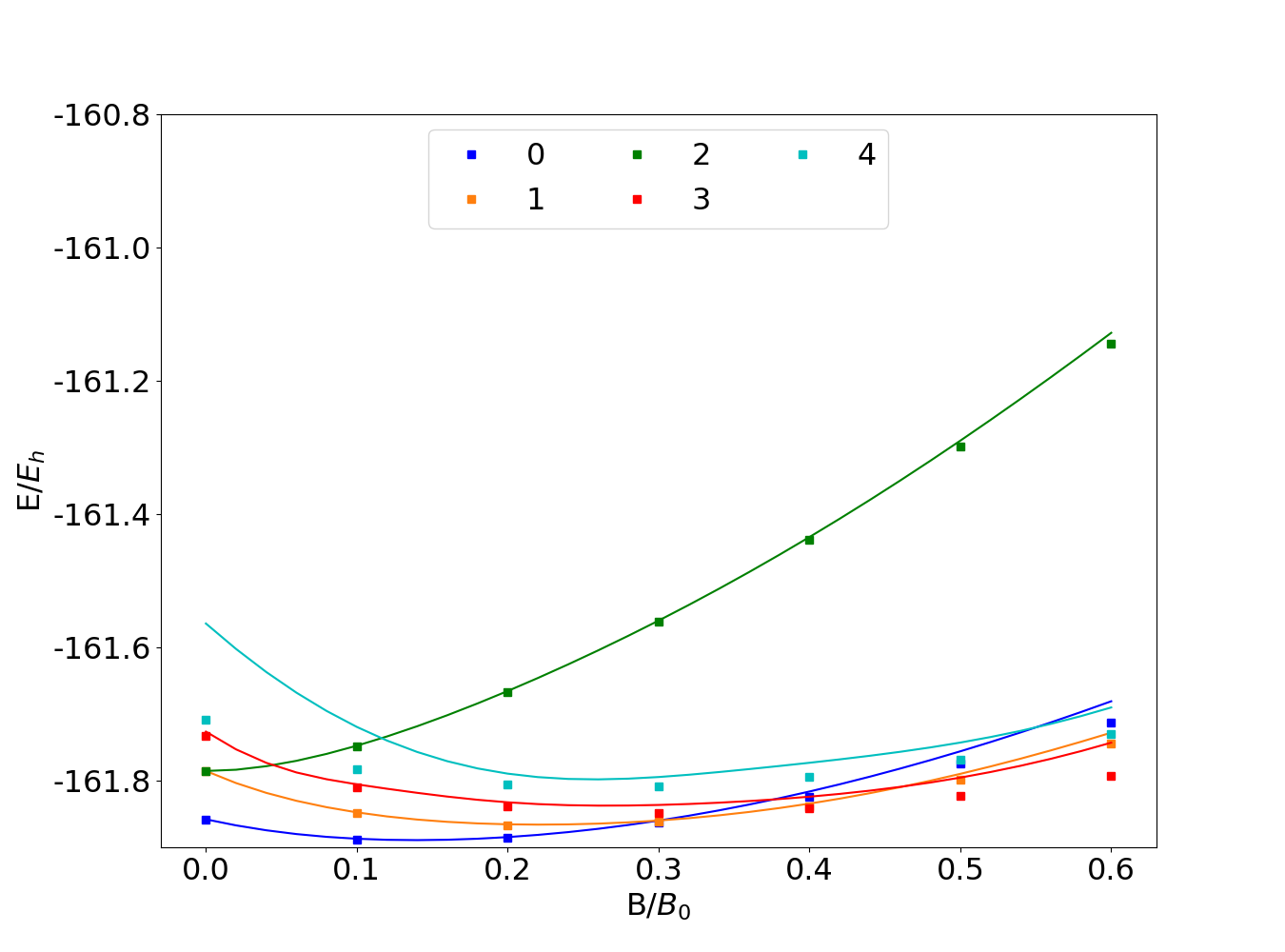}
\includegraphics[width=0.45\linewidth]{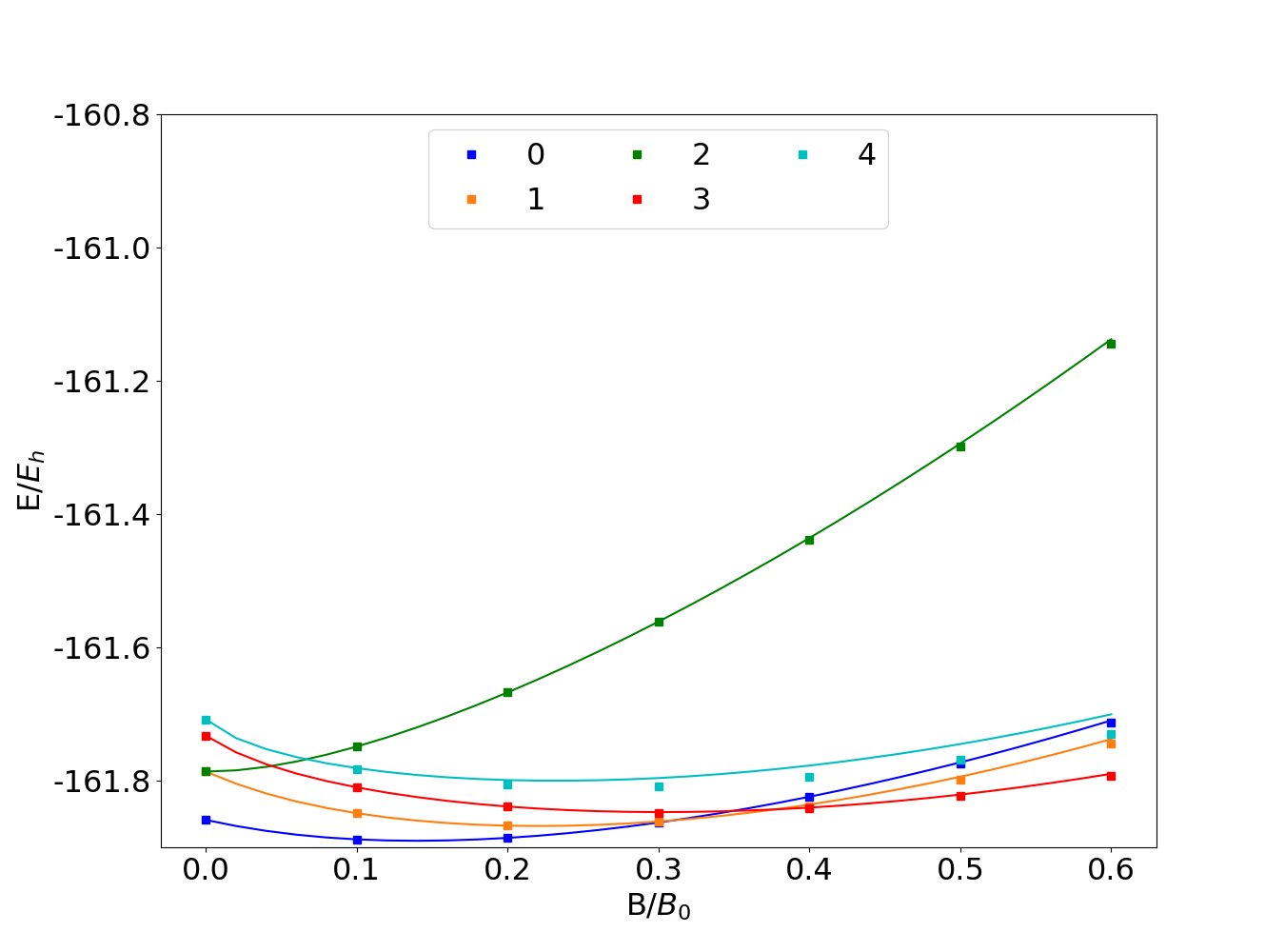}
\caption{Total energy of the Na atom as a function of the magnetic field strength $B$ in the aug-cc-pVTZ (left) and AHGBSP3-9 (right) basis sets.}
\label{fig:Na}
\end{center}
\end{figure*}

\begin{table}
\centering
\small
\begin{tabular}{llcc}
\hline
 & state & aug-cc-pVTZ & AHGBSP3-9\\ 
\hline \hline
0 & $\sigma^{4,3}\pi_+^{1,1}\pi_-^{1,1}$ & \color{blue}{ $ 9.089 $ } & \color{blue}{ $ 0.646 $ }\\ 
1 & $\sigma^{3,3}\pi_+^{1,1}\pi_-^{2,1}$ & \color{red}{ $ 5.002 $ } & \color{red}{ $ 1.991 $ }\\ 
2 & $\sigma^{3,3}\pi_+^{2,1}\pi_-^{1,1}$ & \color{red}{ $ 5.002 $ } & \color{red}{ $ 1.991 $ }\\ 
3 & $\sigma^{3,3}\pi_+^{1,1}\pi_-^{1,1}\delta_-^{1,0}$ & \color{blue}{ $ 17.781 $ } & \color{blue}{ $ 1.160 $ }\\ 
4 & $\sigma^{3,3}\pi_+^{1,1}\pi_-^{1,1}\phi_-^{1,0}$ & \color{blue}{ $ 46.463 $ } & \color{blue}{ $ 13.071 $ }\\ 
\hline
\end{tabular}
\caption{MAEDs between GTO and FEM energies in m$E_h$ for Na in the fully uncontracted aug-cc-pVTZ and AHGBSP3-9 basis sets.}
\label{tab:Na-mean-differ}
\end{table}

The energies of the low-lying states of the Na atom are shown as a
function of the field strength in \cref{fig:Na}. The mean differences
between the FEM and GTO energies are shown in \cref{tab:Na-mean-differ}.

We observe two ground state crossings: the ground state changes
briefly from $\sigma^{4,3}\pi_+^{1,1}\pi_-^{1,1}$ to
$\sigma^{3,3}\pi_+^{1,1}\pi_-^{2,1}$ around $B\approx0.3B_0$, before
changing again to $\sigma^{3,3}\pi_+^{1,1}\pi_-^{1,1}\delta_-^{1,0}$
around $B\approx0.4B_0$. All these $\Pi$ states are quite well
described by aug-cc-pVTZ, with the AHGBSP3-9 basis set exhibiting
strongly reduced MAEDs.

The states with occupied $\delta$ or $\varphi$ orbitals again have
significant MAEDs in the aug-cc-pVTZ basis, while the corresponding
MAEDs are orders of magnitude smaller in the AHGBSP3-9 basis set.

\paragraph{Mg \label{sec:Mg}}

\begin{figure*}
\begin{center}
\includegraphics[width=0.45\linewidth]{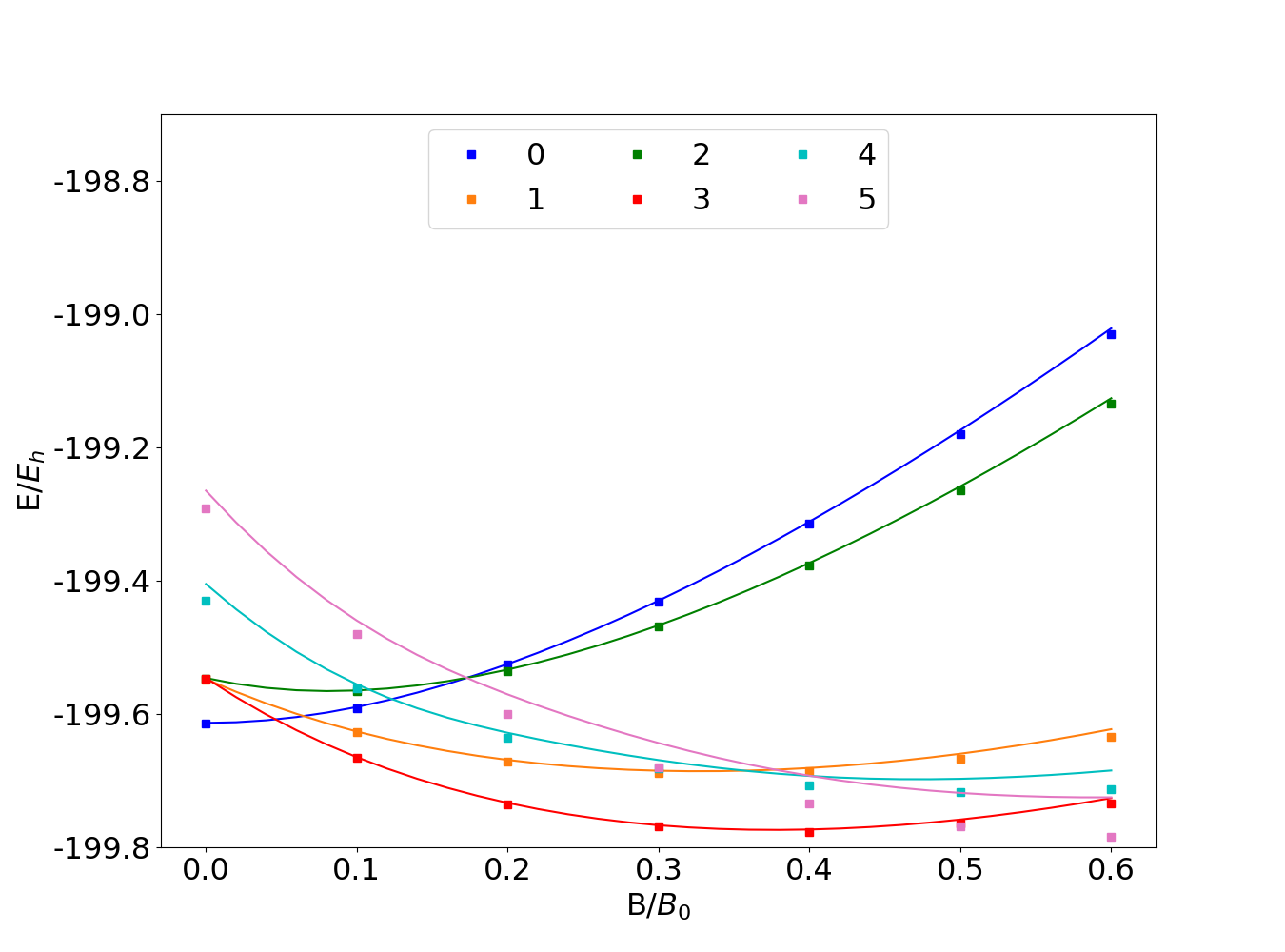}
\includegraphics[width=0.45\linewidth]{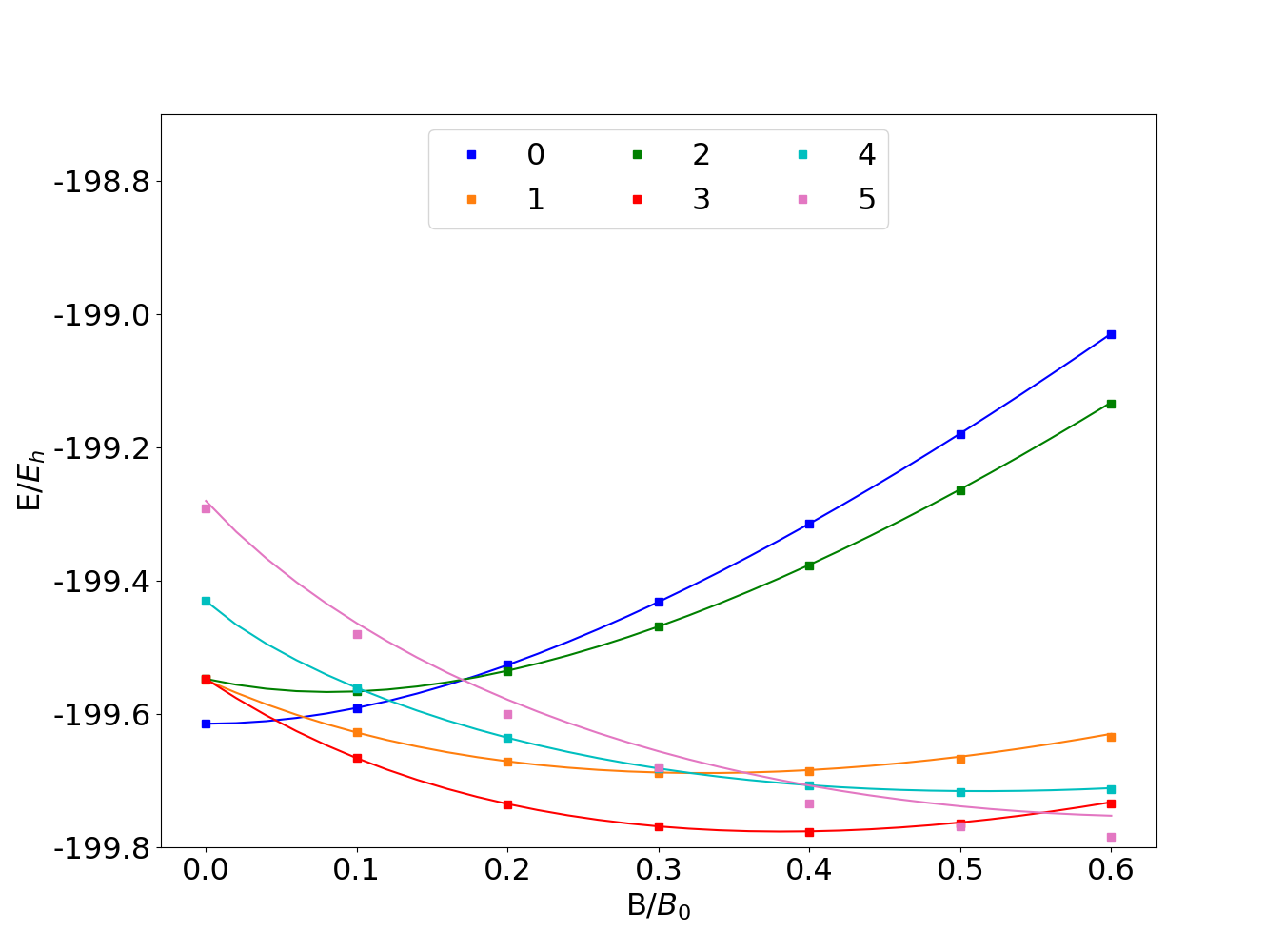}
\caption{Total energy of the Mg atom as a function of the magnetic field strength $B$ in the aug-cc-pVTZ (left) and AHGBSP3-9 (right) basis sets.}
\label{fig:Mg}
\end{center}
\end{figure*}

\begin{table}
\centering
\small
\begin{tabular}{llcc}
\hline
 & state & aug-cc-pVTZ & AHGBSP3-9\\ 
\hline \hline
0 & $\sigma^{4,4}\pi_+^{1,1}\pi_-^{1,1}$ & \color{blue}{ $ 3.261 $ } & \color{blue}{ $ 0.186 $ }\\ 
1 & $\sigma^{5,3}\pi_+^{1,1}\pi_-^{1,1}$ & \color{blue}{ $ 4.436 $ } & \color{blue}{ $ 1.283 $ }\\ 
2 & $\sigma^{4,3}\pi_+^{2,1}\pi_-^{1,1}$ & \color{red}{ $ 3.202 $ } & \color{red}{ $ 0.430 $ }\\ 
3 & $\sigma^{4,3}\pi_+^{1,1}\pi_-^{2,1}$ & \color{red}{ $ 3.202 $ } & \color{red}{ $ 0.430 $ }\\ 
4 & $\sigma^{4,3}\pi_+^{1,1}\pi_-^{1,1}\delta_-^{1,0}$ & \color{blue}{ $ 16.205 $ } & \color{blue}{ $ 0.547 $ }\\ 
5 & $\sigma^{3,3}\pi_+^{1,1}\pi_-^{2,1}\delta_-^{1,0}$ & \color{blue}{ $ 37.781 $ } & \color{blue}{ $ 23.214 $ }\\ 
\hline
\end{tabular}
\caption{MAEDs between GTO and FEM energies in m$E_h$ for Mg in the fully uncontracted aug-cc-pVTZ and AHGBSP3-9 basis sets.}
\label{tab:Mg-mean-differ}
\end{table}

The energies of the low-lying states of the Mg atom are shown as a
function of the field strength in \cref{fig:Mg}. The mean differences
between the FEM and GTO energies are shown in
\cref{tab:Mg-mean-differ}.

We see the ground state changing from
$\sigma^{4,4}\pi_+^{1,1}\pi_-^{1,1}$ to
$\sigma^{4,3}\pi_+^{1,1}\pi_-^{2,1}$ around $B\approx0.05B_0$. The
ground state changes again to
$\sigma^{3,3}\pi_+^{1,1}\pi_-^{2,1}\delta_-^{1,0}$ around
$B\approx0.5B_0$.

The states with occupied $\sigma$ and $\pi$ orbitals are well
described by the aug-cc-pVTZ basis. The
$\sigma^{4,3}\pi_+^{1,1}\pi_-^{1,1}\delta_-^{1,0}$ state has an error
of over 10 m$E_h$ in aug-cc-pVTZ, which is reduced by almost a factor
of 30 in the AHGBSP3-9 basis set. The
$\sigma^{3,3}\pi_+^{1,1}\pi_-^{2,1}\delta_-^{1,0}$ state still shows a
large MAED, which likely arises from the use of the real-orbital
approximation in \cref{sec:real-approx}.

\paragraph{Al \label{sec:Al}}

\begin{figure*}
\begin{center}
\includegraphics[width=0.45\linewidth]{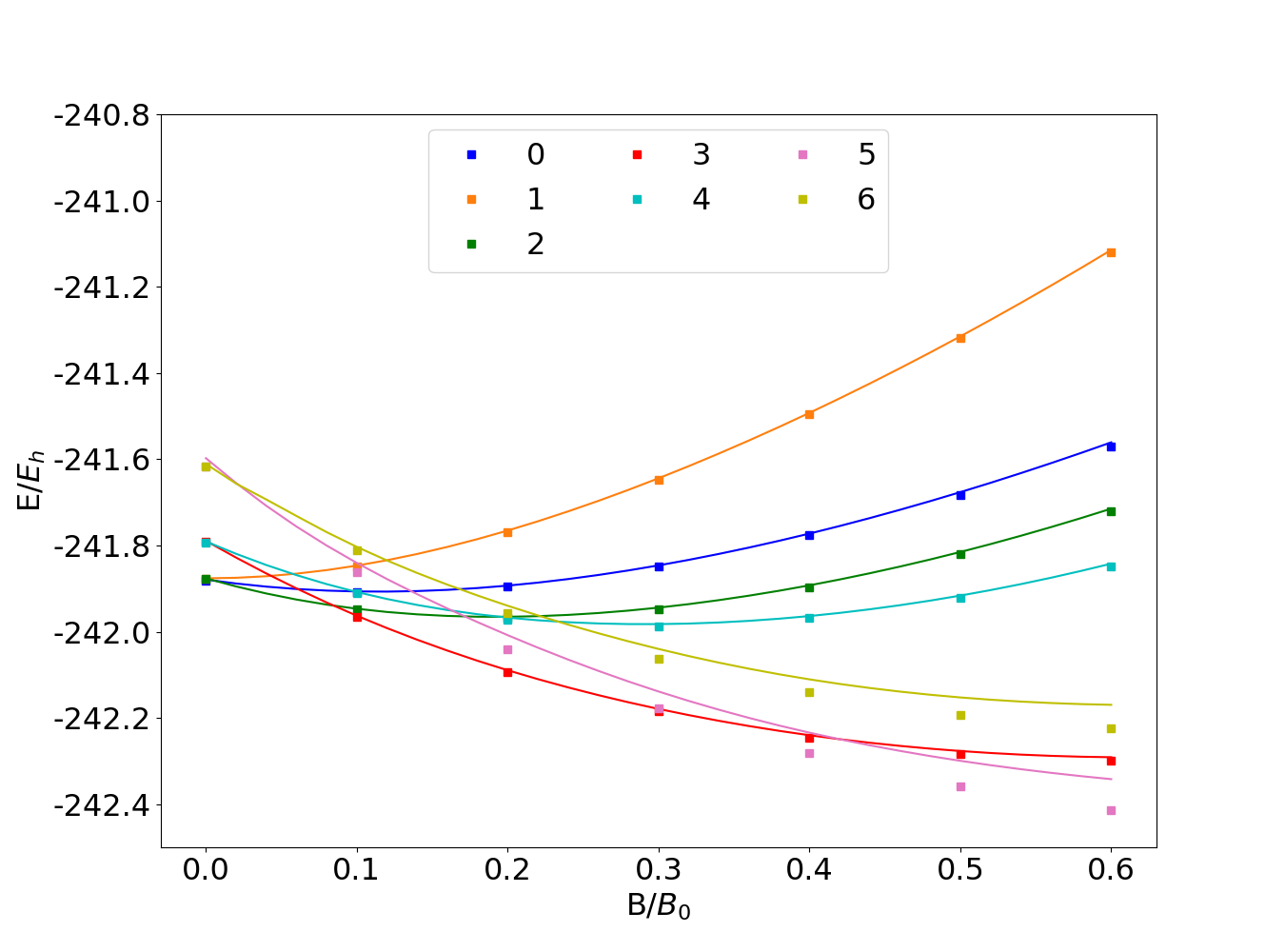}
\includegraphics[width=0.45\linewidth]{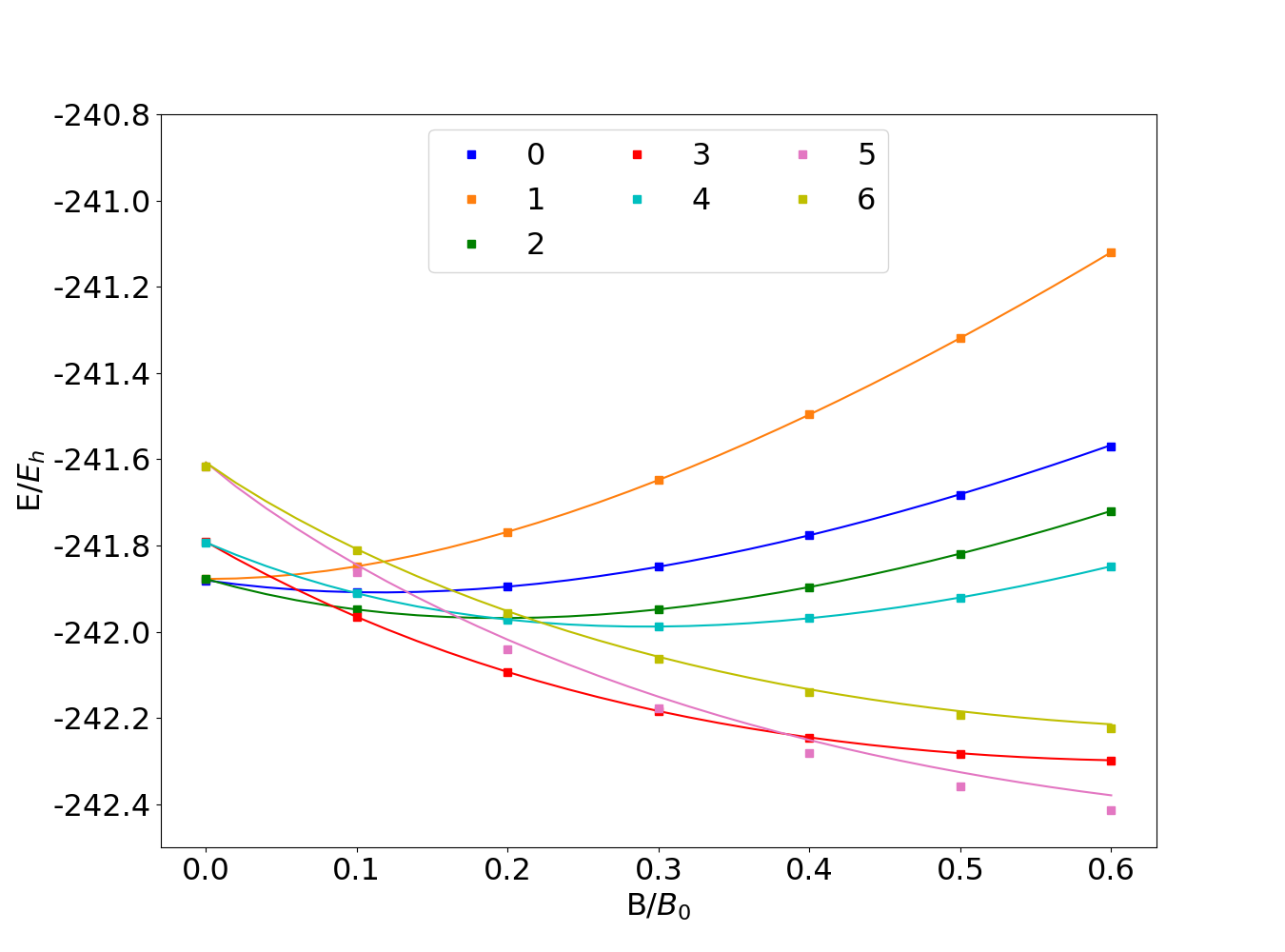}
\caption{Total energy of the Al atom as a function of the magnetic field strength $B$ in the aug-cc-pVTZ (left) and AHGBSP3-9 (right) basis sets.}
\label{fig:Al}
\end{center}
\end{figure*}

\begin{table}
\centering
\small
\begin{tabular}{llcc}
\hline
 & state & aug-cc-pVTZ & AHGBSP3-9\\ 
\hline \hline
0 & $\sigma^{5,4}\pi_+^{1,1}\pi_-^{1,1}$ & \color{blue}{ $ 4.016 $ } & \color{blue}{ $ 0.576 $ }\\ 
1 & $\sigma^{4,4}\pi_+^{2,1}\pi_-^{1,1}$ & \color{red}{ $ 3.474 $ } & \color{red}{ $ 0.139 $ }\\ 
2 & $\sigma^{4,4}\pi_+^{1,1}\pi_-^{2,1}$ & \color{red}{ $ 3.474 $ } & \color{red}{ $ 0.139 $ }\\ 
3 & $\sigma^{5,3}\pi_+^{1,1}\pi_-^{2,1}$ & \color{red}{ $ 4.925 $ } & \color{red}{ $ 0.464 $ }\\ 
4 & $\sigma^{4,3}\pi_+^{2,1}\pi_-^{2,1}$ & \color{blue}{ $ 4.698 $ } & \color{blue}{ $ 0.235 $ }\\ 
5 & $\sigma^{4,3}\pi_+^{1,1}\pi_-^{2,1}\delta_-^{1,0}$ & \color{blue}{ $ 41.080 $ } & \color{blue}{ $ 24.224 $ }\\ 
6 & $\sigma^{4,3}\pi_+^{1,1}\pi_-^{3,1}$ & \color{blue}{ $ 25.320 $ } & \color{blue}{ $ 6.287 $ }\\ 
\hline
\end{tabular}
\caption{MAEDs between GTO and FEM energies in m$E_h$ for Al in the fully uncontracted aug-cc-pVTZ and AHGBSP3-9 basis sets.}
\label{tab:Al-mean-differ}
\end{table}

The energies of the low-lying states of the Al atom are shown as a
function of the field strength in \cref{fig:Al}. The mean differences
between the FEM and GTO energies are shown in
\cref{tab:Al-mean-differ}.

Al changes ground state twice: from the zero-field configuration
$\sigma^{4,4}\pi_+^{1,1}\pi_-^{2,1}$ to
$\sigma^{5,3}\pi_+^{1,1}\pi_-^{2,1}$ around $B\approx0.08B_0$, and
then again to $\sigma^{4,3}\pi_+^{1,1}\pi_-^{2,1}\delta_-^{1,0}$
around $B\approx0.35B_0$.

The states with occupied $\sigma$ and $\pi$ orbitals are adequately
described by the aug-cc-pVTZ basis set, although the AHGBSP3-9 basis
set affords considerably smaller MAEDs. The only exception is the
$\sigma^{4,3}\pi_+^{1,1}\pi_-^{3,1}$ state, which exhibits large
MAEDs, which can likely be attributed to the use of the real-orbital
approximation of \cref{sec:real-approx}.

The MAED for the state with the occupied $\delta$ orbital decreases by
almost 60\% when going from aug-cc-pVTZ to the AHGBSP3-9 basis
set. The large remaining MAED is again likely attributable to the
real-orbital approximation.

\paragraph{Si \label{sec:Si}}

\begin{figure*}
\begin{center}
\includegraphics[width=0.45\linewidth]{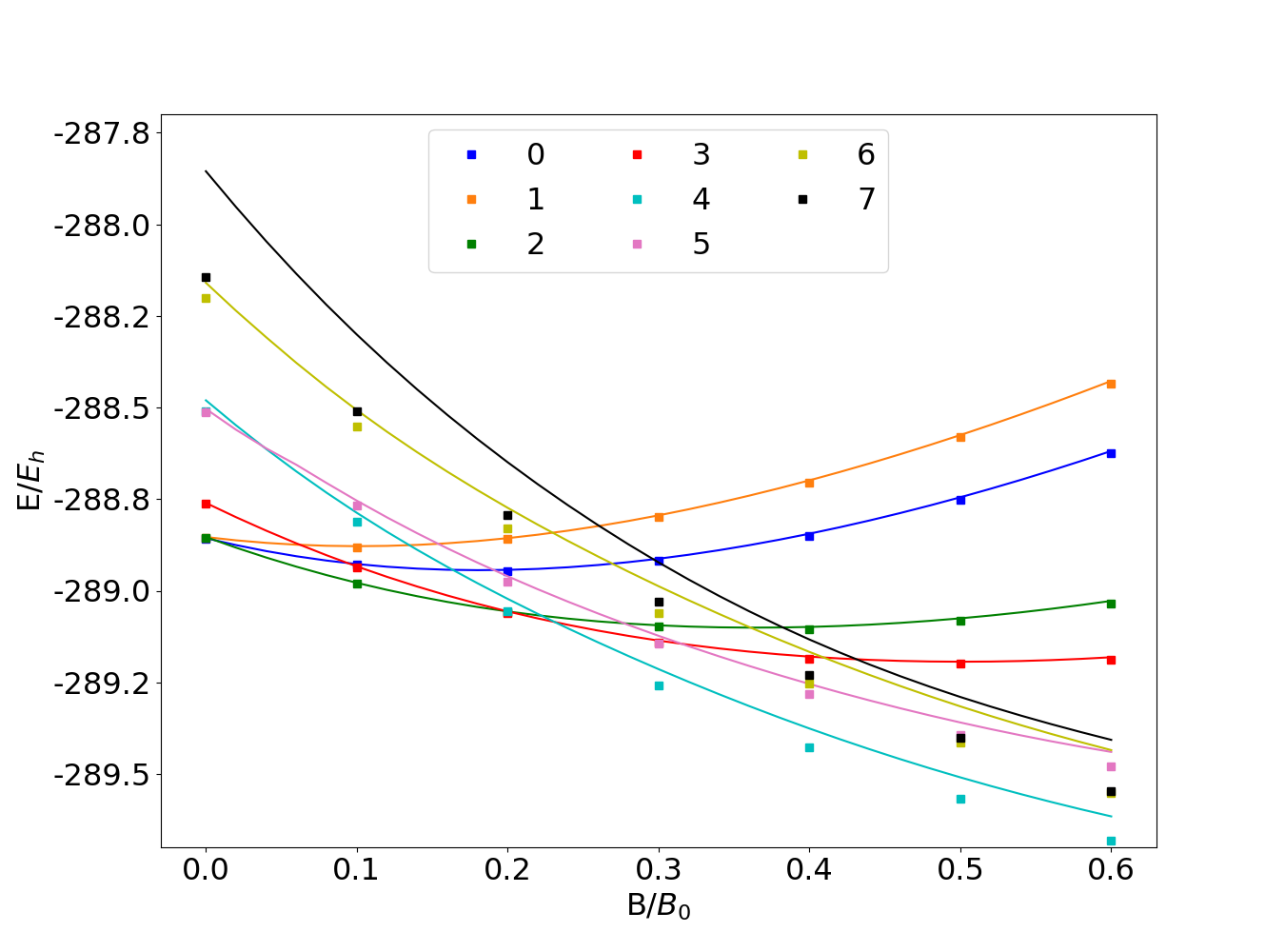}
\includegraphics[width=0.45\linewidth]{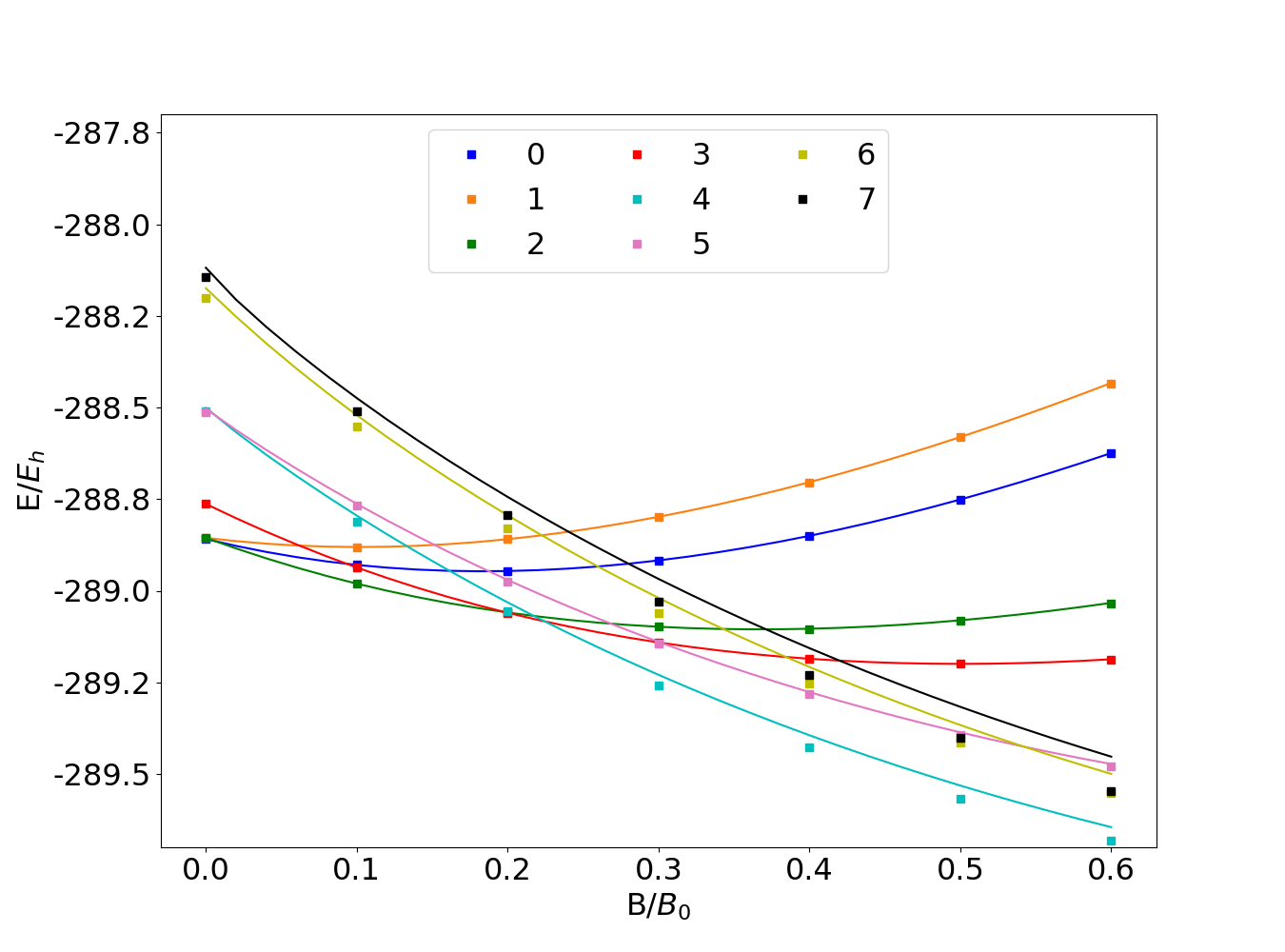}
\caption{Total energy of the Si atom as a function of the magnetic field strength $B$ in the aug-cc-pVTZ (left) and AHGBSP3-9 (right) basis sets.}
\label{fig:Si}
\end{center}
\end{figure*}

\begin{table}
\centering
\small
\begin{tabular}{llcc}
\hline
 & state & aug-cc-pVTZ & AHGBSP3-9\\ 
\hline \hline
0 & $\sigma^{4,4}\pi_+^{2,1}\pi_-^{2,1}$ & \color{blue}{ $ 4.342 $ } & \color{blue}{ $ 0.085 $ }\\ 
1 & $\sigma^{5,4}\pi_+^{2,1}\pi_-^{1,1}$ & \color{red}{ $ 4.019 $ } & \color{red}{ $ 0.203 $ }\\ 
2 & $\sigma^{5,4}\pi_+^{1,1}\pi_-^{2,1}$ & \color{red}{ $ 4.019 $ } & \color{red}{ $ 0.203 $ }\\ 
3 & $\sigma^{5,3}\pi_+^{2,1}\pi_-^{2,1}$ & \color{blue}{ $ 4.743 $ } & \color{blue}{ $ 0.186 $ }\\ 
4 & $\sigma^{5,3}\pi_+^{1,1}\pi_-^{2,1}\delta_-^{1,0}$ & \color{blue}{ $ 43.846 $ } & \color{blue}{ $ 26.202 $ }\\ 
5 & $\sigma^{5,3}\pi_+^{1,1}\pi_-^{3,1}$ & \color{blue}{ $ 23.209 $ } & \color{blue}{ $ 6.468 $ }\\ 
6 & $\sigma^{4,3}\pi_+^{1,1}\pi_-^{3,1}\delta_-^{1,0}$ & \color{blue}{ $ 74.118 $ } & \color{blue}{ $ 40.056 $ }\\ 
7 & $\sigma^{4,3}\pi_+^{1,1}\pi_-^{2,1}\delta_-^{1,0}\phi_-^{1,0}$ & \color{blue}{ $ 157.267 $ } & \color{blue}{ $ 61.168 $ }\\ 
\hline
\end{tabular}
\caption{MAEDs between GTO and FEM energies in m$E_h$ for Si in the fully uncontracted aug-cc-pVTZ and AHGBSP3-9 basis sets.}
\label{tab:Si-mean-differ}
\end{table}

The energies of the low-lying states of the Si atom are shown as a
function of the field strength in \cref{fig:Si}. The mean differences
between the FEM and GTO energies are shown in
\cref{tab:Si-mean-differ}.

Si has a ground state crossing from
$\sigma^{5,4}\pi_+^{1,1}\pi_-^{2,1}$ to
$\sigma^{5,3}\pi_+^{1,1}\pi_-^{2,1}\delta_-^{1,0}$ around
$B\approx0.2B_0$.

The states with occupied $\sigma$ and $\pi$ orbitals appear to be
adequately described in the aug-cc-pVTZ basis set with MAEDs around 4
m$E_h$. The exception is the $\sigma^{4,3}\pi_+^{1,1}\pi_-^{3,1}$
state that has a MAED of over 20 m$E_h$, while the AHGBSP3-9 basis
affords a smaller MAED which is likely dominated by artifacts from the
real-orbital approximation. The state is decently described also by the
aug-cc-pV5Z basis set.

The states with occupied $\delta$ and $\varphi$ orbitals are all badly
described in the aug-cc-pVTZ basis set. The energy differences are
still significant in the AHGBSP3-9 basis set, indicating artefacts
from the real-orbital approximation, even though the improvement in the
MAEDs over aug-cc-pVTZ is clear.

\paragraph{P \label{sec:P}}

\begin{figure*}
\begin{center}
\includegraphics[width=0.45\linewidth]{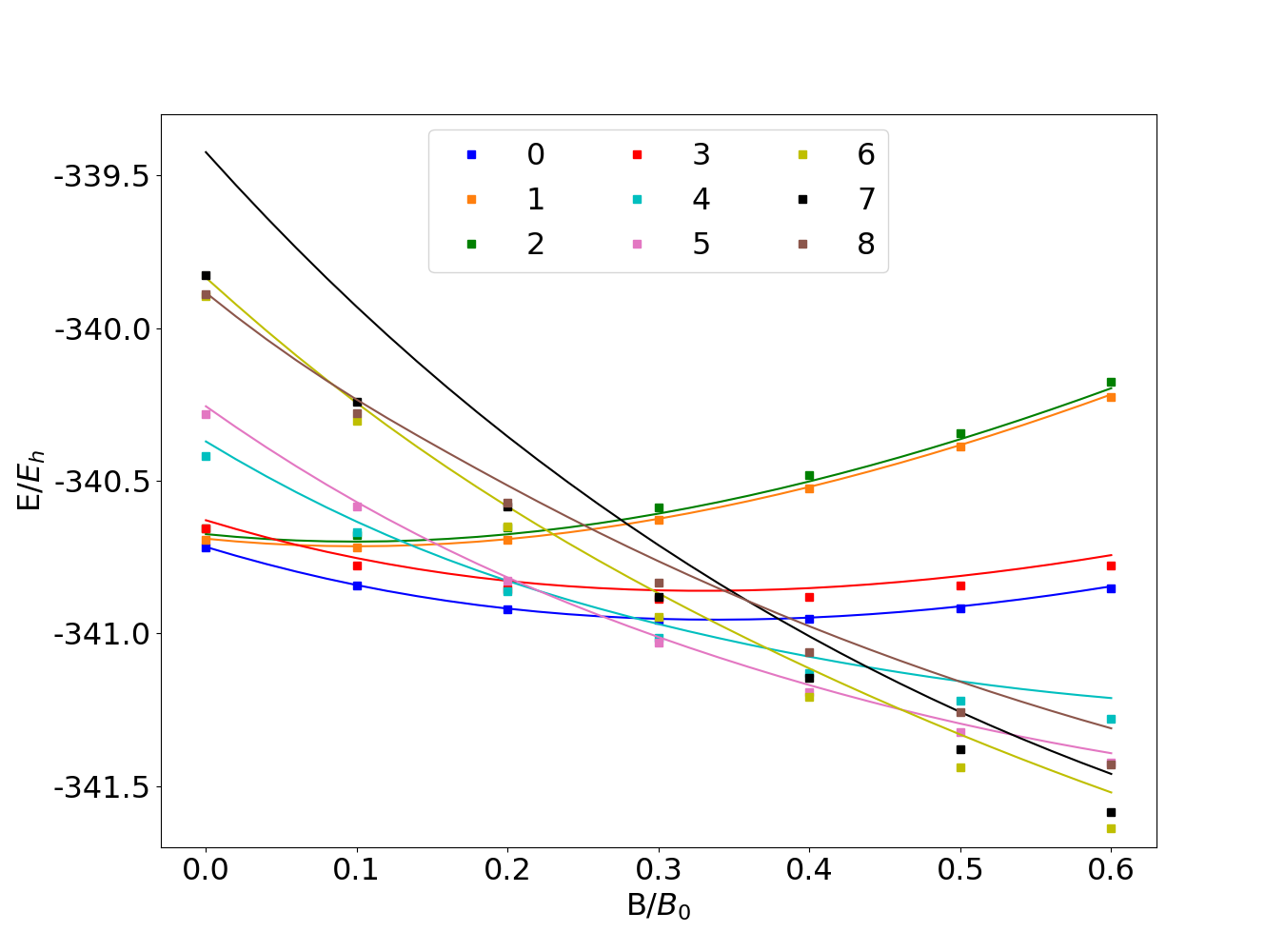}
\includegraphics[width=0.45\linewidth]{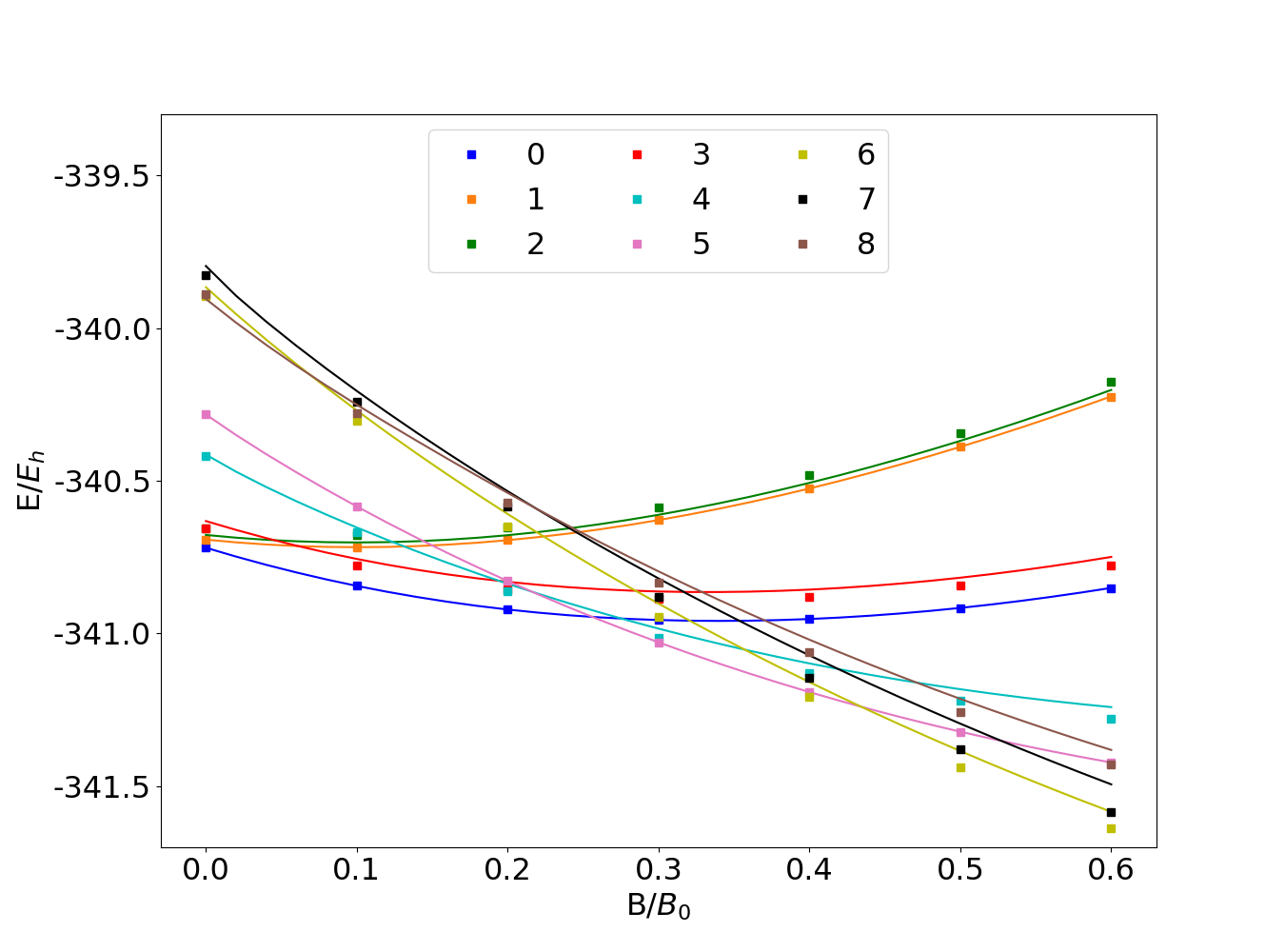}
\caption{Total energy of the P atom as a function of the magnetic field strength $B$ in the aug-cc-pVTZ (left) and AHGBSP3-9 (right) basis sets.}
\label{fig:P}
\end{center}
\end{figure*}

\begin{table}
\centering
\small
\begin{tabular}{llcc}
\hline
 & state & aug-cc-pVTZ & AHGBSP3-9\\ 
\hline \hline
0 & $\sigma^{5,4}\pi_+^{2,1}\pi_-^{2,1}$ & \color{blue}{ $ 3.930 $ } & \color{blue}{ $ 0.085 $ }\\ 
1 & $\sigma^{4,5}\pi_+^{2,1}\pi_-^{2,1}$ & \color{blue}{ $ 4.585 $ } & \color{blue}{ $ 0.126 $ }\\ 
2 & $\sigma^{5,4}\pi_+^{1,2}\pi_-^{2,1}$ & \color{red}{ $ 20.651 $ } & \color{red}{ $ 24.685 $ }\\ 
3 & $\sigma^{5,4}\pi_+^{1,1}\pi_-^{2,2}$ & \color{blue}{ $ 28.453 $ } & \color{blue}{ $ 24.353 $ }\\ 
4 & $\sigma^{5,4}\pi_+^{1,1}\pi_-^{2,1}\delta_-^{1,0}$ & \color{blue}{ $ 49.091 $ } & \color{blue}{ $ 25.662 $ }\\ 
5 & $\sigma^{5,3}\pi_+^{2,1}\pi_-^{2,1}\delta_-^{1,0}$ & \color{blue}{ $ 21.605 $ } & \color{blue}{ $ 0.309 $ }\\ 
6 & $\sigma^{5,3}\pi_+^{1,1}\pi_-^{3,1}\delta_-^{1,0}$ & \color{blue}{ $ 82.647 $ } & \color{blue}{ $ 43.272 $ }\\ 
7 & $\sigma^{5,3}\pi_+^{1,1}\pi_-^{2,1}\delta_-^{1,0}\phi_-^{1,0}$ & \color{blue}{ $ 213.458 $ } & \color{blue}{ $ 60.514 $ }\\ 
8 & $\sigma^{6,3}\pi_+^{1,1}\pi_-^{2,1}\delta_-^{1,0}$ & \color{red}{ $ 67.681 $ } & \color{red}{ $ 34.548 $ }\\ 
\hline
\end{tabular}
\caption{MAEDs between GTO and FEM energies in m$E_h$ for P in the fully uncontracted aug-cc-pVTZ and AHGBSP3-9 basis sets.}
\label{tab:P-mean-differ}
\end{table}

The energies of the low-lying states of the P atom are shown as a
function of the field strength in \cref{fig:P}. The mean differences
between the FEM and GTO energies are shown in
\cref{tab:P-mean-differ}.

P exhibits two ground state crossings: from
$\sigma^{5,4}\pi_+^{2,1}\pi_-^{2,1}$ to
$\sigma^{5,3}\pi_+^{2,1}\pi_-^{2,1}\delta_-^{1,0}$ around
$B\approx0.25B_0$ and then again to
$\sigma^{5,3}\pi_+^{1,1}\pi_-^{3,1}\delta_-^{1,0}$ around
$B\approx0.4B_0$.

The only states well described by aug-cc-pVTZ are
$\sigma^{5,4}\pi_+^{2,1}\pi_-^{2,1}$ and
$\sigma^{4,5}\pi_+^{2,1}\pi_-^{2,1}$. The significant differences
observed for the $\sigma^{5,4}\pi_+^{1,2}\pi_-^{2,1}$ and
$\sigma^{5,4}\pi_+^{1,1}\pi_-^{2,2}$ states are of similar magnitude
for both GTO basis sets and likely arise from the real-orbital
approximation of \cref{sec:real-approx}.

The improvement for all states with occupied $\delta$ and $\varphi$
orbitals when going from aug-cc-pVTZ to AHGBSP3-9 is clear. However,
only the $\sigma^{5,3}\pi_+^{2,1}\pi_-^{2,1}\delta_-^{1,0}$ state has
a small MAED in AHGBSP3-9, the MAEDs for the other states likely
arising from the real-orbital approximation used in the GTO
calculations.

\paragraph{S \label{sec:S}}

\begin{figure*}
\begin{center}
\includegraphics[width=0.45\linewidth]{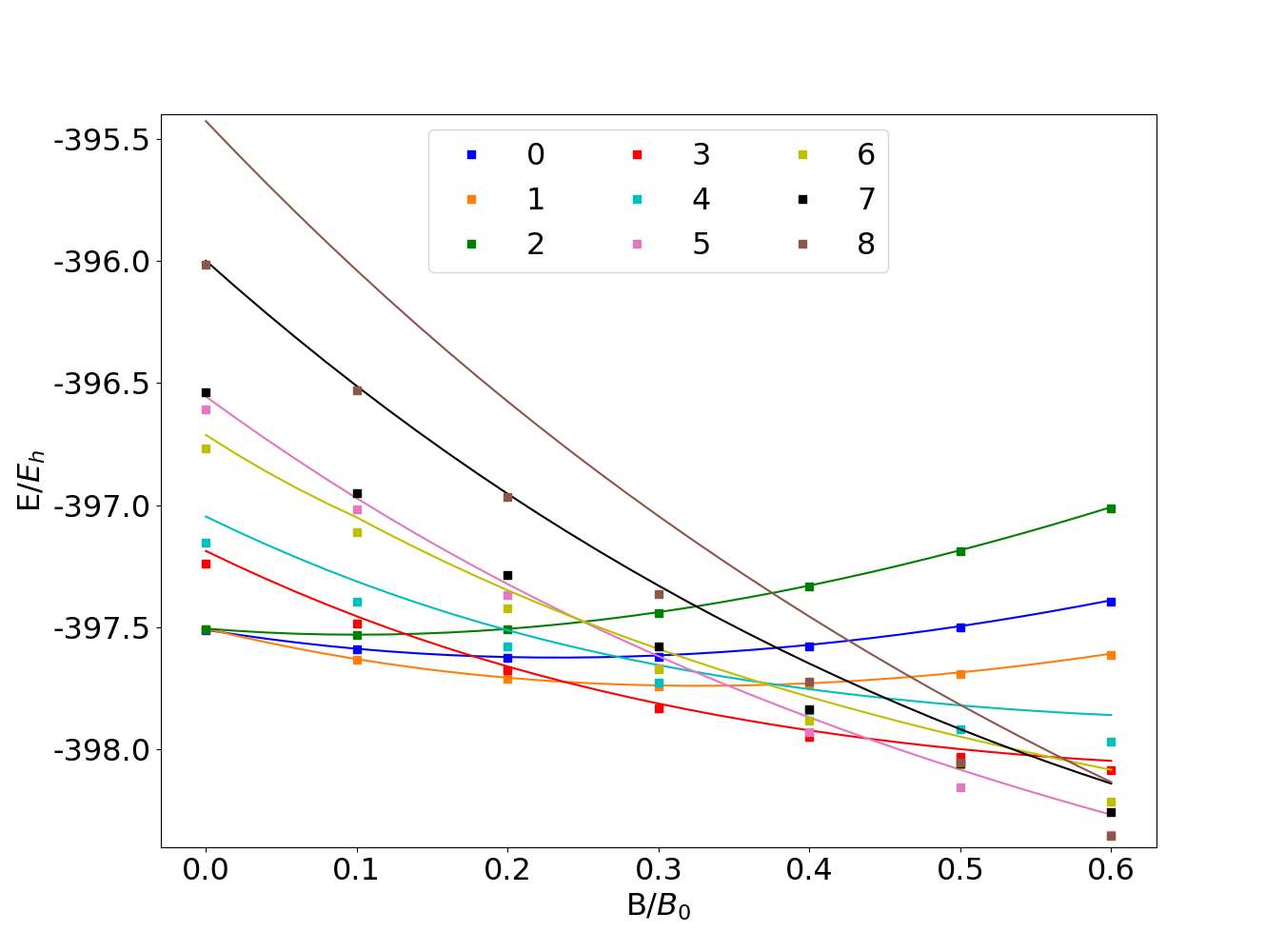}
\includegraphics[width=0.45\linewidth]{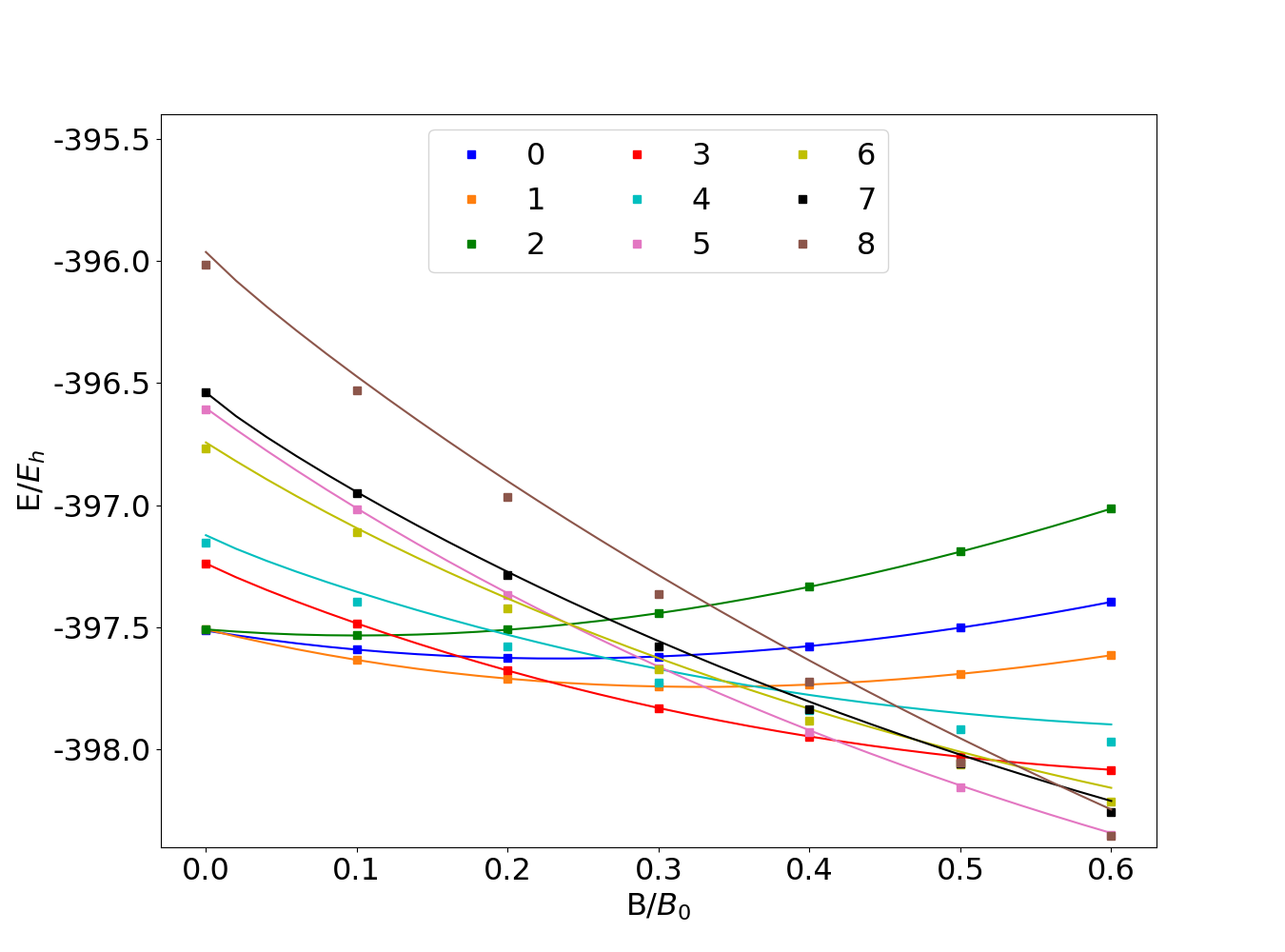}
\caption{Total energy of the S atom as a function of the magnetic field strength $B$ in the aug-cc-pVTZ (left) and AHGBSP3-9 (right) basis sets.}
\label{fig:S}
\end{center}
\end{figure*}

\begin{table}
\centering
\small
\begin{tabular}{llcc}
\hline
 & state & aug-cc-pVTZ & AHGBSP3-9\\ 
\hline \hline
0 & $\sigma^{5,5}\pi_+^{2,1}\pi_-^{2,1}$ & \color{blue}{ $ 4.813 $ } & \color{blue}{ $ 0.065 $ }\\ 
1 & $\sigma^{5,4}\pi_+^{2,1}\pi_-^{2,2}$ & \color{red}{ $ 4.399 $ } & \color{red}{ $ 0.194 $ }\\ 
2 & $\sigma^{5,4}\pi_+^{2,2}\pi_-^{2,1}$ & \color{red}{ $ 4.399 $ } & \color{red}{ $ 0.194 $ }\\ 
3 & $\sigma^{5,4}\pi_+^{2,1}\pi_-^{2,1}\delta_-^{1,0}$ & \color{blue}{ $ 29.830 $ } & \color{blue}{ $ 0.259 $ }\\ 
4 & $\sigma^{5,4}\pi_+^{1,1}\pi_-^{2,2}\delta_-^{1,0}$ & \color{blue}{ $ 88.120 $ } & \color{blue}{ $ 52.820 $ }\\ 
5 & $\sigma^{5,3}\pi_+^{2,1}\pi_-^{3,1}\delta_-^{1,0}$ & \color{blue}{ $ 57.719 $ } & \color{blue}{ $ 6.280 $ }\\ 
6 & $\sigma^{5,4}\pi_+^{1,1}\pi_-^{3,1}\delta_-^{1,0}$ & \color{blue}{ $ 88.238 $ } & \color{blue}{ $ 41.133 $ }\\ 
7 & $\sigma^{5,3}\pi_+^{2,1}\pi_-^{2,1}\delta_-^{1,0}\phi_-^{1,0}$ & \color{blue}{ $ 286.259 $ } & \color{blue}{ $ 22.000 $ }\\ 
8 & $\sigma^{5,3}\pi_+^{1,1}\pi_-^{3,1}\delta_-^{1,0}\phi_-^{1,0}$ & \color{blue}{ $ 358.692 $ } & \color{blue}{ $ 77.808 $ }\\ 
\hline
\end{tabular}
\caption{MAEDs between GTO and FEM energies in m$E_h$ for S in the fully uncontracted aug-cc-pVTZ and AHGBSP3-9 basis sets.}
\label{tab:S-mean-differ}
\end{table}

The energies of the low-lying states of the S atom are shown as a
function of the field strength in \cref{fig:S}. The mean differences
between the FEM and GTO energies are shown in
\cref{tab:S-mean-differ}.

S is the only element to have three ground state crossings: the ground
state first changes from $\sigma^{5,4}\pi_+^{2,1}\pi_-^{2,2}$ to
$\sigma^{5,4}\pi_+^{2,1}\pi_-^{2,1}\delta_-^{1,0}$ around
$B\approx0.225B_0$, then to
$\sigma^{5,3}\pi_+^{2,1}\pi_-^{3,1}\delta_-^{1,0}$ around
$B\approx0.4B_0$, and finally to
$\sigma^{5,3}\pi_+^{1,1}\pi_-^{3,1}\delta_-^{1,0}\phi_-^{1,0}$ around
$B\approx0.6B_0$.

All the states with occupied $\sigma$ and $\pi$ orbitals are again
well described by both GTO basis sets. The
$\sigma^{5,4}\pi_+^{2,1}\pi_-^{2,1}\delta_-^{1,0}$ and
$\sigma^{5,3}\pi_+^{2,1}\pi_-^{3,1}\delta_-^{1,0}$ states are
ill-described by aug-cc-pVTZ but well recovered by AHGBSP3-9,
suggesting room to improve on standard basis sets.

The $\sigma^{5,4}\pi_+^{1,1}\pi_-^{2,2}\delta_-^{1,0}$ and
$\sigma^{5,4}\pi_+^{1,1}\pi_-^{3,1}\delta_-^{1,0}$ states show
considerable improvement going from aug-cc-pVTZ to AHGBSP3-9, even
though these remaining MAEDs are still significant and likely caused
by the real-orbital approximation.

The improvement for the states with occupied $\varphi$ orbitals is
drastic when going from aug-cc-pVTZ to AHGBSP3-9, again indicating
room to improve on standard basis sets, even though the remaining
MAEDs are large, which is likely an artefact of the real-orbital
approximation.

\paragraph{Cl \label{sec:Cl}}

\begin{figure*}
\begin{center}
\includegraphics[width=0.45\linewidth]{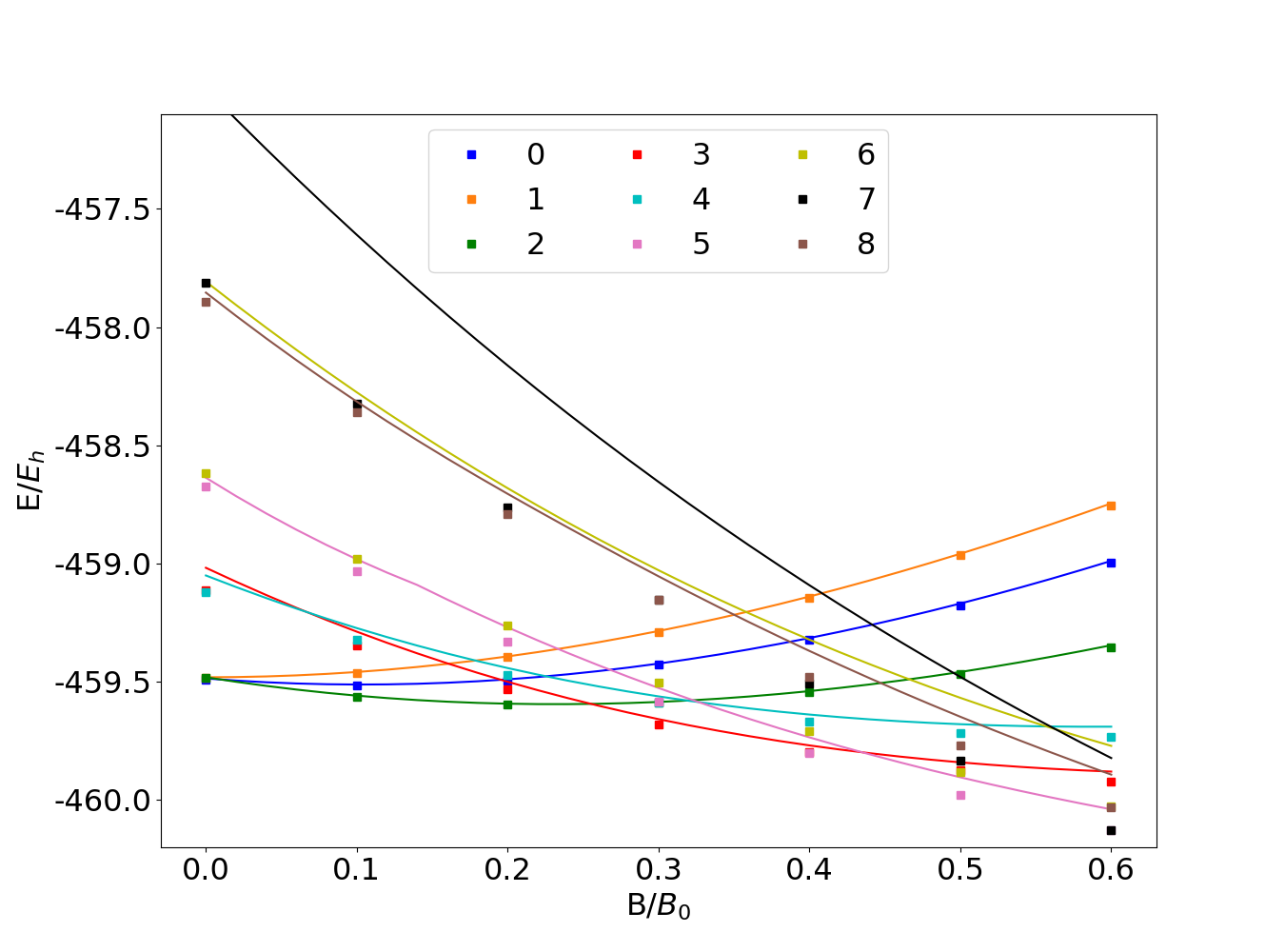}
\includegraphics[width=0.45\linewidth]{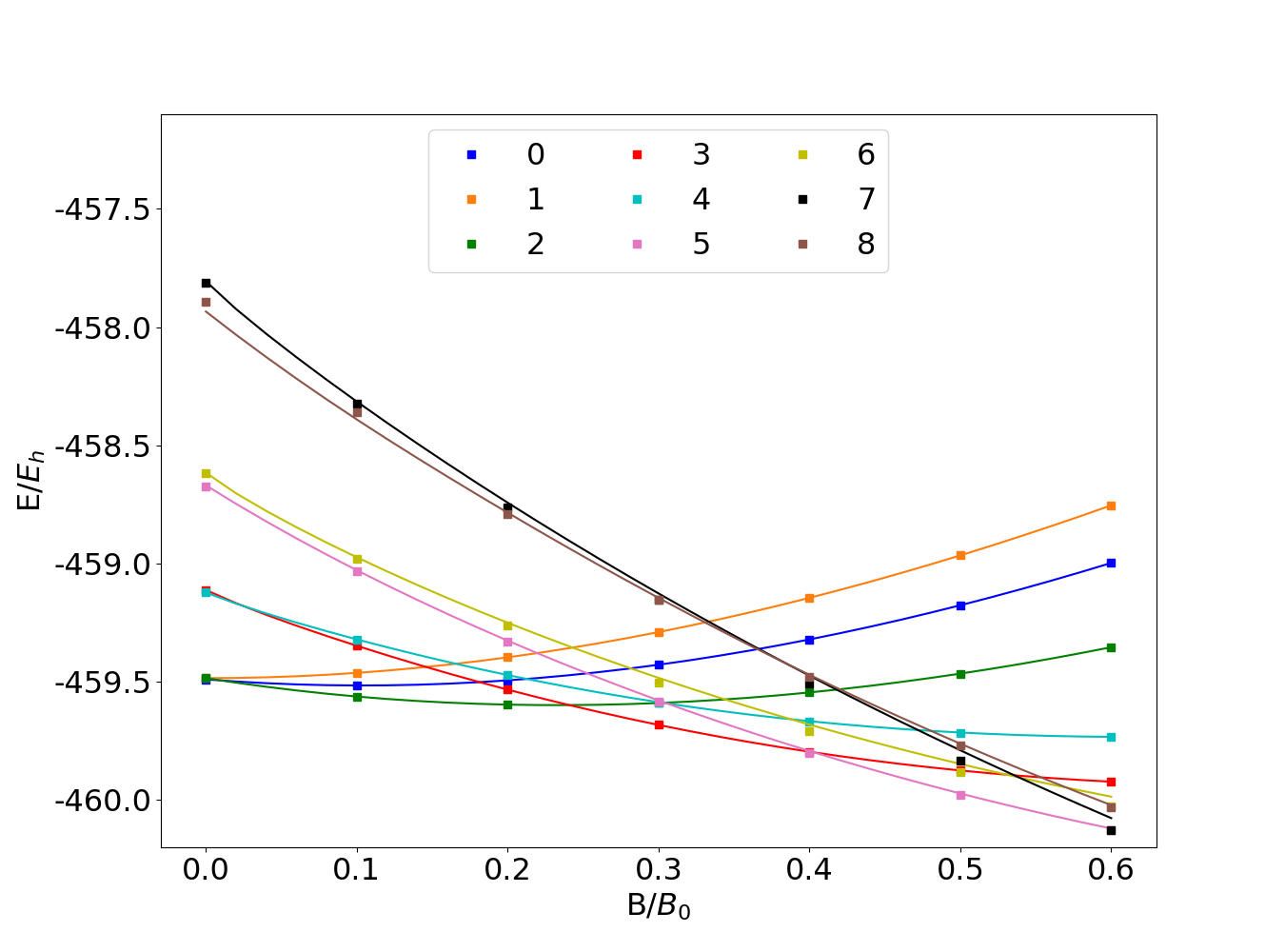}
\caption{Total energy of the Cl atom as a function of the magnetic field strength $B$ in the aug-cc-pVTZ (left) and AHGBSP3-9 (right) basis sets.}
\label{fig:Cl}
\end{center}
\end{figure*}

\begin{table}
\centering
\small
\begin{tabular}{llcc}
\hline
 & state & aug-cc-pVTZ & AHGBSP3-9\\ 
\hline \hline
0 & $\sigma^{5,4}\pi_+^{2,2}\pi_-^{2,2}$ & \color{blue}{ $ 5.634 $ } & \color{blue}{ $ 0.033 $ }\\ 
1 & $\sigma^{5,5}\pi_+^{2,2}\pi_-^{2,1}$ & \color{red}{ $ 5.057 $ } & \color{red}{ $ 0.223 $ }\\ 
2 & $\sigma^{5,5}\pi_+^{2,1}\pi_-^{2,2}$ & \color{red}{ $ 5.057 $ } & \color{red}{ $ 0.223 $ }\\ 
3 & $\sigma^{5,4}\pi_+^{2,1}\pi_-^{2,2}\delta_-^{1,0}$ & \color{red}{ $ 44.993 $ } & \color{red}{ $ 0.095 $ }\\ 
4 & $\sigma^{5,5}\pi_+^{2,1}\pi_-^{2,1}\delta_-^{1,0}$ & \color{blue}{ $ 40.283 $ } & \color{blue}{ $ 0.219 $ }\\ 
5 & $\sigma^{5,4}\pi_+^{2,1}\pi_-^{3,1}\delta_-^{1,0}$ & \color{blue}{ $ 62.755 $ } & \color{blue}{ $ 6.355 $ }\\ 
6 & $\sigma^{5,4}\pi_+^{2,1}\pi_-^{2,1}\delta_-^{1,0}\phi_-^{1,0}$ & \color{blue}{ $ 504.280 $ } & \color{blue}{ $ 20.978 $ }\\ 
7 & $\sigma^{5,3}\pi_+^{2,1}\pi_-^{3,1}\delta_-^{1,0}\phi_-^{1,0}$ & \color{blue}{ $ 529.770 $ } & \color{blue}{ $ 27.293 $ }\\ 
8 & $\sigma^{6,3}\pi_+^{2,1}\pi_-^{3,1}\delta_-^{1,0}$ & \color{red}{ $ 91.583 $ } & \color{red}{ $ 16.412 $ }\\ 
\hline
\end{tabular}
\caption{MAEDs between GTO and FEM energies in m$E_h$ for Cl in the fully uncontracted aug-cc-pVTZ and AHGBSP3-9 basis sets.}
\label{tab:Cl-mean-differ}
\end{table}

The energies of the low-lying states of the Cl atom are shown as a
function of the field strength in \cref{fig:Cl}. The mean differences
between the FEM and GTO energies are shown in
\cref{tab:Cl-mean-differ}.

Cl exhibits two ground state crossings: from the field-free ground
state configuration $\sigma^{5,5}\pi_+^{2,1}\pi_-^{2,2}$ to
$\sigma^{5,4}\pi_+^{2,1}\pi_-^{2,2}\delta_-^{1,0}$ around
$B\approx0.25B_0$, and then to
$\sigma^{5,4}\pi_+^{2,1}\pi_-^{3,1}\delta_-^{1,0}$ around
$B\approx0.4B_0$. We again see that all the states with occupied
$\sigma$ and $\pi$ orbitals are well described by both GTO basis
sets.

We likewise again observe that the description of the states with
occupied $\delta$ and $\varphi$ orbitals can be significantly improved
by going from aug-cc-pVTZ to AHGBSP3-9, showing room to improve on
standard basis sets for finite field calculations.

\paragraph{Ar \label{sec:Ar}}

\begin{figure*}
\begin{center}
\includegraphics[width=0.45\linewidth]{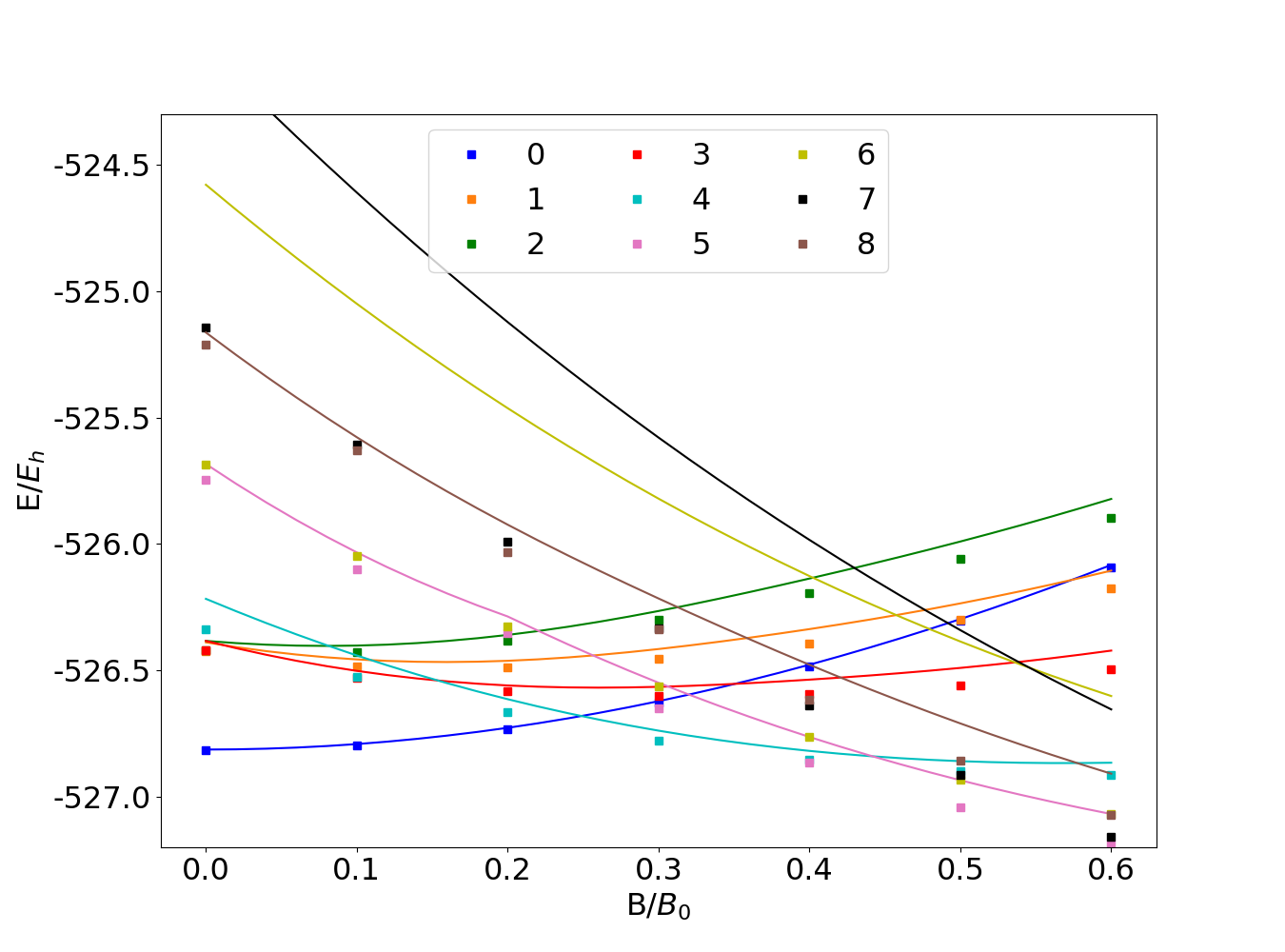}
\includegraphics[width=0.45\linewidth]{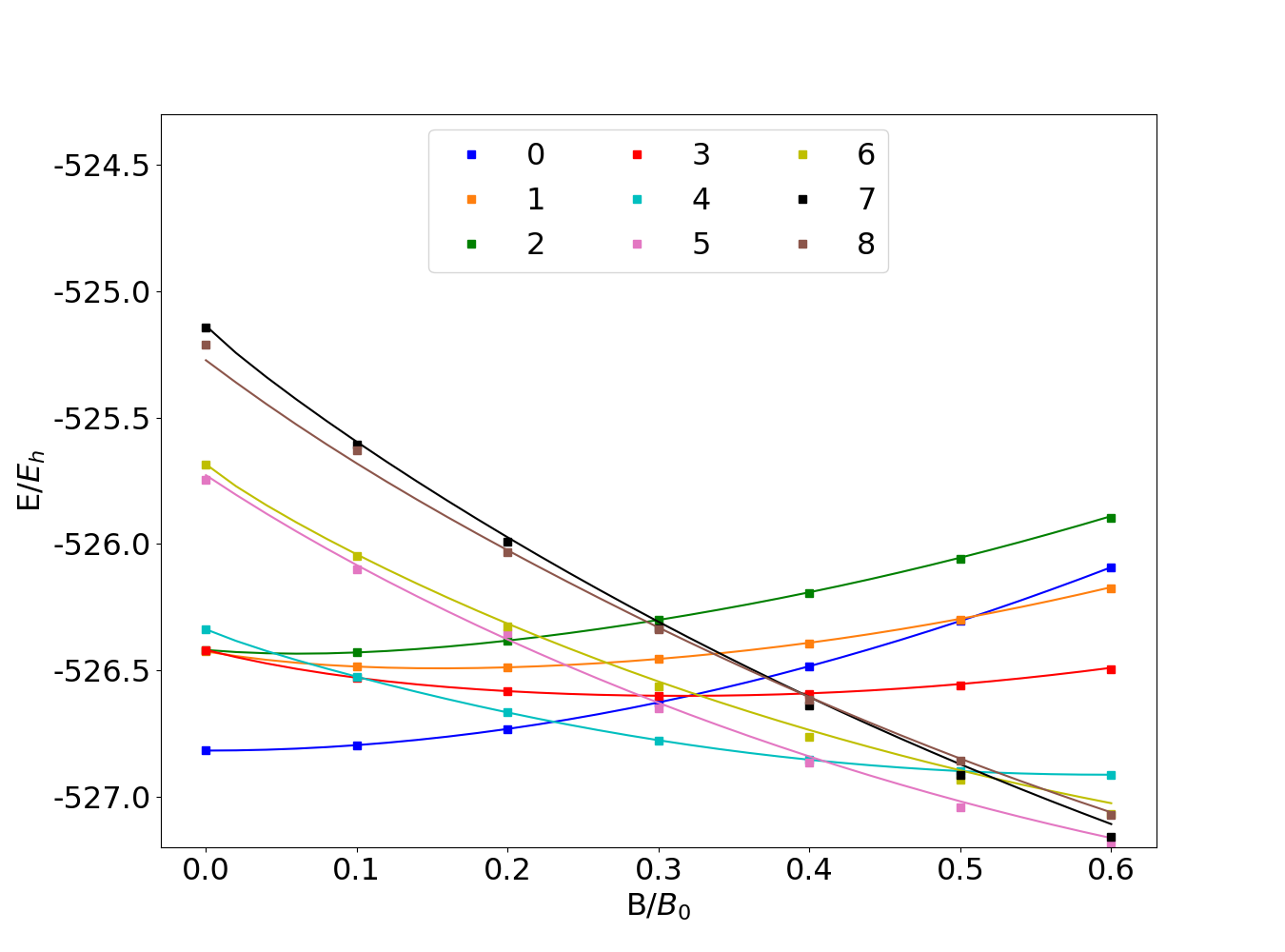}
\caption{Total energy of the Ar atom as a function of the magnetic field strength $B$ in the aug-cc-pVTZ (left) and AHGBSP3-9 (right) basis sets.}
\label{fig:Ar}
\end{center}
\end{figure*}

\begin{table}
\centering
\small
\begin{tabular}{llcc}
\hline
 & state & aug-cc-pVTZ & AHGBSP3-9\\ 
\hline \hline
0 & $\sigma^{5,5}\pi_+^{2,2}\pi_-^{2,2}$ & \color{blue}{ $ 5.710 $ } & \color{blue}{ $ 0.016 $ }\\ 
1 & $\sigma^{6,4}\pi_+^{2,2}\pi_-^{2,2}$ & \color{blue}{ $ 46.060 $ } & \color{blue}{ $ 1.752 $ }\\ 
2 & $\sigma^{6,5}\pi_+^{2,2}\pi_-^{2,1}$ & \color{red}{ $ 45.779 $ } & \color{red}{ $ 2.119 $ }\\ 
3 & $\sigma^{6,5}\pi_+^{2,1}\pi_-^{2,2}$ & \color{red}{ $ 45.779 $ } & \color{red}{ $ 2.119 $ }\\ 
4 & $\sigma^{5,5}\pi_+^{2,1}\pi_-^{2,2}\delta_-^{1,0}$ & \color{red}{ $ 59.510 $ } & \color{red}{ $ 0.062 $ }\\ 
5 & $\sigma^{5,4}\pi_+^{2,1}\pi_-^{3,2}\delta_-^{1,0}$ & \color{red}{ $ 88.711 $ } & \color{red}{ $ 21.867 $ }\\ 
6 & $\sigma^{5,4}\pi_+^{2,1}\pi_-^{2,2}\delta_-^{1,0}\phi_-^{1,0}$ & \color{blue}{ $ 765.591 $ } & \color{blue}{ $ 20.442 $ }\\ 
7 & $\sigma^{5,4}\pi_+^{2,1}\pi_-^{3,1}\delta_-^{1,0}\phi_-^{1,0}$ & \color{blue}{ $ 779.603 $ } & \color{blue}{ $ 26.642 $ }\\ 
8 & $\sigma^{6,4}\pi_+^{2,1}\pi_-^{3,1}\delta_-^{1,0}$ & \color{red}{ $ 111.424 $ } & \color{red}{ $ 22.786 $ }\\ 
\hline
\end{tabular}
\caption{MAEDs between GTO and FEM energies in m$E_h$ for Ar in the fully uncontracted aug-cc-pVTZ and AHGBSP3-9 basis sets.}
\label{tab:Ar-mean-differ}
\end{table}

The energies of the low-lying states of the Ar atom are shown as a
function of the field strength in \cref{fig:Ar}. The mean differences
between the FEM and GTO energies are shown in \cref{tab:Ar-mean-differ}.

Ar has two ground state crossings: from
$\sigma^{5,5}\pi_+^{2,2}\pi_-^{2,2}$ to
$\sigma^{5,5}\pi_+^{2,1}\pi_-^{2,2}\delta_-^{1,0}$ around
$B\approx0.225B_0$, and then again to
$\sigma^{5,4}\pi_+^{2,1}\pi_-^{3,2}\delta_-^{1,0}$ around
$B\approx0.4B_0$.

Only the $\sigma^{5,5}\pi_+^{2,2}\pi_-^{2,2}$ state is well described
by aug-cc-pVTZ. $\sigma^{6,4}\pi_+^{2,2}\pi_-^{2,2}$ exhibits a large
MAED in aug-cc-pVTZ, which is significantly reduced in AHGBSP3-9; the
other states with occupied $\sigma$ and $\pi$ orbitals exhibit similar
behavior.

The MAED of the $\sigma^{5,5}\pi_+^{2,1}\pi_-^{2,2}\delta_-^{1,0}$
state is large in aug-cc-pVTZ, and small in AHGBSP3-9. The other
states with occupied $\delta$ orbitals also see drastic improvements
when going to AHGBSP3-9, proving that these states can be described
significantly better by improved basis sets.

The states with occupied $\varphi$ orbitals are extremely badly
described in aug-cc-pVTZ, and much better described in AHGBSP3-9 whose
MAEDs are likely dominated by the real-orbital approximation.

\section{Summary and Conclusions}
\label{sec:summary}

We have determined complete basis set (CBS) limit energies of the low
lying states of H--Ar in magnetic fields of $B\in[0,0.6B_0]$ at the
unrestricted Hartree--Fock (UHF) level of theory with fully numerical
calculations with the finite element method (FEM), employing complex
wave functions.

We have also suggested a real-orbital approximation for calculations
with Gaussian-type orbital (GTO) basis sets, which we have employed to
carry out calculations in a large variety of GTO basis sets.

We have computed energy differences between the GTO basis set and FEM
calculations to identify atomic states that are poorly described by
the standard GTO basis sets optimized for zero field, and indicated
several states that could be described significantly more accurately
with GTO basis sets optimized for finite field calculations with
London atomic orbitals (LAOs), also known as gauge-including atomic
orbitals (GIAOs).

In general we observe that states with high $\langle
-\hat{L}_z\rangle$ that become important at stronger magnetic fields
due to the orbital Zeeman term are poorly described in the aug-cc-pVTZ
basis set. We notice that the benchmark quality AHGBSP3-9 basis can
often recover these states to high accuracy.

Larger errors are also encountered by higher spin states, which
similarly couple to the magnetic field by the spin Zeeman term, and
which become the ground state at stronger fields. These larger errors
are likely caused by the differences in the spatial form of the
orbitals: the higher spin state has more electrons of the same spin,
which have to obey Pauli's exclusion principle. This results in a
difference in the spatial form that is not taken into account in
standard basis sets optimized at zero field. Also these states are
well described by the benchmark quality AHGBSP3-9 basis set.


Some states appear to be ill-described even by the very large
AHGBSP3-9 basis set.  We believe these discrepancies to stem from the
real-orbital approximation employed in this work, which was described
in \cref{sec:real-approx}. Even when large mean absolute energy
differences (MAED) are observed for the AHGBSP3-9 basis set, we do
observe significant reductions of MAED from the aug-cc-pVTZ basis set.

The use of complex GTOs could be visited in later work, as they will
enable apples-to-apples studies of the MAED, affording direct access
into the basis set truncation error (BSTE). The physical Hamiltonian
could also be recovered with the use of real spherical harmonics by
employing a complex Hamiltonian matrix. Although such complex wave
functions may be supported in \Erkale{} in the future, the results of
this work are already sufficient to serve as a basis for developing
improved Gaussian basis sets for calculations at finite magnetic
fields: our results indicate that basis sets taylored for calculations
at finite magnetic fields can be constructed with the approximate
method employed in this work.

\begin{acknowledgement}
We thank the Academy of Finland for financial support under project
numbers 350282 and 353749. We also thank the Finnish Society for
Sciences and Letters for financial support.
\end{acknowledgement}

\begin{tocentry}
\includegraphics[width=3.25in,height=1.75in,keepaspectratio]{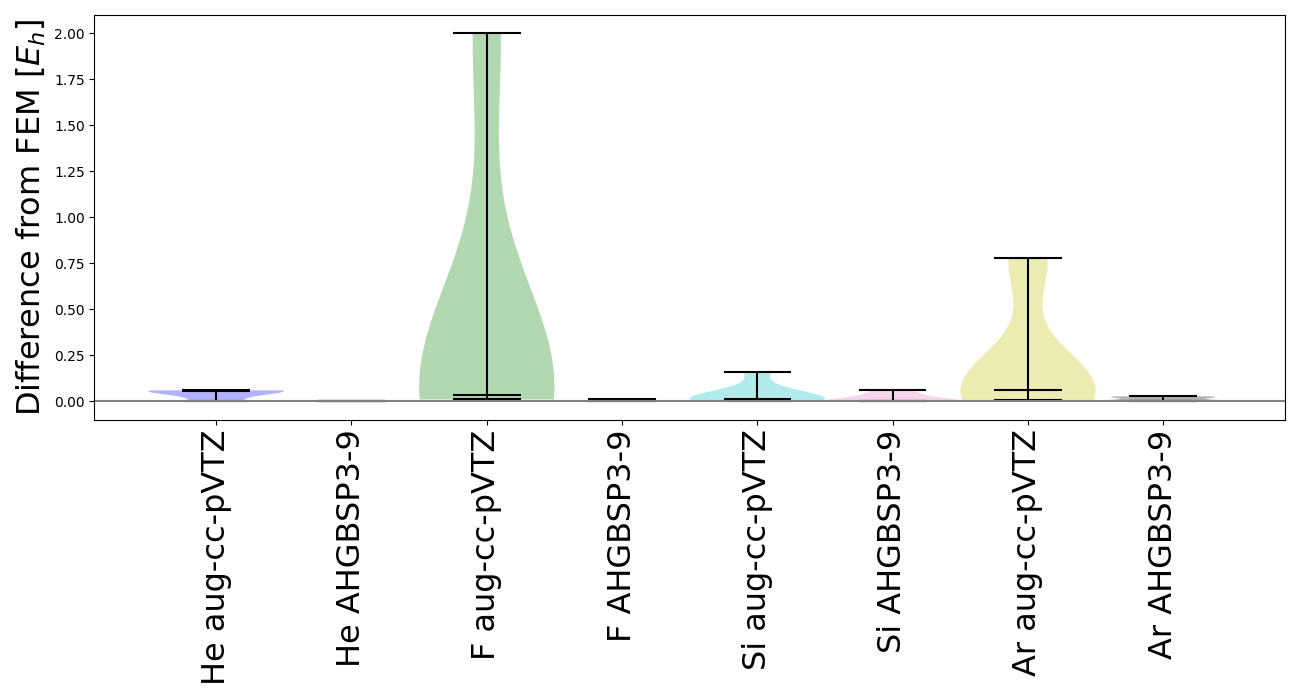}
\end{tocentry}

\section*{Supporting Information Available}

Plots demonstrating the convergence of the numerical basis sets to the
CBS limit, and tables of the numerical CBS limit energies. Mean
absolute energy differences for each state of each atom in each of the
studied basis sets. Plots of the total energies of each state of each
atom as a function of the magnetic field strength in each of the
studied basis sets. Tables of total energies for all states of all
atoms over the studied range of magnetic field strengths.

\bibliography{citations}

\providecommand{\latin}[1]{#1}
\makeatletter
\providecommand{\doi}
  {\begingroup\let\do\@makeother\dospecials
  \catcode`\{=1 \catcode`\}=2 \doi@aux}
\providecommand{\doi@aux}[1]{\endgroup\texttt{#1}}
\makeatother
\providecommand*\mcitethebibliography{\thebibliography}
\csname @ifundefined\endcsname{endmcitethebibliography}
  {\let\endmcitethebibliography\endthebibliography}{}
\begin{mcitethebibliography}{79}
\providecommand*\natexlab[1]{#1}
\providecommand*\mciteSetBstSublistMode[1]{}
\providecommand*\mciteSetBstMaxWidthForm[2]{}
\providecommand*\mciteBstWouldAddEndPuncttrue
  {\def\EndOfBibitem{\unskip.}}
\providecommand*\mciteBstWouldAddEndPunctfalse
  {\let\EndOfBibitem\relax}
\providecommand*\mciteSetBstMidEndSepPunct[3]{}
\providecommand*\mciteSetBstSublistLabelBeginEnd[3]{}
\providecommand*\EndOfBibitem{}
\mciteSetBstSublistMode{f}
\mciteSetBstMaxWidthForm{subitem}{(\alph{mcitesubitemcount})}
\mciteSetBstSublistLabelBeginEnd
  {\mcitemaxwidthsubitemform\space}
  {\relax}
  {\relax}

\bibitem[Angel(1977)]{Angel1977_AJ_1}
Angel,~J. R.~P. {Magnetism in white dwarfs}. \emph{Astrophys. J.}
  \textbf{1977}, \emph{216}, 1--17\relax
\mciteBstWouldAddEndPuncttrue
\mciteSetBstMidEndSepPunct{\mcitedefaultmidpunct}
{\mcitedefaultendpunct}{\mcitedefaultseppunct}\relax
\EndOfBibitem
\bibitem[Schmidt \latin{et~al.}(1995)Schmidt, Bergeron, and
  Fegley]{Schmidt1995_AJ_274}
Schmidt,~G.~D.; Bergeron,~P.; Fegley,~B. {On the nature of spectral features in
  peculiar DQ white dwarfs}. \emph{Astrophys. J.} \textbf{1995}, \emph{443},
  274\relax
\mciteBstWouldAddEndPuncttrue
\mciteSetBstMidEndSepPunct{\mcitedefaultmidpunct}
{\mcitedefaultendpunct}{\mcitedefaultseppunct}\relax
\EndOfBibitem
\bibitem[Jordan \latin{et~al.}(1998)Jordan, Schmelcher, Becken, and
  Schweizer]{Jordan1998__}
Jordan,~S.; Schmelcher,~P.; Becken,~W.; Schweizer,~W. Evidence for helium in
  the magnetic white dwarf {GD229}. 1998\relax
\mciteBstWouldAddEndPuncttrue
\mciteSetBstMidEndSepPunct{\mcitedefaultmidpunct}
{\mcitedefaultendpunct}{\mcitedefaultseppunct}\relax
\EndOfBibitem
\bibitem[Wickramasinghe and Ferrario(2000)Wickramasinghe, and
  Ferrario]{Wickramasinghe2000_PASP_873}
Wickramasinghe,~D.~T.; Ferrario,~L. Magnetism in Isolated and Binary White
  Dwarfs. \emph{Publ. Astron. Soc. Pacific} \textbf{2000}, \emph{112},
  873--924\relax
\mciteBstWouldAddEndPuncttrue
\mciteSetBstMidEndSepPunct{\mcitedefaultmidpunct}
{\mcitedefaultendpunct}{\mcitedefaultseppunct}\relax
\EndOfBibitem
\bibitem[Liebert \latin{et~al.}(2003)Liebert, Harris, Dahn, Schmidt, Kleinman,
  Nitta, Krzesiski, Eisenstein, Smith, Szkody, Hawley, Anderson, Brinkmann,
  Collinge, Fan, Hall, Knapp, Lamb, Margon, Schneider, and
  Silvestri]{Liebert2003_AJ_2521}
Liebert,~J.; Harris,~H.~C.; Dahn,~C.~C.; Schmidt,~G.~D.; Kleinman,~S.~J.;
  Nitta,~A.; Krzesiski,~J.; Eisenstein,~D.; Smith,~J.~A.; Szkody,~P.;
  Hawley,~S.; Anderson,~S.~F.; Brinkmann,~J.; Collinge,~M.~J.; Fan,~X.;
  Hall,~P.~B.; Knapp,~G.~R.; Lamb,~D.~Q.; Margon,~B.; Schneider,~D.~P.;
  Silvestri,~N. {SDSS} White Dwarfs with Spectra Showing Atomic Oxygen and/or
  Carbon Lines. \emph{Astron. J.} \textbf{2003}, \emph{126}, 2521--2528\relax
\mciteBstWouldAddEndPuncttrue
\mciteSetBstMidEndSepPunct{\mcitedefaultmidpunct}
{\mcitedefaultendpunct}{\mcitedefaultseppunct}\relax
\EndOfBibitem
\bibitem[Tellgren \latin{et~al.}(2008)Tellgren, Soncini, and
  Helgaker]{Tellgren2008_JCP_154114}
Tellgren,~E.~I.; Soncini,~A.; Helgaker,~T. {Nonperturbative ab initio
  calculations in strong magnetic fields using London orbitals}. \emph{J. Chem.
  Phys.} \textbf{2008}, \emph{129}, 154114\relax
\mciteBstWouldAddEndPuncttrue
\mciteSetBstMidEndSepPunct{\mcitedefaultmidpunct}
{\mcitedefaultendpunct}{\mcitedefaultseppunct}\relax
\EndOfBibitem
\bibitem[Tellgren \latin{et~al.}(2012)Tellgren, Reine, and
  Helgaker]{Tellgren2012_PCCP_9492}
Tellgren,~E.~I.; Reine,~S.~S.; Helgaker,~T. {Analytical GIAO and hybrid-basis
  integral derivatives: application to geometry optimization of molecules in
  strong magnetic fields}. \emph{Phys. Chem. Chem. Phys.} \textbf{2012},
  \emph{14}, 9492\relax
\mciteBstWouldAddEndPuncttrue
\mciteSetBstMidEndSepPunct{\mcitedefaultmidpunct}
{\mcitedefaultendpunct}{\mcitedefaultseppunct}\relax
\EndOfBibitem
\bibitem[Furness \latin{et~al.}(2015)Furness, Verbeke, Tellgren, Stopkowicz,
  Ekstr{\"{o}}m, Helgaker, and Teale]{Furness2015_JCTC_4169}
Furness,~J.~W.; Verbeke,~J.; Tellgren,~E.~I.; Stopkowicz,~S.;
  Ekstr{\"{o}}m,~U.; Helgaker,~T.; Teale,~A.~M. {Current Density Functional
  Theory Using Meta-Generalized Gradient Exchange-Correlation Functionals}.
  \emph{J. Chem. Theory Comput.} \textbf{2015}, \emph{11}, 4169--4181\relax
\mciteBstWouldAddEndPuncttrue
\mciteSetBstMidEndSepPunct{\mcitedefaultmidpunct}
{\mcitedefaultendpunct}{\mcitedefaultseppunct}\relax
\EndOfBibitem
\bibitem[Reynolds and Shiozaki(2015)Reynolds, and
  Shiozaki]{Reynolds2015_PCCP_14280}
Reynolds,~R.~D.; Shiozaki,~T. {Fully relativistic self-consistent field under a
  magnetic field}. \emph{Phys. Chem. Chem. Phys.} \textbf{2015}, \emph{17},
  14280--14283\relax
\mciteBstWouldAddEndPuncttrue
\mciteSetBstMidEndSepPunct{\mcitedefaultmidpunct}
{\mcitedefaultendpunct}{\mcitedefaultseppunct}\relax
\EndOfBibitem
\bibitem[Irons \latin{et~al.}(2017)Irons, Zemen, and
  Teale]{Irons2017_JCTC_3636}
Irons,~T. J.~P.; Zemen,~J.; Teale,~A.~M. Efficient Calculation of Molecular
  Integrals over {London} Atomic Orbitals. \emph{J. Chem. Theory Comput.}
  \textbf{2017}, \emph{13}, 3636--3649\relax
\mciteBstWouldAddEndPuncttrue
\mciteSetBstMidEndSepPunct{\mcitedefaultmidpunct}
{\mcitedefaultendpunct}{\mcitedefaultseppunct}\relax
\EndOfBibitem
\bibitem[Reimann \latin{et~al.}(2017)Reimann, Borgoo, Tellgren, Teale, and
  Helgaker]{Reimann2017_JCTC_4089}
Reimann,~S.; Borgoo,~A.; Tellgren,~E.~I.; Teale,~A.~M.; Helgaker,~T.
  {Magnetic-Field Density-Functional Theory (BDFT): Lessons from the Adiabatic
  Connection}. \emph{J. Chem. Theory Comput.} \textbf{2017}, \emph{13},
  4089--4100\relax
\mciteBstWouldAddEndPuncttrue
\mciteSetBstMidEndSepPunct{\mcitedefaultmidpunct}
{\mcitedefaultendpunct}{\mcitedefaultseppunct}\relax
\EndOfBibitem
\bibitem[Hampe and Stopkowicz(2017)Hampe, and Stopkowicz]{Hampe2017_JCP_154105}
Hampe,~F.; Stopkowicz,~S. {Equation-of-motion coupled-cluster methods for atoms
  and molecules in strong magnetic fields}. \emph{J. Chem. Phys.}
  \textbf{2017}, \emph{146}, 154105\relax
\mciteBstWouldAddEndPuncttrue
\mciteSetBstMidEndSepPunct{\mcitedefaultmidpunct}
{\mcitedefaultendpunct}{\mcitedefaultseppunct}\relax
\EndOfBibitem
\bibitem[Reynolds \latin{et~al.}(2018)Reynolds, Yanai, and
  Shiozaki]{Reynolds2018_JCP_14106}
Reynolds,~R.~D.; Yanai,~T.; Shiozaki,~T. {Large-scale relativistic complete
  active space self-consistent field with robust convergence}. \emph{J. Chem.
  Phys.} \textbf{2018}, \emph{149}, 014106\relax
\mciteBstWouldAddEndPuncttrue
\mciteSetBstMidEndSepPunct{\mcitedefaultmidpunct}
{\mcitedefaultendpunct}{\mcitedefaultseppunct}\relax
\EndOfBibitem
\bibitem[Sen and Tellgren(2018)Sen, and Tellgren]{Sen2018_JCP_184112}
Sen,~S.; Tellgren,~E.~I. {Non-perturbative calculation of orbital and spin
  effects in molecules subject to non-uniform magnetic fields}. \emph{J. Chem.
  Phys.} \textbf{2018}, \emph{148}, 184112\relax
\mciteBstWouldAddEndPuncttrue
\mciteSetBstMidEndSepPunct{\mcitedefaultmidpunct}
{\mcitedefaultendpunct}{\mcitedefaultseppunct}\relax
\EndOfBibitem
\bibitem[Sun \latin{et~al.}(2018)Sun, Williams-Young, Stetina, and
  Li]{Sun2018_JCTC_348}
Sun,~S.; Williams-Young,~D.~B.; Stetina,~T.~F.; Li,~X. Generalized
  Hartree--Fock with Nonperturbative Treatment of Strong Magnetic Fields:
  Application to Molecular Spin Phase Transitions. \emph{J. Chem. Theory
  Comput.} \textbf{2018}, \emph{15}, 348--356\relax
\mciteBstWouldAddEndPuncttrue
\mciteSetBstMidEndSepPunct{\mcitedefaultmidpunct}
{\mcitedefaultendpunct}{\mcitedefaultseppunct}\relax
\EndOfBibitem
\bibitem[Holzer \latin{et~al.}(2019)Holzer, Teale, Hampe, Stopkowicz, Helgaker,
  and Klopper]{Holzer2019_JCP_214112}
Holzer,~C.; Teale,~A.~M.; Hampe,~F.; Stopkowicz,~S.; Helgaker,~T.; Klopper,~W.
  {GW} quasiparticle energies of atoms in strong magnetic fields. \emph{J.
  Chem. Phys.} \textbf{2019}, \emph{150}, 214112\relax
\mciteBstWouldAddEndPuncttrue
\mciteSetBstMidEndSepPunct{\mcitedefaultmidpunct}
{\mcitedefaultendpunct}{\mcitedefaultseppunct}\relax
\EndOfBibitem
\bibitem[Sun \latin{et~al.}(2019)Sun, Williams-Young, and
  Li]{Sun2019_JCTC_3162}
Sun,~S.; Williams-Young,~D.; Li,~X. An ab Initio Linear Response Method for
  Computing Magnetic Circular Dichroism Spectra with Nonperturbative Treatment
  of Magnetic Field. \emph{J. Chem. Theory Comput.} \textbf{2019}, \emph{15},
  3162--3169\relax
\mciteBstWouldAddEndPuncttrue
\mciteSetBstMidEndSepPunct{\mcitedefaultmidpunct}
{\mcitedefaultendpunct}{\mcitedefaultseppunct}\relax
\EndOfBibitem
\bibitem[Hampe and Stopkowicz(2019)Hampe, and Stopkowicz]{Hampe2019_JCTC_4036}
Hampe,~F.; Stopkowicz,~S. Transition-Dipole Moments for Electronic Excitations
  in Strong Magnetic Fields Using Equation-of-Motion and Linear Response
  Coupled-Cluster Theory. \emph{J. Chem. Theory Comput.} \textbf{2019},
  \emph{15}, 4036--4043\relax
\mciteBstWouldAddEndPuncttrue
\mciteSetBstMidEndSepPunct{\mcitedefaultmidpunct}
{\mcitedefaultendpunct}{\mcitedefaultseppunct}\relax
\EndOfBibitem
\bibitem[Williams-Young \latin{et~al.}(2020)Williams-Young, Petrone, Sun,
  Stetina, Lestrange, Hoyer, Nascimento, Koulias, Wildman, Kasper, Goings,
  Ding, DePrince, Valeev, and Li]{WilliamsYoung2020_WCMS_1436}
Williams-Young,~D.~B.; Petrone,~A.; Sun,~S.; Stetina,~T.~F.; Lestrange,~P.;
  Hoyer,~C.~E.; Nascimento,~D.~R.; Koulias,~L.; Wildman,~A.; Kasper,~J.;
  Goings,~J.~J.; Ding,~F.; DePrince,~A.~E.; Valeev,~E.~F.; Li,~X. The {Chronus}
  {Quantum} software package. \emph{Wiley Interdiscip. Rev. Comput. Mol. Sci.}
  \textbf{2020}, \emph{10}, e1436\relax
\mciteBstWouldAddEndPuncttrue
\mciteSetBstMidEndSepPunct{\mcitedefaultmidpunct}
{\mcitedefaultendpunct}{\mcitedefaultseppunct}\relax
\EndOfBibitem
\bibitem[Holzer \latin{et~al.}(2021)Holzer, Pausch, and
  Klopper]{Holzer2021_FC_746162}
Holzer,~C.; Pausch,~A.; Klopper,~W. The {GW}/{BSE} Method in Magnetic Fields.
  \emph{Front. Chem.} \textbf{2021}, \emph{9}, 746162\relax
\mciteBstWouldAddEndPuncttrue
\mciteSetBstMidEndSepPunct{\mcitedefaultmidpunct}
{\mcitedefaultendpunct}{\mcitedefaultseppunct}\relax
\EndOfBibitem
\bibitem[Pausch \latin{et~al.}(2022)Pausch, Holzer, and
  Klopper]{Pausch2022_JCTC_3747}
Pausch,~A.; Holzer,~C.; Klopper,~W. Efficient Calculation of Magnetic Circular
  Dichroism Spectra Using Spin-Noncollinear Linear-Response Time-Dependent
  Density Functional Theory in Finite Magnetic Fields. \emph{J. Chem. Theory
  Comput.} \textbf{2022}, \emph{18}, 3747--3758\relax
\mciteBstWouldAddEndPuncttrue
\mciteSetBstMidEndSepPunct{\mcitedefaultmidpunct}
{\mcitedefaultendpunct}{\mcitedefaultseppunct}\relax
\EndOfBibitem
\bibitem[Pausch and Holzer(2022)Pausch, and Holzer]{Pausch2022_JPCL_4335}
Pausch,~A.; Holzer,~C. Linear Response of Current-Dependent Density Functional
  Approximations in Magnetic Fields. \emph{J. Phys. Chem. Lett.} \textbf{2022},
  \emph{13}, 4335--4341\relax
\mciteBstWouldAddEndPuncttrue
\mciteSetBstMidEndSepPunct{\mcitedefaultmidpunct}
{\mcitedefaultendpunct}{\mcitedefaultseppunct}\relax
\EndOfBibitem
\bibitem[Monzel \latin{et~al.}(2022)Monzel, Pausch, Peters, Tellgren, Helgaker,
  and Klopper]{Monzel2022_JCP_54106}
Monzel,~L.; Pausch,~A.; Peters,~L. D.~M.; Tellgren,~E.~I.; Helgaker,~T.;
  Klopper,~W. Molecular dynamics of linear molecules in strong magnetic fields.
  \emph{J. Chem. Phys.} \textbf{2022}, \emph{157}, 054106\relax
\mciteBstWouldAddEndPuncttrue
\mciteSetBstMidEndSepPunct{\mcitedefaultmidpunct}
{\mcitedefaultendpunct}{\mcitedefaultseppunct}\relax
\EndOfBibitem
\bibitem[Blaschke and Stopkowicz(2022)Blaschke, and
  Stopkowicz]{Blaschke2022_JCP_44115}
Blaschke,~S.; Stopkowicz,~S. Cholesky decomposition of complex two-electron
  integrals over {GIAOs}: Efficient {MP}2 computations for large molecules in
  strong magnetic fields. \emph{J. Chem. Phys.} \textbf{2022}, \emph{156},
  044115\relax
\mciteBstWouldAddEndPuncttrue
\mciteSetBstMidEndSepPunct{\mcitedefaultmidpunct}
{\mcitedefaultendpunct}{\mcitedefaultseppunct}\relax
\EndOfBibitem
\bibitem[Speake \latin{et~al.}(2022)Speake, Irons, Wibowo, Johnson, David, and
  Teale]{Speake2022_JCTC_7412}
Speake,~B.~T.; Irons,~T. J.~P.; Wibowo,~M.; Johnson,~A.~G.; David,~G.;
  Teale,~A.~M. An Embedded Fragment Method for Molecules in Strong Magnetic
  Fields. \emph{J. Chem. Theory Comput.} \textbf{2022}, \emph{18},
  7412--7427\relax
\mciteBstWouldAddEndPuncttrue
\mciteSetBstMidEndSepPunct{\mcitedefaultmidpunct}
{\mcitedefaultendpunct}{\mcitedefaultseppunct}\relax
\EndOfBibitem
\bibitem[Culpitt \latin{et~al.}(2023)Culpitt, Peters, Tellgren, and
  Helgaker]{Culpitt2023_JCP_114115}
Culpitt,~T.; Peters,~L. D.~M.; Tellgren,~E.~I.; Helgaker,~T. Time-dependent
  nuclear-electronic orbital {Hartree}--{Fock} theory in a strong uniform
  magnetic field. \emph{J. Chem. Phys.} \textbf{2023}, \emph{158}, 114115\relax
\mciteBstWouldAddEndPuncttrue
\mciteSetBstMidEndSepPunct{\mcitedefaultmidpunct}
{\mcitedefaultendpunct}{\mcitedefaultseppunct}\relax
\EndOfBibitem
\bibitem[Holzer(2023)]{Holzer2023_JCTC_3131}
Holzer,~C. Practical Post-{Kohn}--{Sham} Methods for Time-Reversal Symmetry
  Breaking References. \emph{J. Chem. Theory Comput.} \textbf{2023}, \emph{19},
  3131--3145\relax
\mciteBstWouldAddEndPuncttrue
\mciteSetBstMidEndSepPunct{\mcitedefaultmidpunct}
{\mcitedefaultendpunct}{\mcitedefaultseppunct}\relax
\EndOfBibitem
\bibitem[Greenstein(1984)]{Greenstein1984_AJ_47}
Greenstein,~J.~L. The identification of hydrogen in {GRW} $+70^\circ8247$.
  \emph{Astrophys. J.} \textbf{1984}, \emph{281}, L47\relax
\mciteBstWouldAddEndPuncttrue
\mciteSetBstMidEndSepPunct{\mcitedefaultmidpunct}
{\mcitedefaultendpunct}{\mcitedefaultseppunct}\relax
\EndOfBibitem
\bibitem[Greenstein \latin{et~al.}(1985)Greenstein, Henry, and
  Oconnell]{Greenstein1985_AJ_25}
Greenstein,~J.~L.; Henry,~R. J.~W.; Oconnell,~R.~F. Futher identifications of
  hydrogen in {GRW} $+70^\circ8247$. \emph{Astrophys. J.} \textbf{1985},
  \emph{289}, L25\relax
\mciteBstWouldAddEndPuncttrue
\mciteSetBstMidEndSepPunct{\mcitedefaultmidpunct}
{\mcitedefaultendpunct}{\mcitedefaultseppunct}\relax
\EndOfBibitem
\bibitem[Hollands \latin{et~al.}(2023)Hollands, Stopkowicz, Kitsaras, Hampe,
  Blaschke, and Hermes]{Hollands2023_MNRAS_3560}
Hollands,~M.~A.; Stopkowicz,~S.; Kitsaras,~M.-P.; Hampe,~F.; Blaschke,~S.;
  Hermes,~J.~J. A {DZ} white dwarf with a 30 {MG} magnetic field. \emph{Mon.
  Not. R. Astron. Soc.} \textbf{2023}, \emph{520}, 3560--3575\relax
\mciteBstWouldAddEndPuncttrue
\mciteSetBstMidEndSepPunct{\mcitedefaultmidpunct}
{\mcitedefaultendpunct}{\mcitedefaultseppunct}\relax
\EndOfBibitem
\bibitem[Lange \latin{et~al.}(2012)Lange, Tellgren, Hoffmann, and
  Helgaker]{Lange2012_S_327}
Lange,~K.~K.; Tellgren,~E.~I.; Hoffmann,~M.~R.; Helgaker,~T. A Paramagnetic
  Bonding Mechanism for Diatomics in Strong Magnetic Fields. \emph{Science}
  \textbf{2012}, \emph{337}, 327--331\relax
\mciteBstWouldAddEndPuncttrue
\mciteSetBstMidEndSepPunct{\mcitedefaultmidpunct}
{\mcitedefaultendpunct}{\mcitedefaultseppunct}\relax
\EndOfBibitem
\bibitem[Davidson and Feller(1986)Davidson, and Feller]{Davidson1986_CR_681}
Davidson,~E.~R.; Feller,~D. {Basis set selection for molecular calculations}.
  \emph{Chem. Rev.} \textbf{1986}, \emph{86}, 681--696\relax
\mciteBstWouldAddEndPuncttrue
\mciteSetBstMidEndSepPunct{\mcitedefaultmidpunct}
{\mcitedefaultendpunct}{\mcitedefaultseppunct}\relax
\EndOfBibitem
\bibitem[Jensen(2013)]{Jensen2013_WIRCMS_273}
Jensen,~F. Atomic orbital basis sets. \emph{Wiley Interdiscip. Rev. Comput.
  Mol. Sci.} \textbf{2013}, \emph{3}, 273--295\relax
\mciteBstWouldAddEndPuncttrue
\mciteSetBstMidEndSepPunct{\mcitedefaultmidpunct}
{\mcitedefaultendpunct}{\mcitedefaultseppunct}\relax
\EndOfBibitem
\bibitem[Hill(2013)]{Hill2013_IJQC_21}
Hill,~J.~G. {Gaussian basis sets for molecular applications}. \emph{Int. J.
  Quantum Chem.} \textbf{2013}, \emph{113}, 21--34\relax
\mciteBstWouldAddEndPuncttrue
\mciteSetBstMidEndSepPunct{\mcitedefaultmidpunct}
{\mcitedefaultendpunct}{\mcitedefaultseppunct}\relax
\EndOfBibitem
\bibitem[Jones \latin{et~al.}(1999)Jones, Ortiz, and
  Ceperley]{Jones1999_PRA_2875}
Jones,~M.~D.; Ortiz,~G.; Ceperley,~D.~M. {Spectrum of neutral helium in strong
  magnetic fields}. \emph{Phys. Rev. A} \textbf{1999}, \emph{59},
  2875--2885\relax
\mciteBstWouldAddEndPuncttrue
\mciteSetBstMidEndSepPunct{\mcitedefaultmidpunct}
{\mcitedefaultendpunct}{\mcitedefaultseppunct}\relax
\EndOfBibitem
\bibitem[Hampe \latin{et~al.}(2020)Hampe, Gross, and
  Stopkowicz]{Hampe2020_PCCP_23522}
Hampe,~F.; Gross,~N.; Stopkowicz,~S. Full triples contribution in
  coupled-cluster and equation-of-motion coupled-cluster methods for atoms and
  molecules in strong magnetic fields. \emph{Phys. Chem. Chem. Phys.}
  \textbf{2020}, \emph{22}, 23522--23529\relax
\mciteBstWouldAddEndPuncttrue
\mciteSetBstMidEndSepPunct{\mcitedefaultmidpunct}
{\mcitedefaultendpunct}{\mcitedefaultseppunct}\relax
\EndOfBibitem
\bibitem[Stopkowicz \latin{et~al.}(2015)Stopkowicz, Gauss, Lange, Tellgren, and
  Helgaker]{Stopkowicz2015_JCP_74110}
Stopkowicz,~S.; Gauss,~J.; Lange,~K.~K.; Tellgren,~E.~I.; Helgaker,~T.
  {Coupled-cluster theory for atoms and molecules in strong magnetic fields}.
  \emph{J. Chem. Phys.} \textbf{2015}, \emph{143}, 074110\relax
\mciteBstWouldAddEndPuncttrue
\mciteSetBstMidEndSepPunct{\mcitedefaultmidpunct}
{\mcitedefaultendpunct}{\mcitedefaultseppunct}\relax
\EndOfBibitem
\bibitem[Detmer \latin{et~al.}(1997)Detmer, Schmelcher, Diakonos, and
  Cederbaum]{Detmer1997_PRA_1825}
Detmer,~T.; Schmelcher,~P.; Diakonos,~F.~K.; Cederbaum,~L.~S. {Hydrogen
  molecule in magnetic fields: The ground states of the $\Sigma$ manifold of
  the parallel configuration}. \emph{Phys. Rev. A} \textbf{1997}, \emph{56},
  1825--1838\relax
\mciteBstWouldAddEndPuncttrue
\mciteSetBstMidEndSepPunct{\mcitedefaultmidpunct}
{\mcitedefaultendpunct}{\mcitedefaultseppunct}\relax
\EndOfBibitem
\bibitem[Detmer \latin{et~al.}(1998)Detmer, Schmelcher, and
  Cederbaum]{Detmer1998_PRA_1767}
Detmer,~T.; Schmelcher,~P.; Cederbaum,~L.~S. {Hydrogen molecule in a magnetic
  field: The lowest states of the $\Pi$ manifold and the global ground state of
  the parallel configuration}. \emph{Phys. Rev. A} \textbf{1998}, \emph{57},
  1767--1777\relax
\mciteBstWouldAddEndPuncttrue
\mciteSetBstMidEndSepPunct{\mcitedefaultmidpunct}
{\mcitedefaultendpunct}{\mcitedefaultseppunct}\relax
\EndOfBibitem
\bibitem[Schmelcher \latin{et~al.}(1999)Schmelcher, Ivanov, and
  Becken]{Schmelcher1999_PRA_3424}
Schmelcher,~P.; Ivanov,~M.~V.; Becken,~W. Exchange and correlation energies of
  ground states of atoms and molecules in strong magnetic fields. \emph{Phys.
  Rev. A} \textbf{1999}, \emph{59}, 3424--3431\relax
\mciteBstWouldAddEndPuncttrue
\mciteSetBstMidEndSepPunct{\mcitedefaultmidpunct}
{\mcitedefaultendpunct}{\mcitedefaultseppunct}\relax
\EndOfBibitem
\bibitem[Becken \latin{et~al.}(1999)Becken, Schmelcher, and
  Diakonos]{Becken1999_JPBAMOP_1557}
Becken,~W.; Schmelcher,~P.; Diakonos,~F.~K. {The helium atom in a strong
  magnetic field}. \emph{J. Phys. B At. Mol. Opt. Phys.} \textbf{1999},
  \emph{32}, 1557--1584\relax
\mciteBstWouldAddEndPuncttrue
\mciteSetBstMidEndSepPunct{\mcitedefaultmidpunct}
{\mcitedefaultendpunct}{\mcitedefaultseppunct}\relax
\EndOfBibitem
\bibitem[Becken and Schmelcher(2000)Becken, and
  Schmelcher]{Becken2000_JPBAMOP_545}
Becken,~W.; Schmelcher,~P. {Non-zero angular momentum states of the helium atom
  in a strong magnetic field}. \emph{J. Phys. B At. Mol. Opt. Phys.}
  \textbf{2000}, \emph{33}, 545--568\relax
\mciteBstWouldAddEndPuncttrue
\mciteSetBstMidEndSepPunct{\mcitedefaultmidpunct}
{\mcitedefaultendpunct}{\mcitedefaultseppunct}\relax
\EndOfBibitem
\bibitem[Becken and Schmelcher(2001)Becken, and
  Schmelcher]{Becken2001_PRA_53412}
Becken,~W.; Schmelcher,~P. {Higher-angular-momentum states of the helium atom
  in a strong magnetic field}. \emph{Phys. Rev. A} \textbf{2001}, \emph{63},
  053412\relax
\mciteBstWouldAddEndPuncttrue
\mciteSetBstMidEndSepPunct{\mcitedefaultmidpunct}
{\mcitedefaultendpunct}{\mcitedefaultseppunct}\relax
\EndOfBibitem
\bibitem[Al-Hujaj and Schmelcher(2004)Al-Hujaj, and
  Schmelcher]{AlHujaj2004_PRA_33411}
Al-Hujaj,~O.-A.; Schmelcher,~P. {Lithium in strong magnetic fields}.
  \emph{Phys. Rev. A} \textbf{2004}, \emph{70}, 033411\relax
\mciteBstWouldAddEndPuncttrue
\mciteSetBstMidEndSepPunct{\mcitedefaultmidpunct}
{\mcitedefaultendpunct}{\mcitedefaultseppunct}\relax
\EndOfBibitem
\bibitem[Al-Hujaj and Schmelcher(2004)Al-Hujaj, and
  Schmelcher]{AlHujaj2004_PRA_23411}
Al-Hujaj,~O.-A.; Schmelcher,~P. {Beryllium in strong magnetic fields}.
  \emph{Phys. Rev. A} \textbf{2004}, \emph{70}, 023411\relax
\mciteBstWouldAddEndPuncttrue
\mciteSetBstMidEndSepPunct{\mcitedefaultmidpunct}
{\mcitedefaultendpunct}{\mcitedefaultseppunct}\relax
\EndOfBibitem
\bibitem[Lehtola \latin{et~al.}(2020)Lehtola, Dimitrova, and
  Sundholm]{Lehtola2020_MP_1597989}
Lehtola,~S.; Dimitrova,~M.; Sundholm,~D. Fully numerical electronic structure
  calculations on diatomic molecules in weak to strong magnetic fields.
  \emph{Mol. Phys.} \textbf{2020}, \emph{118}, e1597989\relax
\mciteBstWouldAddEndPuncttrue
\mciteSetBstMidEndSepPunct{\mcitedefaultmidpunct}
{\mcitedefaultendpunct}{\mcitedefaultseppunct}\relax
\EndOfBibitem
\bibitem[Aldrich and Greene(1979)Aldrich, and Greene]{Aldrich1979_PSS_343}
Aldrich,~C.; Greene,~R.~L. Hydrogen-Like Systems in Arbitrary Magnetic
  Fields---A Variational Approach-. \emph{Phys. Status Solidi} \textbf{1979},
  \emph{93}, 343--350\relax
\mciteBstWouldAddEndPuncttrue
\mciteSetBstMidEndSepPunct{\mcitedefaultmidpunct}
{\mcitedefaultendpunct}{\mcitedefaultseppunct}\relax
\EndOfBibitem
\bibitem[Schmelcher and Cederbaum(1988)Schmelcher, and
  Cederbaum]{Schmelcher1988_PRA_672}
Schmelcher,~P.; Cederbaum,~L.~S. {Molecules in strong magnetic fields:
  Properties of atomic orbitals}. \emph{Phys. Rev. A} \textbf{1988}, \emph{37},
  672--681\relax
\mciteBstWouldAddEndPuncttrue
\mciteSetBstMidEndSepPunct{\mcitedefaultmidpunct}
{\mcitedefaultendpunct}{\mcitedefaultseppunct}\relax
\EndOfBibitem
\bibitem[Kubo(2007)]{Kubo2007_JPCA_5572}
Kubo,~A. {The Hydrogen Molecule in Strong Magnetic Fields: Optimizations of
  Anisotropic Gaussian Basis Sets}. \emph{J. Phys. Chem. A} \textbf{2007},
  \emph{111}, 5572--5581\relax
\mciteBstWouldAddEndPuncttrue
\mciteSetBstMidEndSepPunct{\mcitedefaultmidpunct}
{\mcitedefaultendpunct}{\mcitedefaultseppunct}\relax
\EndOfBibitem
\bibitem[Zhu \latin{et~al.}(2014)Zhu, Zhang, and Trickey]{Zhu2014_PRA_22504}
Zhu,~W.; Zhang,~L.; Trickey,~S.~B. {Comparative studies of density-functional
  approximations for light atoms in strong magnetic fields}. \emph{Phys. Rev.
  A} \textbf{2014}, \emph{90}, 022504\relax
\mciteBstWouldAddEndPuncttrue
\mciteSetBstMidEndSepPunct{\mcitedefaultmidpunct}
{\mcitedefaultendpunct}{\mcitedefaultseppunct}\relax
\EndOfBibitem
\bibitem[Zhu and Trickey(2017)Zhu, and Trickey]{Zhu2017_JCP_244108}
Zhu,~W.; Trickey,~S.~B. {Accurate and balanced anisotropic Gaussian type
  orbital basis sets for atoms in strong magnetic fields}. \emph{J. Chem.
  Phys.} \textbf{2017}, \emph{147}, 244108\relax
\mciteBstWouldAddEndPuncttrue
\mciteSetBstMidEndSepPunct{\mcitedefaultmidpunct}
{\mcitedefaultendpunct}{\mcitedefaultseppunct}\relax
\EndOfBibitem
\bibitem[Lehtola(2019)]{Lehtola2019_IJQC_25945}
Lehtola,~S. Fully numerical {Hartree}--{Fock} and density functional
  calculations. {I}. {Atoms}. \emph{Int. J. Quantum Chem.} \textbf{2019},
  \emph{119}, e25945\relax
\mciteBstWouldAddEndPuncttrue
\mciteSetBstMidEndSepPunct{\mcitedefaultmidpunct}
{\mcitedefaultendpunct}{\mcitedefaultseppunct}\relax
\EndOfBibitem
\bibitem[Lehtola(2019)]{Lehtola2019_IJQC_25968}
Lehtola,~S. A review on non-relativistic, fully numerical electronic structure
  calculations on atoms and diatomic molecules. \emph{Int. J. Quantum Chem.}
  \textbf{2019}, \emph{119}, e25968\relax
\mciteBstWouldAddEndPuncttrue
\mciteSetBstMidEndSepPunct{\mcitedefaultmidpunct}
{\mcitedefaultendpunct}{\mcitedefaultseppunct}\relax
\EndOfBibitem
\bibitem[Ivanov and Schmelcher(1998)Ivanov, and
  Schmelcher]{Ivanov1998_PRA_3793}
Ivanov,~M.~V.; Schmelcher,~P. Ground state of the lithium atom in strong
  magnetic fields. \emph{Phys. Rev. A} \textbf{1998}, \emph{57},
  3793--3800\relax
\mciteBstWouldAddEndPuncttrue
\mciteSetBstMidEndSepPunct{\mcitedefaultmidpunct}
{\mcitedefaultendpunct}{\mcitedefaultseppunct}\relax
\EndOfBibitem
\bibitem[Ivanov and Schmelcher(1999)Ivanov, and
  Schmelcher]{Ivanov1999_PRA_3558}
Ivanov,~M.~V.; Schmelcher,~P. Ground state of the carbon atom in strong
  magnetic fields. \emph{Phys. Rev. A} \textbf{1999}, \emph{60},
  3558--3568\relax
\mciteBstWouldAddEndPuncttrue
\mciteSetBstMidEndSepPunct{\mcitedefaultmidpunct}
{\mcitedefaultendpunct}{\mcitedefaultseppunct}\relax
\EndOfBibitem
\bibitem[Ivanov and Schmelcher(2001)Ivanov, and
  Schmelcher]{Ivanov2001_JPBAMOP_2031}
Ivanov,~M.~V.; Schmelcher,~P. The boron atom and boron positive ion in strong
  magnetic fields. \emph{J. Phys. B: At., Mol. Opt. Phys.} \textbf{2001},
  \emph{34}, 2031--2044\relax
\mciteBstWouldAddEndPuncttrue
\mciteSetBstMidEndSepPunct{\mcitedefaultmidpunct}
{\mcitedefaultendpunct}{\mcitedefaultseppunct}\relax
\EndOfBibitem
\bibitem[Ivanov and Schmelcher(2001)Ivanov, and
  Schmelcher]{Ivanov2001_EPJD_279}
Ivanov,~M.~V.; Schmelcher,~P. The beryllium atom and beryllium positive ion in
  strong magnetic fields. \emph{Eur. Phys. J. D} \textbf{2001}, \emph{14},
  279--288\relax
\mciteBstWouldAddEndPuncttrue
\mciteSetBstMidEndSepPunct{\mcitedefaultmidpunct}
{\mcitedefaultendpunct}{\mcitedefaultseppunct}\relax
\EndOfBibitem
\bibitem[Ivanov and Schmelcher(2000)Ivanov, and
  Schmelcher]{Ivanov2000_PRA_22505}
Ivanov,~M.~V.; Schmelcher,~P. Ground states of {H}, {He}, {\ldots}, {Ne}, and
  their singly positive ions in strong magnetic fields: The high-field regime.
  \emph{Phys. Rev. A} \textbf{2000}, \emph{61}, 022505\relax
\mciteBstWouldAddEndPuncttrue
\mciteSetBstMidEndSepPunct{\mcitedefaultmidpunct}
{\mcitedefaultendpunct}{\mcitedefaultseppunct}\relax
\EndOfBibitem
\bibitem[Ivanov and Schmelcher(2001)Ivanov, and Schmelcher]{Ivanov2001_AQC_361}
Ivanov,~M.~V.; Schmelcher,~P. Finite-difference calculations for atoms and
  diatomic molecules in strong magnetic and static electric fields. \emph{Adv.
  Quantum Chem.} \textbf{2001}, \emph{40}, 361--379\relax
\mciteBstWouldAddEndPuncttrue
\mciteSetBstMidEndSepPunct{\mcitedefaultmidpunct}
{\mcitedefaultendpunct}{\mcitedefaultseppunct}\relax
\EndOfBibitem
\bibitem[Lehtola(2020)]{Lehtola2020_JCP_134108}
Lehtola,~S. Polarized {Gaussian} basis sets from one-electron ions. \emph{J.
  Chem. Phys.} \textbf{2020}, \emph{152}, 134108\relax
\mciteBstWouldAddEndPuncttrue
\mciteSetBstMidEndSepPunct{\mcitedefaultmidpunct}
{\mcitedefaultendpunct}{\mcitedefaultseppunct}\relax
\EndOfBibitem
\bibitem[London(1937)]{London1937_JPlR_397}
London,~F. {Th{\'{e}}orie quantique des courants interatomiques dans les
  combinaisons aromatiques}. \emph{J. Phys. le Radium} \textbf{1937}, \emph{8},
  397--409\relax
\mciteBstWouldAddEndPuncttrue
\mciteSetBstMidEndSepPunct{\mcitedefaultmidpunct}
{\mcitedefaultendpunct}{\mcitedefaultseppunct}\relax
\EndOfBibitem
\bibitem[Pople(1962)]{Pople1962_JCP_53}
Pople,~J.~A. Molecular-Orbital Theory of Diamagnetism. {I}. An Approximate
  {LCAO} Scheme. \emph{J. Chem. Phys.} \textbf{1962}, \emph{37}, 53--59\relax
\mciteBstWouldAddEndPuncttrue
\mciteSetBstMidEndSepPunct{\mcitedefaultmidpunct}
{\mcitedefaultendpunct}{\mcitedefaultseppunct}\relax
\EndOfBibitem
\bibitem[Ditchfield(1974)]{Ditchfield1974_MP_789}
Ditchfield,~R. {Self-consistent perturbation theory of diamagnetism. I. A
  gauge-invariant LCAO method for N.M.R. chemical shifts}. \emph{Mol. Phys.}
  \textbf{1974}, \emph{27}, 789--807\relax
\mciteBstWouldAddEndPuncttrue
\mciteSetBstMidEndSepPunct{\mcitedefaultmidpunct}
{\mcitedefaultendpunct}{\mcitedefaultseppunct}\relax
\EndOfBibitem
\bibitem[Lehtola(2019)]{Lehtola2019_IJQC_25944}
Lehtola,~S. Fully numerical {Hartree}--{Fock} and density functional
  calculations. {II}. {Diatomic} molecules. \emph{Int. J. Quantum Chem.}
  \textbf{2019}, \emph{119}, e25944\relax
\mciteBstWouldAddEndPuncttrue
\mciteSetBstMidEndSepPunct{\mcitedefaultmidpunct}
{\mcitedefaultendpunct}{\mcitedefaultseppunct}\relax
\EndOfBibitem
\bibitem[Lehtola(2023)]{Lehtola2018__}
Lehtola,~S. {HelFEM -- Finite element methods for electronic structure
  calculations on small systems}. 2023;
  \url{http://github.com/susilehtola/HelFEM}, Accessed 26 March 2023.\relax
\mciteBstWouldAddEndPunctfalse
\mciteSetBstMidEndSepPunct{\mcitedefaultmidpunct}
{}{\mcitedefaultseppunct}\relax
\EndOfBibitem
\bibitem[Lehtola(2019)]{Lehtola2019_JCTC_1593}
Lehtola,~S. Assessment of Initial Guesses for Self-Consistent Field
  Calculations. Superposition of Atomic Potentials: Simple yet Efficient.
  \emph{J. Chem. Theory Comput.} \textbf{2019}, \emph{15}, 1593--1604\relax
\mciteBstWouldAddEndPuncttrue
\mciteSetBstMidEndSepPunct{\mcitedefaultmidpunct}
{\mcitedefaultendpunct}{\mcitedefaultseppunct}\relax
\EndOfBibitem
\bibitem[Lehtola \latin{et~al.}(2020)Lehtola, Visscher, and
  Engel]{Lehtola2020_JCP_144105}
Lehtola,~S.; Visscher,~L.; Engel,~E. Efficient implementation of the
  superposition of atomic potentials initial guess for electronic structure
  calculations in {Gaussian} basis sets. \emph{J. Chem. Phys.} \textbf{2020},
  \emph{152}, 144105\relax
\mciteBstWouldAddEndPuncttrue
\mciteSetBstMidEndSepPunct{\mcitedefaultmidpunct}
{\mcitedefaultendpunct}{\mcitedefaultseppunct}\relax
\EndOfBibitem
\bibitem[Lehtola \latin{et~al.}(2012)Lehtola, Hakala, Sakko, and
  H{\"{a}}m{\"{a}}l{\"{a}}inen]{Lehtola2012_JCC_1572}
Lehtola,~J.; Hakala,~M.; Sakko,~A.; H{\"{a}}m{\"{a}}l{\"{a}}inen,~K. {ERKALE}
  -- A flexible program package for X-ray properties of atoms and molecules.
  \emph{J. Comput. Chem.} \textbf{2012}, \emph{33}, 1572--1585\relax
\mciteBstWouldAddEndPuncttrue
\mciteSetBstMidEndSepPunct{\mcitedefaultmidpunct}
{\mcitedefaultendpunct}{\mcitedefaultseppunct}\relax
\EndOfBibitem
\bibitem[Dunning(1989)]{Dunning1989_JCP_1007}
Dunning,~T.~H. {Gaussian basis sets for use in correlated molecular
  calculations. I. The atoms boron through neon and hydrogen}. \emph{J. Chem.
  Phys.} \textbf{1989}, \emph{90}, 1007\relax
\mciteBstWouldAddEndPuncttrue
\mciteSetBstMidEndSepPunct{\mcitedefaultmidpunct}
{\mcitedefaultendpunct}{\mcitedefaultseppunct}\relax
\EndOfBibitem
\bibitem[Kendall \latin{et~al.}(1992)Kendall, Dunning, and
  Harrison]{Kendall1992_JCP_6796}
Kendall,~R.~A.; Dunning,~T.~H.; Harrison,~R.~J. {Electron affinities of the
  first-row atoms revisited. Systematic basis sets and wave functions}.
  \emph{J. Chem. Phys.} \textbf{1992}, \emph{96}, 6796\relax
\mciteBstWouldAddEndPuncttrue
\mciteSetBstMidEndSepPunct{\mcitedefaultmidpunct}
{\mcitedefaultendpunct}{\mcitedefaultseppunct}\relax
\EndOfBibitem
\bibitem[Woon and Dunning(1993)Woon, and Dunning]{Woon1993_JCP_1358}
Woon,~D.~E.; Dunning,~T.~H. {Gaussian basis sets for use in correlated
  molecular calculations. III. The atoms aluminum through argon}. \emph{J.
  Chem. Phys.} \textbf{1993}, \emph{98}, 1358\relax
\mciteBstWouldAddEndPuncttrue
\mciteSetBstMidEndSepPunct{\mcitedefaultmidpunct}
{\mcitedefaultendpunct}{\mcitedefaultseppunct}\relax
\EndOfBibitem
\bibitem[Peterson \latin{et~al.}(1994)Peterson, Woon, and
  Dunning]{Peterson1994_JCP_7410}
Peterson,~K.~A.; Woon,~D.~E.; Dunning,~T.~H. Benchmark calculations with
  correlated molecular wave functions. {IV}. The classical barrier height of
  the \ce{H+H2 <=> H2+H} reaction. \emph{J. Chem. Phys.} \textbf{1994},
  \emph{100}, 7410--7415\relax
\mciteBstWouldAddEndPuncttrue
\mciteSetBstMidEndSepPunct{\mcitedefaultmidpunct}
{\mcitedefaultendpunct}{\mcitedefaultseppunct}\relax
\EndOfBibitem
\bibitem[Weigend and Ahlrichs(2005)Weigend, and Ahlrichs]{Weigend2005_PCCP_305}
Weigend,~F.; Ahlrichs,~R. {Balanced basis sets of split valence, triple zeta
  valence and quadruple zeta valence quality for H to Rn: Design and assessment
  of accuracy.} \emph{Phys. Chem. Chem. Phys.} \textbf{2005}, \emph{7},
  3297--305\relax
\mciteBstWouldAddEndPuncttrue
\mciteSetBstMidEndSepPunct{\mcitedefaultmidpunct}
{\mcitedefaultendpunct}{\mcitedefaultseppunct}\relax
\EndOfBibitem
\bibitem[Krishnan \latin{et~al.}(1980)Krishnan, Binkley, Seeger, and
  Pople]{Krishnan1980_JCP_650}
Krishnan,~R.; Binkley,~J.~S.; Seeger,~R.; Pople,~J.~A. {Self-consistent
  molecular orbital methods. XX. A basis set for correlated wave functions}.
  \emph{J. Chem. Phys.} \textbf{1980}, \emph{72}, 650--654\relax
\mciteBstWouldAddEndPuncttrue
\mciteSetBstMidEndSepPunct{\mcitedefaultmidpunct}
{\mcitedefaultendpunct}{\mcitedefaultseppunct}\relax
\EndOfBibitem
\bibitem[Grev and Schaefer(1989)Grev, and Schaefer]{Grev1989_JCP_7305}
Grev,~R.~S.; Schaefer,~H.~F. 6-311{G} is not of valence triple-zeta quality.
  \emph{J. Chem. Phys.} \textbf{1989}, \emph{91}, 7305--7306\relax
\mciteBstWouldAddEndPuncttrue
\mciteSetBstMidEndSepPunct{\mcitedefaultmidpunct}
{\mcitedefaultendpunct}{\mcitedefaultseppunct}\relax
\EndOfBibitem
\bibitem[Moran \latin{et~al.}(2006)Moran, Simmonett, Leach, Allen, Schleyer,
  and Schaefer]{Moran2006_JACS_9342}
Moran,~D.; Simmonett,~A.~C.; Leach,~F.~E.; Allen,~W.~D.; Schleyer,~P. V.~R.;
  Schaefer,~H.~F. {Popular theoretical methods predict benzene and arenes to be
  nonplanar}. \emph{J. Am. Chem. Soc.} \textbf{2006}, \emph{128},
  9342--9343\relax
\mciteBstWouldAddEndPuncttrue
\mciteSetBstMidEndSepPunct{\mcitedefaultmidpunct}
{\mcitedefaultendpunct}{\mcitedefaultseppunct}\relax
\EndOfBibitem
\bibitem[Lehtola and Karttunen(2022)Lehtola, and
  Karttunen]{Lehtola2022_WIRCMS_1610}
Lehtola,~S.; Karttunen,~A.~J. Free and open source software for computational
  chemistry education. \emph{Wiley Interdiscip. Rev. Comput. Mol. Sci.}
  \textbf{2022}, \emph{12}, e1610\relax
\mciteBstWouldAddEndPuncttrue
\mciteSetBstMidEndSepPunct{\mcitedefaultmidpunct}
{\mcitedefaultendpunct}{\mcitedefaultseppunct}\relax
\EndOfBibitem
\bibitem[Reimann \latin{et~al.}(2019)Reimann, Borgoo, Austad, Tellgren, Teale,
  Helgaker, and Stopkowicz]{Reimann2019_MP_97}
Reimann,~S.; Borgoo,~A.; Austad,~J.; Tellgren,~E.~I.; Teale,~A.~M.;
  Helgaker,~T.; Stopkowicz,~S. {Kohn--Sham energy decomposition for molecules
  in a magnetic field}. \emph{Mol. Phys.} \textbf{2019}, \emph{117},
  97--109\relax
\mciteBstWouldAddEndPuncttrue
\mciteSetBstMidEndSepPunct{\mcitedefaultmidpunct}
{\mcitedefaultendpunct}{\mcitedefaultseppunct}\relax
\EndOfBibitem
\end{mcitethebibliography}

\end{document}